\documentclass[useAMS,usenatbib]{mn2e}
\usepackage{graphicx}
\usepackage{rotating,times,pictex,graphicx,latexsym}
\usepackage{color}
\usepackage{longtable,amsmath}
\usepackage{lscape}

\title{A large scale extinction map of the Galactic Anticenter from 2MASS} 

\author[Froebrich, Murphy, Smith \& Walsh]{D.~Froebrich$^{1,3}$\thanks{E-mail:
df@star.kent.ac.uk}, G.C.~Murphy$^{2,3}$, M.D.~Smith$^{1}$ and J.~Walsh$^{4}$\\
$^1$ Centre for Astrophysics and Planetary Science, University of Kent,
Canterbury, CT2 7NH, UK \\ $^2$ Laboratoire d'Astrophysique, Observatoire de
Grenoble, BP 53, 38041 Grenoble Cedex 9, France \\ $^3$ Dublin Institute for
Advanced Studies, 5 Merrion Square, Dublin 2, Ireland \\ $^4$ School of Computer
Science and Statistics, Trinity College, College Green, Dublin 2, Ireland} 

\begin{document}

\date{Received sooner; accepted later}
\pagerange{\pageref{firstpage}--\pageref{lastpage}} \pubyear{2007}
\maketitle

\label{firstpage}

\begin{abstract}

We present a $127^\circ \times 63^\circ$ extinction map of the Anticenter of the
Galaxy, based on $\left< J-H \right>$ and $\left< H-K \right>$ colour excess
maps from 2MASS. This 8001 square degree map with a resolution of 4 arcminutes
is provided as online material. The colour excess ratio $\left< J-H \right> /
\left< H-K \right>$ is used to determine the power law index of the reddening
law ($\beta$) for individual regions contained in the area (e.g. Orion, Perseus,
Taurus, Auriga, Monoceros, Camelopardalis, Cassiopeia). On average we find a
dominant value of $\beta$\,=\,1.8\,$\pm$\,0.2 for the individual clouds, in
agreement with the canonical value for the interstellar medium. We also show
that there is an internal scatter of $\beta$ values in these regions, and that
in some areas more than one dominant $\beta$ value is present. This indicates
large scale variations in the dust properties. The analysis of the $A_V$
values within individual regions shows a change in the slope of the column
density distribution with distance. This can either be attributed to a change in
the governing physical processes in molecular clouds on spatial scales of about
1\,pc or an $A_V$ dilution with distance in our map.

\end{abstract}


\section{Introduction}

Understanding the formation of stars is inextricably linked to the formation,
evolution and physical properties of molecular clouds. Dust is one of the best
tracers of the distribution of material in giant molecular clouds. Its
optical properties (reddening and dimming of star light) allow us to determine
line of sight column densities of material. The knowledge of how clouds are
structured (e.g. the distribution of mass, size and column density of clumps
within clouds) can be used to obtain constraints about how clouds fragment and
will eventually lead to a better understanding of the initial mass function
(e.g. Padoan, Nordlund, \& Jones \cite{1997MNRAS.288..145P}). 

The mapping of the dust distribution in giant molecular clouds or entire cloud
complexes becomes increasingly achievable due to the availability of all sky
near infrared surveys such as 2MASS (Skrutskie et al.
\cite{2006AJ....131.1163S}) as well as large computer clusters and Grid
technology. Basic techniques to map the column density of dust, i.e. extinction,
are: 1) star counting (e.g. Wolf \cite{1923AN....219..109W}, Bok 
\cite{1956AJ.....61..309B}, Froebrich et al. \cite{2005A&A...432L..67F}); 2)
colour excess (e.g. Lada et al. \cite{1994ApJ...429..694L}, Lombardi \& Alves
\cite{2001A&A...377.1023L}); 3) or a combination thereof (e.g. Lombardi
\cite{2005A&A...438..169L}); While most simple, the star counting techniques
suffer from enhanced noise compared to the colour excess techniques, when
applied to near infrared data. On the other hand, using colour excess methods to
determine extinction requires the knowledge of the reddening law, which might
vary with position and/or column density (e.g. Froebrich \& del Burgo
\cite{2006MNRAS.369.1901F}). In this paper we apply the most basic colour excess
method from Lada et al. \cite{1994ApJ...429..694L} and determine the reddening
law in the near infrared following Froebrich \& del Burgo
\cite{2006MNRAS.369.1901F}.

A large fraction of nearby giant molecular clouds is part of the Gould Belt.
Towards the Anticenter of the Galaxy we find a variety of these clouds (e.g.
Perseus, Taurus, Orion) in relative close proximity and with a relatively small
amount of contamination from other clouds in the Galactic Plane. Additionally, a
number of more distant cloud complexes (e.g. Cassiopeia, Camelopardalis, Auriga,
Monoceros) are found in the same part of the sky. We hence selected an area of
$127^\circ \times 63^\circ$ (in Galactic Coordinates) towards the Galactic
Anticenter to be mapped in detail and to investigate the column density
distribution in the major cloud complexes situated in this region.

The paper is structured as follows. In Sect.\,\ref{extinction} we describe our
method to determine the extinction map. A description and discussion of the
determination of the reddening law is given in Sect.\,\ref{reddening}. Finally
we study the distribution of column densities in Sect.\,\ref{clumps}, followed
by the conclusions in Sect.\,\ref{conclusion}.

\begin{figure*}
\beginpicture
\setcoordinatesystem units <-1.25984252mm,1.25984252mm> point at 179.5 0
\setplotarea x from 243 to 116 , y from -32 to  31
\put {\includegraphics[width=16cm]{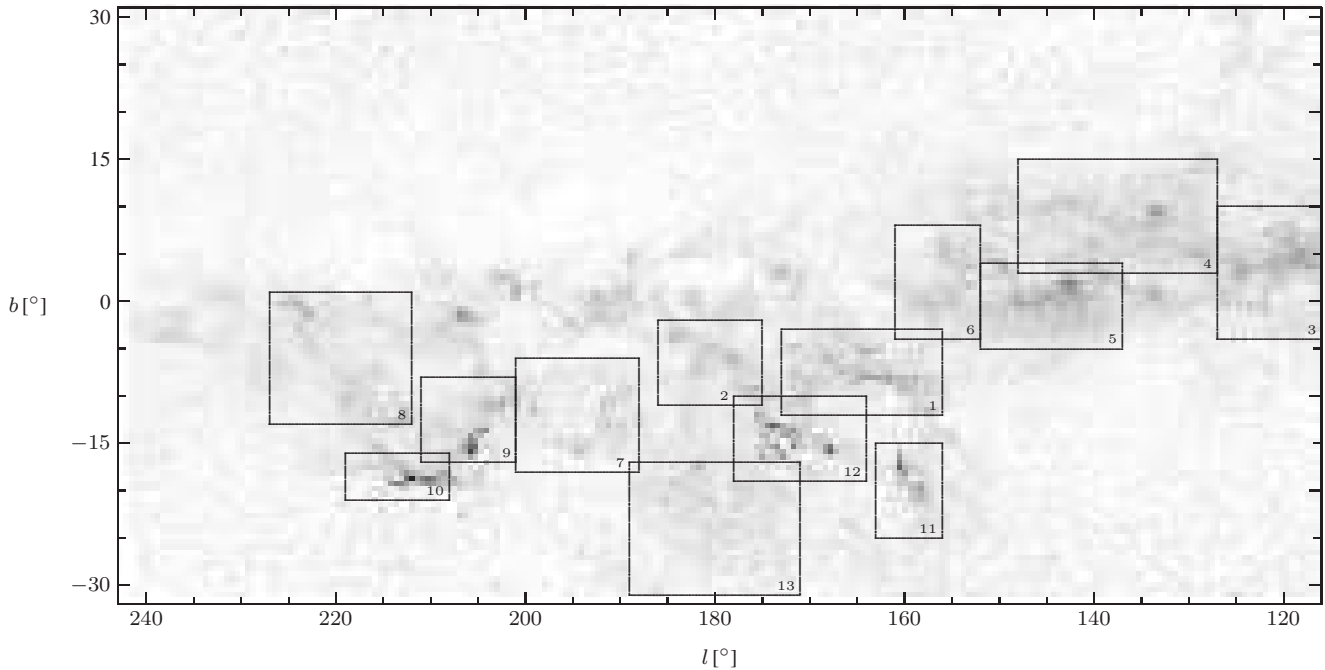}} at 179.5 -0.5
\axis left label {$b$\,[$^\circ$]}
ticks in long numbered from -30 to 30 by 15
      short unlabeled from -30 to 30 by 5 /
\axis right label {}
ticks in long unlabeled from -30 to 30 by 10
      short unlabeled from -30 to 30 by 5 /
\axis bottom label {$l$\,[$^\circ$]}
ticks in long numbered from 120 to 240 by 20
      short unlabeled from 120 to 240 by 5 /
\axis top label {}
ticks in long unlabeled from 120 to 240 by 20
      short unlabeled from 120 to 240 by 5 /

\plot 156 -25 163 -25 163 -15 156 -15 156 -25 /
\plot 164 -19 178 -19 178 -10 164 -10 164 -19 /
\plot 171 -31 189 -31 189 -17 171 -17 171 -31 /
\plot 188 -18 201 -18 201  -6 188  -6 188 -18 /
\plot 201 -17 211 -17 211  -8 201  -8 201 -17 /
\plot 208 -21 219 -21 219 -16 208 -16 208 -21 /
\plot 156 -12 173 -12 173  -3 156  -3 156 -12 /
\plot 175 -11 186 -11 186  -2 175  -2 175 -11 /
\plot 212 -13 227 -13 227   1 212   1 212 -13 /
\plot 116 -4 127 -4 127 10 116 10 116 -4 /
\plot 127  3 148  3 148 15 127 15 127  3 /
\plot 137 -5 152 -5 152  4 137  4 137 -5 /
\plot 152 -4 161 -4 161  8 152  8 152 -4 /

\put {\tiny $ 1$} at 157   -11
\put {\tiny $ 2$} at 179   -10
\put {\tiny $ 3$} at 117   -03
\put {\tiny $ 4$} at 128   +04
\put {\tiny $ 5$} at 138   -04
\put {\tiny $ 6$} at 153   -03
\put {\tiny $ 7$} at 190   -17
\put {\tiny $ 8$} at 213   -12
\put {\tiny $ 9$} at 202   -16
\put {\tiny $10$} at 209.5 -20
\put {\tiny $11$} at 157.5 -24
\put {\tiny $12$} at 165.5 -18
\put {\tiny $13$} at 172.5 -30
	     	 
\endpicture 
\caption{\label{ext_map} Gray scale illustration of our 8001 square degrees
extinction map of the Galactic Anticenter obtained from $\left< J-H \right>$ and
$\left< H-K \right>$ colour excess determinations in 2MASS. Pixel values are
scaled linearly from -0.33\,mag to 10\,mag optical extinction. The rectangles
mark the regions selected for detailed analysis (numbers are identical to their
ID in Table\,\ref{tab_beta}), and magnified extinction maps of these areas can
be found in Appendix\,\ref{app_images}. The full resolution FITS file of this
map is available as online material.} 
\end{figure*}

\section{The Extinction Map}
\label{extinction}

We selected all objects from the 2MASS point source catalogue with a Galactic
Longitude of $116^\circ < l < 243^\circ$ and a Galactic Latitude of $-32^\circ <
b < +31^\circ$. To ensure good photometric accuracy only objects with a quality
flag better than 'E' are used. The area was then decomposed into $1^\circ \times
1^\circ$ sized regions. Independent of the position we determined the median
colour of stars ($J-H$ and $H-K$) in 4'$\times$4' sized boxes with an
oversampling of 2. This leads to a pixel size of 2', 30$\times$30 pixels per
square degree, or 3810$\times$1890 pixel for our 8001 square degree sized final
map. The calculations required for such a project are embarrassingly parallel
(i.e. the problem domain and the associated computational algorithms used to
solve this problem can be optimally decomposed to use the available computer
resources) and have been solved in a distributed fashion. We used the
Grid-Ireland infrastructure (e.g. Coghlan et al. \cite{Coghlan2.05.EGC}) to
perform these calculations. 

The correct determination of extinction from colour excess methods requires that
we remove the foreground stars to the cloud and that we reject young stellar
objects. Both types of objects systematically influence the determined
extinction values. Foreground stars lead to on average bluer colours, hence the
extinction is underestimated. Young stars are intrinsically red and hence lead
to an overestimate of the extinction. Due to the large range of distances of the
clouds in our map (ranging from 140\,pc to at least 1\,kpc) the removal of
foreground stars is very difficult. In principle the number of foreground stars
has to be evaluated for each individual cloud and removed statistically. The
large number of clouds in the map with partly unknown distances, or overlapping
clouds close to the Galactic Plane makes this an almost impossible task.
However, if the stars in each box are not dominated by foreground objects and
one uses the median colour of stars instead of the average, then the foreground
stars are removed automatically. Froebrich \& del Burgo
\cite{2006MNRAS.369.1901F} have shown (see their Fig.\,9) that such an approach
indeed reproduces the intrinsic extinction values of clouds extremely well, as
long as the cloud is not too distant, or has a very high extinction. For very
distant clouds or very high extinction regions, however, this approach detects
zero extinction. Hence, locally more sophisticated procedures such as described
in Lombardi \cite{2005A&A...438..169L} or suggested in Froebrich \& del Burgo
\cite{2006MNRAS.369.1901F} have to be used to obtain correct extinction values
for the highest extinction regions. Since only a very small portion of the map
area is affected and we are investigating only the large scale distribution of
clouds and their properties in this paper, the main  conclusions of this
paper are not significantly changed by this effect. However, when using the
presented extinction map one has to keep in mind that the extinction values in
the very dense regions of clouds are underestimated and that there is a
small $A_V$ dilution for more distant clouds. Similarly to the removal of
foreground stars, the use of the median colour also removes intrinsically red
YSOs, as long as they do not dominate (as e.g. in the Orion Nebula cluster).

\begin{figure*}
\beginpicture
\setcoordinatesystem units <-1.25984252mm,1.25984252mm> point at 179.5 0
\setplotarea x from 243 to 116 , y from -32 to  31
\put {\includegraphics[width=16cm]{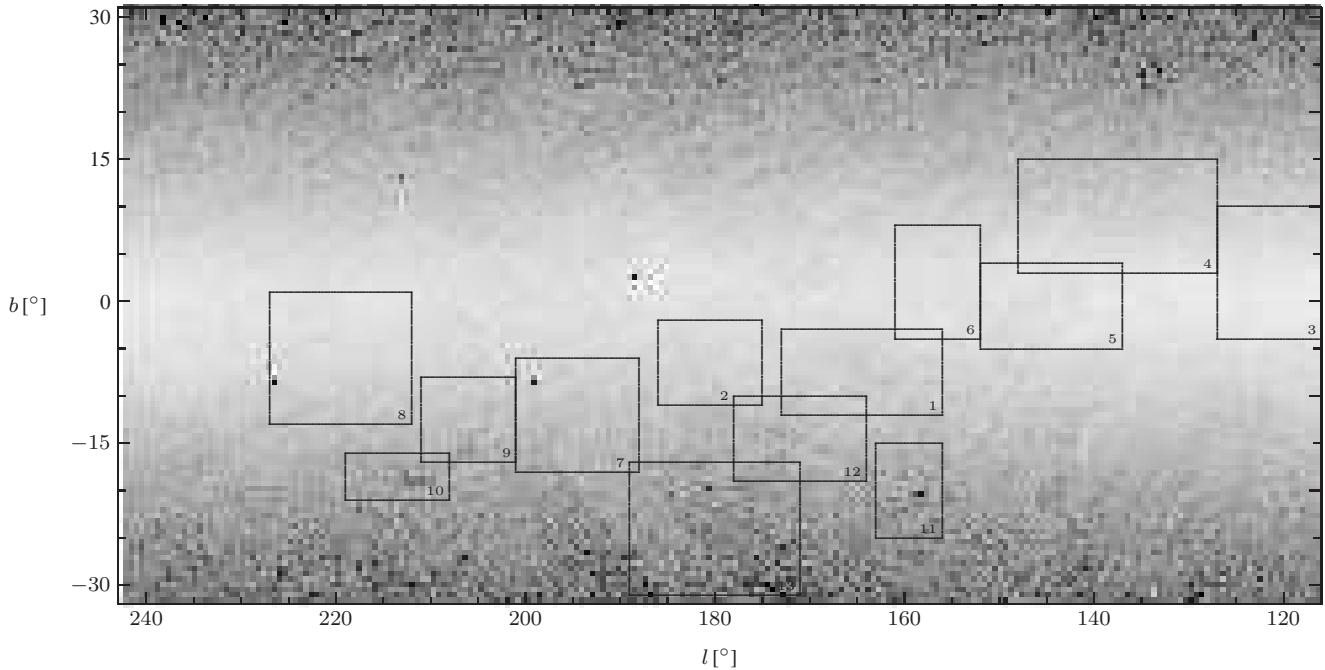}} at 179.5 -0.5
\axis left label {$b$\,[$^\circ$]}
ticks in long numbered from -30 to 30 by 15
      short unlabeled from -30 to 30 by 5 /
\axis right label {}
ticks in long unlabeled from -30 to 30 by 10
      short unlabeled from -30 to 30 by 5 /
\axis bottom label {$l$\,[$^\circ$]}
ticks in long numbered from 120 to 240 by 20
      short unlabeled from 120 to 240 by 5 /
\axis top label {}
ticks in long unlabeled from 120 to 240 by 20
      short unlabeled from 120 to 240 by 5 /

\plot 156 -25 163 -25 163 -15 156 -15 156 -25 /
\plot 164 -19 178 -19 178 -10 164 -10 164 -19 /
\plot 171 -31 189 -31 189 -17 171 -17 171 -31 /
\plot 188 -18 201 -18 201  -6 188  -6 188 -18 /
\plot 201 -17 211 -17 211  -8 201  -8 201 -17 /
\plot 208 -21 219 -21 219 -16 208 -16 208 -21 /
\plot 156 -12 173 -12 173  -3 156  -3 156 -12 /
\plot 175 -11 186 -11 186  -2 175  -2 175 -11 /
\plot 212 -13 227 -13 227   1 212   1 212 -13 /
\plot 116 -4 127 -4 127 10 116 10 116 -4 /
\plot 127  3 148  3 148 15 127 15 127  3 /
\plot 137 -5 152 -5 152  4 137  4 137 -5 /
\plot 152 -4 161 -4 161  8 152  8 152 -4 /

\put {\tiny $ 1$} at 157   -11
\put {\tiny $ 2$} at 179   -10
\put {\tiny $ 3$} at 117   -03
\put {\tiny $ 4$} at 128   +04
\put {\tiny $ 5$} at 138   -04
\put {\tiny $ 6$} at 153   -03
\put {\tiny $ 7$} at 190   -17
\put {\tiny $ 8$} at 213   -12
\put {\tiny $ 9$} at 202   -16
\put {\tiny $10$} at 209.5 -20
\put {\tiny $11$} at 157.5 -24
\put {\tiny $12$} at 165.5 -18
\put {\tiny $13$} at 172.5 -30
	     	 
\endpicture 
\caption{\label{noisemap} Gray scale illustration the noise map. Pixel values
show the one sigma noise level scaled linearly from 0.0\,mag (white) to 0.6\,mag
(black) of optical extinction. The rectangles mark the regions selected for
detailed analysis (numbers are identical to their ID in Table\,\ref{tab_beta})} 
\end{figure*}

The two median $J-H$ and $H-K$ colour maps of the entire field show small but
significant large-scale gradients of the colour values of background (i.e.
extinction free) regions. These gradients are caused by a change of the stellar
population with distance from the Galactic Plane. In particular, the median
colour of stars becomes bluer at lower Galactic Latitudes and (less prominent)
redder towards Galactic Longitudes approaching the Anticenter of the Galaxy. To
determine the $\left< J-H \right>$ and $\left< H-K \right>$ colour excess maps,
we thus fitted a polynomial of 3rd order in $l$ and $b$ to the extinction free
regions and subtracted it from the median colour maps. 

The resulting $\left< J-H \right>$ and $\left< H-K \right>$ colour excess maps
are converted separately into H-band extinction maps using Equ.\,8 and 9 of
Froebrich \& del Burgo \cite{2006MNRAS.369.1901F}, respectively. A power law
index of the reddening law of $\beta$\,=\,1.73 is applied (see
Sect.\,\ref{reddening} for details). The resulting two H-band extinction maps
are averaged and converted into our final optical extinction map by
$A_V$\,=\,5.689\,$\times$\,$A_H$, using the reddening law from Mathis
\cite{1990ARA&A..28...37M}. In Fig.\,\ref{ext_map} we show this map in
gray-scale. Indicated in this figure are the regions of which we show magnified
versions of the extinction map in Appendix\,\ref{app_images}. These are also the
regions we analyse in detail in the forthcoming sections. The full resolution
FITS version of our entire map will be provided as online material to this
paper. Note that the construction of the final extinction maps as average of the
$A_H$ extinction maps from $\left< J-H \right>$ and $\left< H-K \right>$ colour
excess maps leads to further dilution of the very high extinction cores. This
will, however, effect the same areas that are dominated by foreground stars, as
described above. Thus, the number of high column density regions in our analysis
will be underestimated.

The noise in the extinction map can in principle only be measured in extinction
free regions as the standard deviation of extinction values from zero. To
estimate the noise in all areas we have determined a noise map (see
Fig.\,\ref{noisemap}). Since we used the median colour of stars in each
pixel, we need to determine the standard error of the median for our noise map.
For a large sample and normal distribution this can in principle be done by
$\sigma_{med}$ = $1.25 \cdot \sigma / \sqrt{N}$, where $\sigma$ is the error of
the colour of the individual stars and $N$ the number of stars in each box.
Since we use stars of different photometric quality (see below) and we do not
know if the sample of stars in each pixel follows a normal distribution, we
cannot use this equation. However, the fraction of stars for any given
photometric quality is very constant (see below). Thus, we can conclude that the
noise in each pixel is proportional to $ 1 / \sqrt{N}$ for both, the $\left< J-H
\right>$ and $\left< H-K \right>$ maps. We have thus scaled and co-added the $ 1
/ \sqrt{N}$ maps of stars with (J and H) and (H and K) photometry, in order to
obtain the final noise map. The scaling was done to match the measured noise in
the $A_V$ map in extinction free regions. As can be seen, the noise in the map
varies significantly with Galactic Latitude, as well as with extinction. The
3\,$\sigma$ noise ranges from 0.36\,mag optical extinction, close to the
Galactic Plane, to 1.0\,mag $A_V$ in the most northern and southern parts of the
map as well as in high extinction regions. In Table\,\ref{tab_beta} we list for
all individual regions the average number of stars with detections in J and H,
as well as the average noise in the area. There are pixels in the map
(0.0066\,\%) where the noise is extremely large, i.e, there are no stars to
determine the colour. These regions are: 1) around very bright stars; 2) at
$|b|>30^\circ$; 3) high extinction cores in Orion, Taurus, Perseus. For Ori\,A
we determine an average noise of 0.24\,mag, which can be compared to the
1\,$\sigma$ noise level of about 0.2\,mag $A_V$ obtained by Lombardi \& Alves
\cite{2001A&A...377.1023L} (see their Fig.\,6) in this region using the
optimised NICER method. 

There are further areas in the map where the extinction values are not very
reliable. These are positions close to dense star clusters. Here, the high
number of stars in the cluster dominates the extinction determination. Hence,
depending on the colour of the cluster members we measure either a too low
(mostly blue stars) or too high (mostly young, red stars) extinction. Since this
is a common effect, we over plotted circles at the positions of all known star
clusters (found in SIMBAD) in the gray scale maps of our selected regions in
Appendix\,\ref{app_images}. Additionally we over plot $+$ signs at the positions
of all possible new star clusters with $|b| < 20^\circ$ found by Froebrich et
al. \cite{2006MNRAS.subm.F}.

Does the use of 2MASS quality BCD stars instead of only quality A stars
influence the noise in our map? Independent of the position about 40--45\,\% of
all stars are of quality AAA. Given the larger errors in photometry for the
quality BCD objects, one can estimate that the noise in the extinction map is
about 10\,\% (i.e. 0.03\,mag) higher, when including the lower quality objects.
However, as has been pointed out by Froebrich \& del Burgo
\cite{2006MNRAS.369.1901F}, systematic errors of extinction determinations based
on colour excess methods can be much higher than this, since the population of
stars seen through a cloud can differ from the population of stars in a control
field (even if close-by). The more than twice as large number of stars per
resolution element when including the lower quality objects will lead to a much
better representation of the population of stars and, as a result, to smaller
systematic errors. This out weights in our opinion the slightly higher resulting
statistical noise in our maps and justifies the use of the lower quality 2MASS
sources.

Note that the galaxies M\,31 ($l$=121.2, $b$=-21.6) and M\,33 ($l$=133.6,
$b$=-31.3) show up as apparently high extinction regions. In M\,31 the structure
in our extinction map resembles quite closely the 170\,$\mu$m emission map from
Haas et al. \cite{1998A&A...338L..33H} and in particular we are able to identify
the dominant ring of material in the disc of the Andromeda Galaxy. 

\begin{table*}
\caption{\label{tab_beta} Names and areas of our individual regions, average
numbers of stars in the 4'\,x\,4' box, average noise in the field, peak in the
distribution of $\beta$, intrinsic scatter of the $\beta$ distribution,
probability that the measured distribution is in agreement with the predicted
one for the assumption of a constant $\beta_0$ value and intrinsic scatter in
the whole region and slope of the column density distribution in the logN vs.
$A_V$ plot.}   
\begin{center}
\begin{tabular}{llllcccccc}
ID & Name & \multicolumn{1}{c}{$l$-range} & \multicolumn{1}{c}{$b$-range} &
\multicolumn{1}{c}{$\#$stars/box} & 
\multicolumn{1}{c}{$\sigma_{A_V}$\,[mag]} & \multicolumn{1}{c}{$\beta_0$} &
\multicolumn{1}{c}{$\sigma_{\beta_{intr}}$\,[mag]} & \multicolumn{1}{c}{P\,[\%]} &
\multicolumn{1}{c}{slope} \\ \hline 
 1 & Auriga\,1         & 156 -- 173 & $-$12 -- $-$03 & 47 & 0.15 & 1.99 & 0.45 & 98.5 & -0.68 \\
 2 & Auriga\,2         & 175 -- 186 & $-$11 -- $-$02 & 53 & 0.14 & 2.05 & 0.39 & 99.9 & -0.83 \\
 3 & Cassiopeia        & 116 -- 127 & $-$04 -- $+$10 & 90 & 0.11 & 1.67 & 0.43 & 99.9 & -0.72 \\
 4 & Camelopardalis\,1 & 127 -- 148 & $+$03 -- $+$15 & 53 & 0.15 & 1.96 & 0.49 & 98.4 & -0.91 \\
 5 & Camelopardalis\,2 & 137 -- 152 & $-$05 -- $+$04 & 78 & 0.12 & 1.40 & 0.54 & 98.5 & -0.71 \\
 6 & Camelopardalis\,3 & 152 -- 161 & $-$04 -- $+$08 & 74 & 0.12 & 1.66 & 0.49 & 95.2 & -1.20 \\ 
 7 & $\lambda$-Ori     & 188 -- 201 & $-$18 -- $-$06 & 36 & 0.18 & 1.87 & 0.24 & 96.3 & -0.81 \\
 8 & Monoceros         & 212 -- 227 & $-$13 -- $+$01 & 67 & 0.14 & 1.52 & 0.53 & 82.7 & -0.68 \\
 9 & Orion\,B          & 201 -- 211 & $-$17 -- $-$08 & 34 & 0.18 & 1.77 & 0.41 & 91.2 & -0.38 \\
10 & Orion\,A          & 208 -- 219 & $-$21 -- $-$16 & 20 & 0.24 & 1.66 & 0.59 & 99.8 & -0.28 \\
11 & Perseus           & 156 -- 163 & $-$25 -- $-$15 & 18 & 0.25 & 1.97 & 0.51 & 99.9 & -0.30 \\
12 & Taurus            & 164 -- 178 & $-$19 -- $-$10 & 27 & 0.20 & 1.99 & 0.47 & 99.9 & -0.32 \\
13 & Taurusextended    & 171 -- 189 & $-$31 -- $-$17 & 14 & 0.29 & 1.92 & 0.43 & 99.9 & -0.49 \\
14 & Entire Field      & 116 -- 243 & $-$32 -- $+$31 & ---& ---  & 1.73 & 0.52 & 98.9 & ---   \\                \\
\end{tabular}
\end{center}
\end{table*}

\section{The Reddening Law}
\label{reddening}

\begin{figure}
\includegraphics[height=8cm,angle=-90]{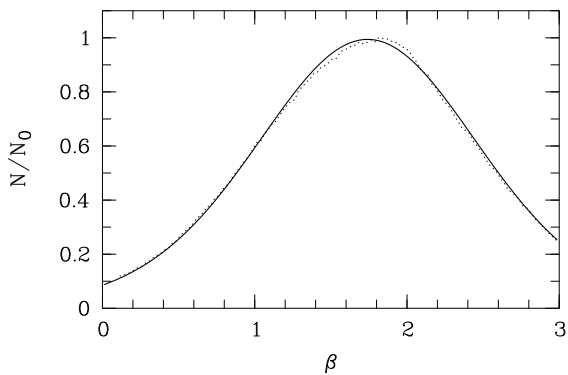}
\caption{\label{beta_dist} Histogram of the distribution of $\beta$ values in
the entire field (dotted line) measured for all pixels where the $\left< J-H
\right>$ and $\left< H-K \right>$ colour excess values are at least 3\,$\sigma$
above the noise. The solid line represents the predicted distribution for a
value of $\beta$=1.73, $\sigma_{\beta_{intr}}$=0.52 and the measured
distribution of $A_V$ values in the entire field (see text for details). A
Kolmogorov Smirnov test shows that both distributions are identical with a
probability of 98.9\,\%.}
\end{figure}

The conversion of colour excess maps into extinction maps requires the knowledge
of the extinction law. Our approach of using $\left< J-H \right>$ and $\left<
H-K \right>$ colour excess maps enables us to determine the power law index of
the extinction law using the colour excess ratio $R$\,$\equiv$\,$\left< J-H
\right>$/$\left< H-K \right>$ and 
\begin{equation}
0 = \left( \frac{\lambda_2}{\lambda_1} \right)^{\beta} + R \cdot \left(
\frac{\lambda_3}{\lambda_2} \right)^{-\beta} - R - 1
\label{equ1}
\end{equation}
from Froebrich \& del Burgo \cite{2006MNRAS.369.1901F}. We now can analyse the
resulting distribution of $\beta$ values in our maps. For all the following
analysis of $\beta$ we use only areas where both, the $\left< J-H \right>$ and
$\left< H-K \right>$ colour excess values are 3\,$\sigma$ above the noise. In
Fig.\,\ref{beta_dist} we show the histogram of the measured distribution of all
$\beta$ values ($N_{o}(\beta)$) as dotted line. It peaks at 1.73, in good
agreement with the canonical value for the interstellar medium (Draine
\cite{2003ARA&A..41..241D}). We hence use $\beta$\,=\,1.73 to convert our colour
excess maps into extinction. The distribution of $\beta$ values shows a rather
broad peak with a FWHM of 1.6. This, seemingly, rather large value is in good
agreement for the predicted scatter of about $\pm$\,40\,\% when using the colour
excess ratio to determine $\beta$ (see e.g. Fig.\,1 in Froebrich \& del Burgo
\cite{2006MNRAS.369.1901F}).  

It is possible to investigate in more detail, if the measured distribution of
$\beta$ values $N_0(\beta)$ is in agreement with a single value of
$\beta_0$=1.73 in the entire field or not. This can be done by convolving the
observed distribution of signal to noise values, $N_{o}(\left< J-H \right>)$, in
our extinction map from Fig.\,\ref{ext_map} with the predictions for the width
of the distribution of $\beta$. We have, however, to allow for the possibility
that there is an intrinsic scatter in the $\beta$ values
($\sigma^2_{\beta_{int}}$) around $\beta_0$. We hence have to adapt Equ.\,11
from Froebrich \& del Burgo \cite{2006MNRAS.369.1901F} and obtain:

\begin{equation}
\sigma^2_\beta = \frac{\sigma^2_{\left< \lambda_1 - \lambda_2 \right>}}{\left<
\lambda_1 - \lambda_2 \right>^2} R^2 \left[ 1 + \frac{R^2}{\alpha_{\rm co}} - 2
\frac{R}{\gamma_{\rm co}} \right] \left( \frac{\partial {\bf B}(R)}{\partial R}
\right)^2 + \sigma^2_{\beta_{int}}
\label{equ2}
\end{equation}
To determine the predicted distribution of $\beta$ values we use:
\begin{equation}
N_{p}(\beta) = N_0 \int N_{o}(\left< J-H \right>) \cdot e^{ - \frac{1}{2} \left(
\frac{\beta - \beta_0}{\sigma_\beta(\left< J-H \right>)} \right)^2} \cdot d
\left< J-H \right> 
\label{equ3}
\end{equation}
were $N_0$ is a normalisation factor and $\beta_0 = 1.73$ the peak of the
measured $\beta$ distribution. A Kolmogorov Smirnov test can then be performed
with $N_{o}(\beta)$ and $N_{p}(\beta)$ to determine the probability ($P$)
that the two distributions are drawn from the same sample.

In order to determine the predicted $\beta$ distribution we require to estimate
the parameters $\alpha_{co}$ and $\gamma_{co}$. The parameter $\alpha_{co}$
represents the square of the ratio of the noise in the two colour excess maps
and $\gamma_{co}$ represents the ratio of the square of the noise in one colour
excess map and the covariance of the two colour excess maps (see
Equ.\,\ref{equ2} and Froebrich \& del Burgo \cite{2006MNRAS.369.1901F} for
details). For our dataset we measure values of: $\alpha_{co}$=3.6 and
$\gamma_{co}$=3.7. Note that the values for both parameters might slightly vary
with position or hence from cloud to cloud. However, we have measured here the
same values for $\alpha_{co}$ and $\gamma_{co}$ as were obtained for 2MASS data
in IC\,1396\,W by Froebrich \& del Burgo \cite{2006MNRAS.369.1901F}. Due to this
fact, and since the dependence of the predicted $\beta$ distribution on
$\alpha_{co}$ and $\gamma_{co}$ is rather weak, we fix these values for all our
determinations.

This gives us the possibility to vary the intrinsic scatter
$\sigma^2_{\beta_{int}}$ and the $\beta_0$ value until we find the highest
probability $P$ that the observed and predicted distributions of $\beta$ values
match. We list the obtained values of those parameters as well as the
probabilities for the individual regions in Table\,\ref{tab_beta}.

We over plot the resulting best predicted distribution for $\beta$ values of the
entire field as solid line in Fig.\,\ref{beta_dist}. We find that with a
98.9\,\% probability the two distributions are equal, when a value of $\beta_0 =
1.73$ and an intrinsic scatter of $\sigma^2_{\beta_{int}} = 0.52$ is assumed.
The large intrinsic scatter shows, that there is a good chance that indeed
variable $\beta$ values are present in different parts of the field. Hence,
repeating the same analysis for selected regions should lead to different
$\beta$ values in different clouds. If locally those clouds have more uniform
dust properties, then a lower intrinsic scatter should be found.

Thus, similar to the entire area, we investigate the peak value of the $\beta$
distribution, the intrinsic scatter and how well the distribution matches the
assumption of a constant $\beta$ value for our individual regions. In
Table\,\ref{tab_beta} we summarise the obtained results. We find an average
value of $\beta$\,=\,1.8\,$\pm$\,0.2. Almost in all investigated regions the
$\beta$ value is in agreement with the estimate for the entire field. Only in
Auriga\,2 ($\beta$=2.05), Monoceros ($\beta$=1.52) and Camelopardalis\,2
($\beta$=1.40) significantly different values are found, hinting to different
dust properties such as the grain size distribution in these areas. Notably,
regions with smaller $\beta$ values tend to be closer to the Galactic Plane.
However, it is not clear if this is caused by different dust properties or by
the overlap of clouds in this area. Furthermore, small systematic off-sets
in either $\left< J-H \right>$ or $\left< H-K \right>$ can cause changes in the
determined $\beta$ value. Thus, the procedure of fitting a polynomial to the
extinction free regions in order to determine the colour excess maps might also
contribute to these smaller $\beta$ values. Since very close to the Galactic
Plane there are no extinction free regions, the procedure might have introduced
a small off-set in the colour excess maps and hence to the $\beta$ values.
Clearly, more detailed investigations are required to find the cause of the
smaller $\beta$ values very close to the Galactic plane. 

We also find that for most regions the intrinsic scatter of the $\beta$ values
is smaller or in the order of the value for the entire field. Only in Ori\,A is
a significantly larger value found. The probabilities that measured and
predicted values are drawn from the same sample are listed in
Table\,\ref{tab_beta} and the individual distributions are shown in
Appendix\,\ref{app_plots}. We find that in some regions the proposal of constant
$\beta$ values with the corresponding internal scatter is more probable than in
others. Especially in the regions Monoceros and Ori\,B low probabilities
($P<95\,\%$) are found. We can conclude that in these regions we do not see a
dominant fraction of the dust possessing constant optical properties. This might
indicate large scale variations of the optical dust properties within these
clouds.

\section{The Column Density Distribution}
\label{clumps}
 
In order to analyse and interpret the distribution of column density values in
the entire map, as well as in the individual regions, we first have to
recapitulate the selection effects induced by our method. As outlined in detail
in Sect.\,\ref{extinction}, our method of determining the extinction maps leads
to three biases. (1) Very distant clouds, as well as very high extinction
regions will not be detected by our method, since these areas are dominated by
foreground stars. (2) There is a small $A_V$ dilution with distance. (3)
The averaging of extinction maps from $\left< J-H \right>$ and $\left< H-K
\right>$ colour excess maps leads to dilution of the high extinction cores. (4)
Regions close to rich star clusters will be dominated by the colours of the
cluster members. However, due to the use of the median stellar colours instead
of the mean colours, all not too distant clouds without rich clusters and
extremely dense cores, will have no systematically influenced extinction values.

The general way of analysing the structure of molecular clouds is to determine
the distribution of clumps within them. This includes the clump mass, size and
column density distribution. We have hence used the $clumpfind$-2D algorithm
(Williams et al. \cite{1994ApJ...428..693W}) to extract all detectable clumps in
our final extinction map and determined their size and mass distribution.
Contrary to line or dust emission maps, extinction maps have real, very smooth
large scale profiles which make it very difficult to extract structure
information using an automated thresholding technique. This is reflected by the
fact that the determined clump properties from our map depended, partly
significantly, on the chosen levels and also on the pixel size of the map.

Hence, instead of using the clump properties in the map, we will analyse the
column density distribution in the individual regions. Since the distance of the
clouds will influence the measured slope of the column density distribution (see
below), we will not analyse the distribution of the entire map as a whole. We
show in Appendix\,\ref{app_plots} the $A_V$ distributions for all individual
regions (solid line). For each region we fit a linear polynomial to the log(N)
vs. $A_V$ plots (dashed line) for extinction values larger than the 5\,$\sigma$
noise (vertical line). The measured slopes are indicated in the individual
diagrams as well as in Table\,\ref{tab_beta}.

The individual $A_V$ distributions nicely show the above discussed biases for
high extinction values. Some show a clear drop-off for regions with very high
column densities. Depending on the distance of the cloud and its position, this
drop-off occurs at different values. For Ori\,A and Ori\,B we see a clear drop
at $A_V$\,=\,10\,mag, while for Perseus ($A_V$\,=\,8\,mag) and Camelopardalis\,2
($A_V$\,=\,6\,mag) lower values are found. The values of the slopes range from
about -0.3 to almost -1.0, with the exception of Camelopardalis\,3 where -1.2 is
found. 

The interpretation of these values is, however, rather difficult. Two effects
could influence the measured slope: (1) the distance of the cloud; (2) the
intrinsic distribution of material. The distance could influence the slope since
in more distant clouds the extinction per resolution element is averaged over a
larger physical area in the cloud. Hence, for more distant clouds the column
density distribution on larger physical scales is measured. One can artificially
move a cloud to larger distances by using a larger box-size for the extinction
measurements. Indeed one finds that this results in a steeper column density
distribution. This is in agreement with the values obtained from our data, where
more distant clouds possess on average steeper column density distributions. We
basically find that in all clouds where our resolution element is smaller than
1\,pc the slope of the distribution is -0.3. For all clouds were we are not able
to resolve 1\,pc (cloud distance larger than 850\,pc) the slope in the column
density distribution is about -0.75. This leads to the conclusion that there
might be a change in the governing physics determining the column density
distribution in molecular clouds on a scale of about 1\,pc. Interestingly, this
is about the Jeans length for a column density that is required for
self-shielding and molecular hydrogen formation in molecular clouds (Hartmann et
al. \cite{2001ApJ...562..852H}). The already discussed small $A_V$ dilution
with distance could as well contribute to the change in the slope. Furthermore,
there could also be intrinsic differences between clouds, as reflected by the
different slopes measured for e.g. Ori\,A and Ori\,B, which are believed to be
at the same distance. We note however, that if measured only for $b < -13^\circ$
the slope for Ori\,B is -0.32, hence in very good agreement with Ori\,A and the
other close-by clouds Taurus and Perseus. For the region with $b > -13^\circ$ in
Ori\,B the slope is -0.75, indicating that these clouds are more distant. We
further note that an overlap of different clouds, as may happen close to the
Galactic plane, does not change the measured slope if the individual clouds are
at about the same distance and have intrinsically the same slope.

\section{Conclusions}
\label{conclusion}

Using Grid technology and 2MASS data we created the largest extinction map to
date based on near infrared colour excess. Our 127$^\circ \times 63^\circ$ sized
map of the Galactic Anticenter contains the molecular cloud complexes of Orion,
Taurus, Cepheus, Monoceros, Auriga, Camelopardalis and Cassiopeia. The three
sigma noise level ranges from 0.36\,mag optical extinction close to the Galactic
Plane to 1.0\,mag $A_V$ at $|b| \approx 30^\circ$ and in high extinction
regions. Our final optical extinction map has a resolution of 4' and a pixel
size of 2'.

The colour excess ratio $\left< J-H \right>$/$\left< H-K \right>$ is used to
determine the power law index $\beta$ of the wavelength dependence of the
extinction in the near infrared. In the entire field the distribution of $\beta$
peaks at a value of 1.73. The average value of the investigated cloud complexes
is determined to $\beta = 1.8 \pm 0.2$. Both values are in agreement with the
assumed values for the interstellar medium. We analyse in detail the
distribution of $\beta$ values in the entire field and individual regions. It is
found that there is an internal scatter in the $\beta$ values of about
0.4\,--\,0.5. A comparison of the measured and predicted $\beta$ distribution by
means of a Kolmogorov Smirnov tests shows that in most regions the proposal of a
constant $\beta$ with an internal scatter is valid. In some regions (e.g.
Monoceros, Ori\,B) low probabilities that observations and predictions match are
found. Hence, there is no dominant $\beta$ value in the cloud, a clear
indication of large scale variations of the optical dust properties in these
areas.

We further analysed the column density distribution of the individual regions.
Due to our technique, the highest extinction cores ($A_V >$10\,mag) cannot be
identified reliably. For extinction values up to 10\,mag $A_V$ we find that the
column density distributions show a linear behavior in the log(N) vs. $A_V$
plot. On average a steeper column density distribution for more distant clouds
is found, i.e. the $A_V$ distribution is steeper when measured on larger spatial
scales. This seems to indicate a change in the governing physics of the
column density at scales of about 1\,pc or an $A_V$ dilution with distance in
our map.

\section*{acknowledgments}
 
We are grateful to the anonymous referee for helpful comments which
significantly improved the paper. We would like to thank D.~O'Callaghan and
S.~Childs for very helpful discussions on Grid computing. D.F. and G.C.M
received funding by the Cosmo Grid project, funded by the Program for Research
in Third Level Institutions under the National Development Plan and with
assistance from the European Regional Development Fund. This publication makes
use of data products from the Two Micron All Sky Survey, which is a joint
project of the University of Massachusetts and the Infrared Processing and
Analysis Center/California Institute of Technology, funded by the National
Aeronautics and Space Administration and the National Science Foundation. This
research has made use of the SIMBAD database, operated at CDS, Strasbourg,
France.

\label{lastpage}


\begin{appendix}

\section{Individual Regions}

\subsection{Extinction and $\beta$ maps}
\label{app_images}

In the following we show for all individual regions investigated in this paper
larger versions of the extinction and $\beta$ maps. Note that the full
resolution FITS file of the extinction map is provided as online material to the
paper. We overplot in the maps the position of all known star clusters in the
field found in SIMBAD (circles) as well as the new star cluster candidates ($+$
signs) found in Froebrich et al. \cite{2006MNRAS.subm.F}, in order to mark
regions were potentially the cluster members might dominate the extinction
determination (see Sect.\,\ref{extinction} for details).

\begin{figure*}
\beginpicture
\setcoordinatesystem units <-4.1176mm,4.1176mm> point at 0 0
\setplotarea x from 173 to 156 , y from -12 to -3
\put {\includegraphics[width=7cm]{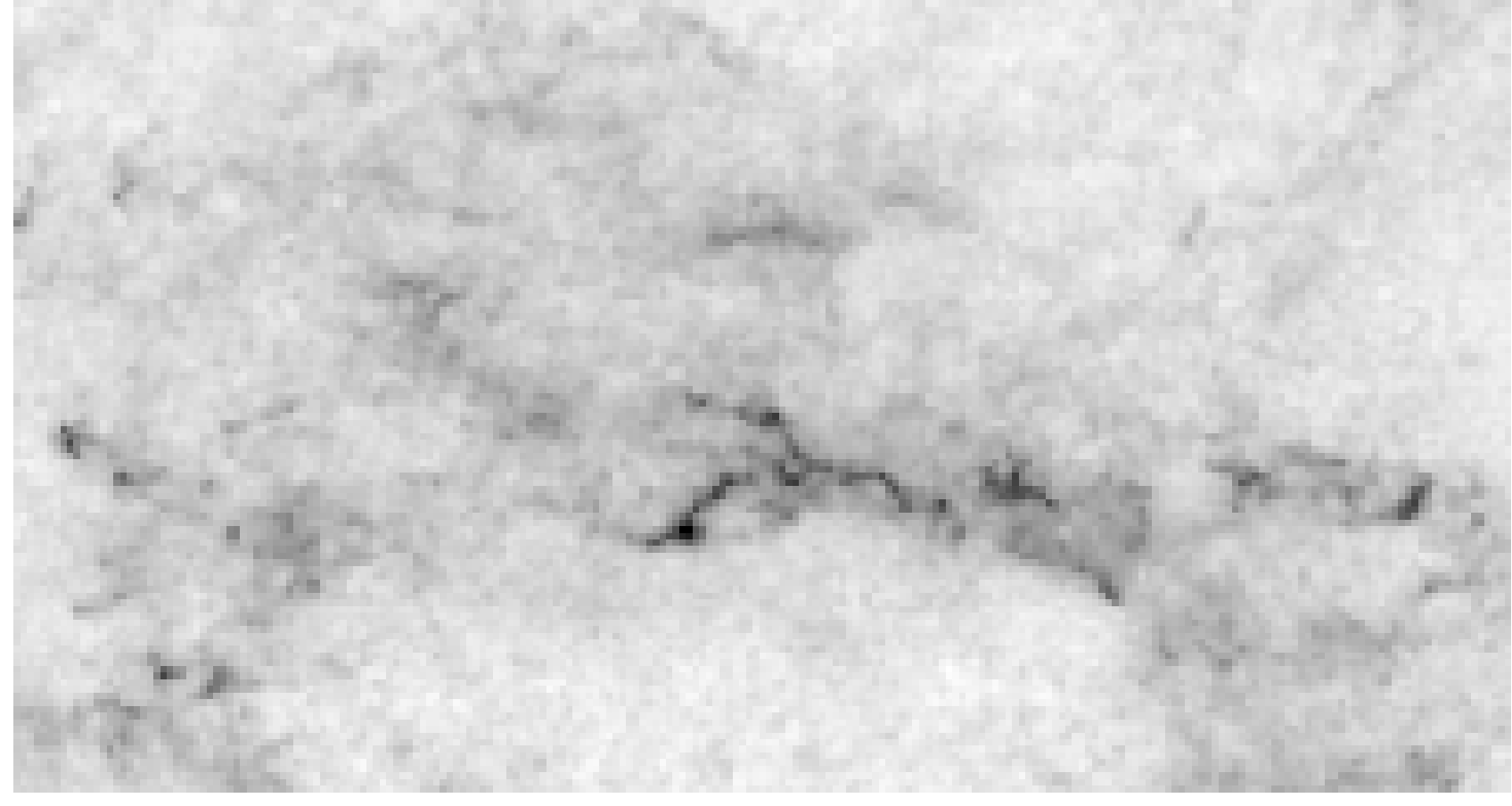}} at 164.5 -7.5
\axis left label {$b$\,[$^\circ$]}
ticks in long numbered from -12 to -3 by 4
      short unlabeled from -12 to -3 by 1 /
\axis right label {}
ticks in long unlabeled from -12 to -3 by 4
      short unlabeled from -12 to -3 by 1 /
\axis bottom label {$l$\,[$^\circ$]}
ticks in long numbered from 156 to 173 by 4
      short unlabeled from 156 to 173 by 1 /
\axis top label {}
ticks in long unlabeled from 156 to 173 by 4
      short unlabeled from 156 to 173 by 1 /
\put {\tiny $+$} at 157.246  -8.449	 
\put {\tiny $+$} at 158.475  -8.902	 
\put {\tiny $+$} at 159.055  -9.697	 
\put {\tiny $+$} at 159.224  -6.276	 
\put {\tiny $+$} at 159.225  -9.605	 
\put {\tiny $+$} at 159.71   -9.018	 
\put {\tiny $+$} at 161.167  -7.748	
\put {\tiny $+$} at 161.557  -8.756	
\put {\tiny $+$} at 162.923  -6.881	
\put {\tiny $+$} at 163.945  -3.808	 
\put {\tiny $+$} at 165.496  -7.655	
\put {\tiny $+$} at 165.668  -11.205	
\put {\tiny $+$} at 165.882  -4.043	
\put {\tiny $+$} at 167.794  -5.155	
\put {\tiny $+$} at 169.042  -3.863	 
\put {\tiny $+$} at 170.62   -8.699	 
\put {\tiny $+$} at 170.933  -11.153	 
\put {\tiny $+$} at 172.285  -5.651	 
\put {\tiny $\circ$} at 157.08  -3.64  
\put {\tiny $\circ$} at 162.83  -9.24  
\put {\tiny $\circ$} at 165.34  -9.01  
\put {\tiny $\circ$} at 167.58  -4.12  
\put {\tiny $\circ$} at 170.19  -4.01  
\put {\tiny $\circ$} at 170.53  -9.07  
\setcoordinatesystem units <-4.1176mm,4.1176mm> point at 19.4288 0
\setplotarea x from 173 to 156 , y from -12 to -3
\put {\includegraphics[width=7cm]{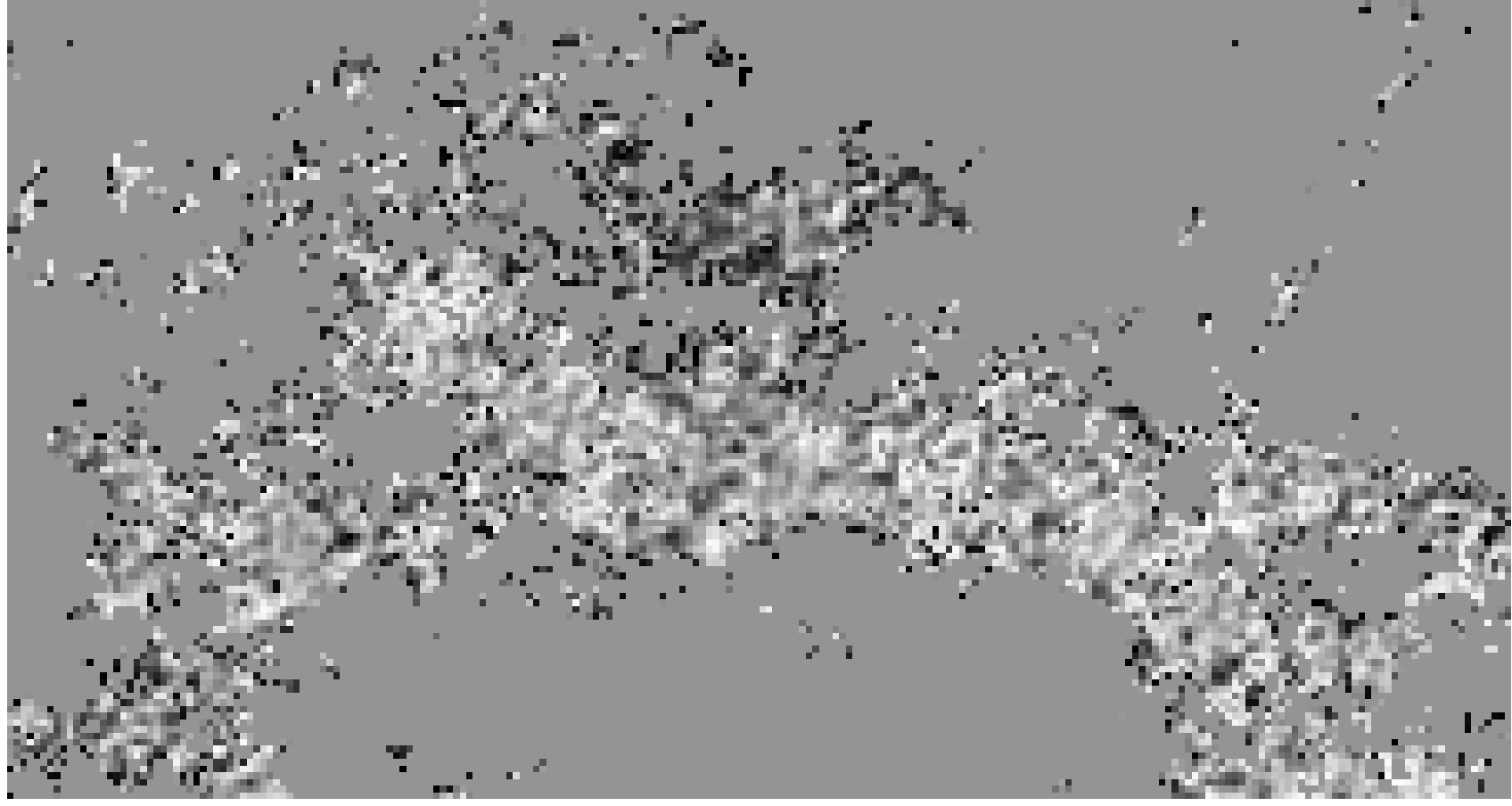}} at 164.5 -7.5
\axis left label {}
ticks in long unlabeled from -12 to -3 by 4
      short unlabeled from -12 to -3 by 1 /
\axis right label {}
ticks in long numbered from -12 to -3 by 4
      short unlabeled from -12 to -3 by 1 /
\axis bottom label {$l$\,[$^\circ$]}
ticks in long numbered from 156 to 173 by 4
      short unlabeled from 156 to 173 by 1 /
\axis top label {}
ticks in long unlabeled from 156 to 173 by 4
      short unlabeled from 156 to 173 by 1 /
\put {\tiny $+$} at 157.246  -8.449	 
\put {\tiny $+$} at 158.475  -8.902	 
\put {\tiny $+$} at 159.055  -9.697	 
\put {\tiny $+$} at 159.224  -6.276	 
\put {\tiny $+$} at 159.225  -9.605	 
\put {\tiny $+$} at 159.71   -9.018	 
\put {\tiny $+$} at 161.167  -7.748	
\put {\tiny $+$} at 161.557  -8.756	
\put {\tiny $+$} at 162.923  -6.881	
\put {\tiny $+$} at 163.945  -3.808	 
\put {\tiny $+$} at 165.496  -7.655	
\put {\tiny $+$} at 165.668  -11.205	
\put {\tiny $+$} at 165.882  -4.043	
\put {\tiny $+$} at 167.794  -5.155	
\put {\tiny $+$} at 169.042  -3.863	 
\put {\tiny $+$} at 170.62   -8.699	 
\put {\tiny $+$} at 170.933  -11.153	 
\put {\tiny $+$} at 172.285  -5.651	 
\put {\tiny $\circ$} at 157.08  -3.64  
\put {\tiny $\circ$} at 162.83  -9.24  
\put {\tiny $\circ$} at 165.34  -9.01  
\put {\tiny $\circ$} at 167.58  -4.12  
\put {\tiny $\circ$} at 170.19  -4.01  
\put {\tiny $\circ$} at 170.53  -9.07  
\endpicture 
\caption{\label{map_aurigae1} {\bf Left panel:} Gray scale extinction of the
region Auriga\,1. Extinction values are scaled linearly from -0.33 to 7\,mag of
optical extinction. {\bf Right panel:} Gray scale map of $\beta$ values in the
region Auriga\,1 for areas where the $<J-H>$ and $<H-K>$ values are more than
3$\sigma$ above the noise (for the remaining areas a value of $\beta$=1.73 is
adopted). The $\beta$ values are scaled linearly from zero (black) to three
(white). In both panels we indicate the positions of known star clusters
from the 2MASS database as circles and from the list from Froebrich et al.
(2006) as crosses.} 
\end{figure*}

\begin{figure*}
\beginpicture
\setcoordinatesystem units <-6.3636mm,6.3636mm> point at 0 0
\setplotarea x from 186 to 175 , y from -11 to -2
\put {\includegraphics[width=7cm]{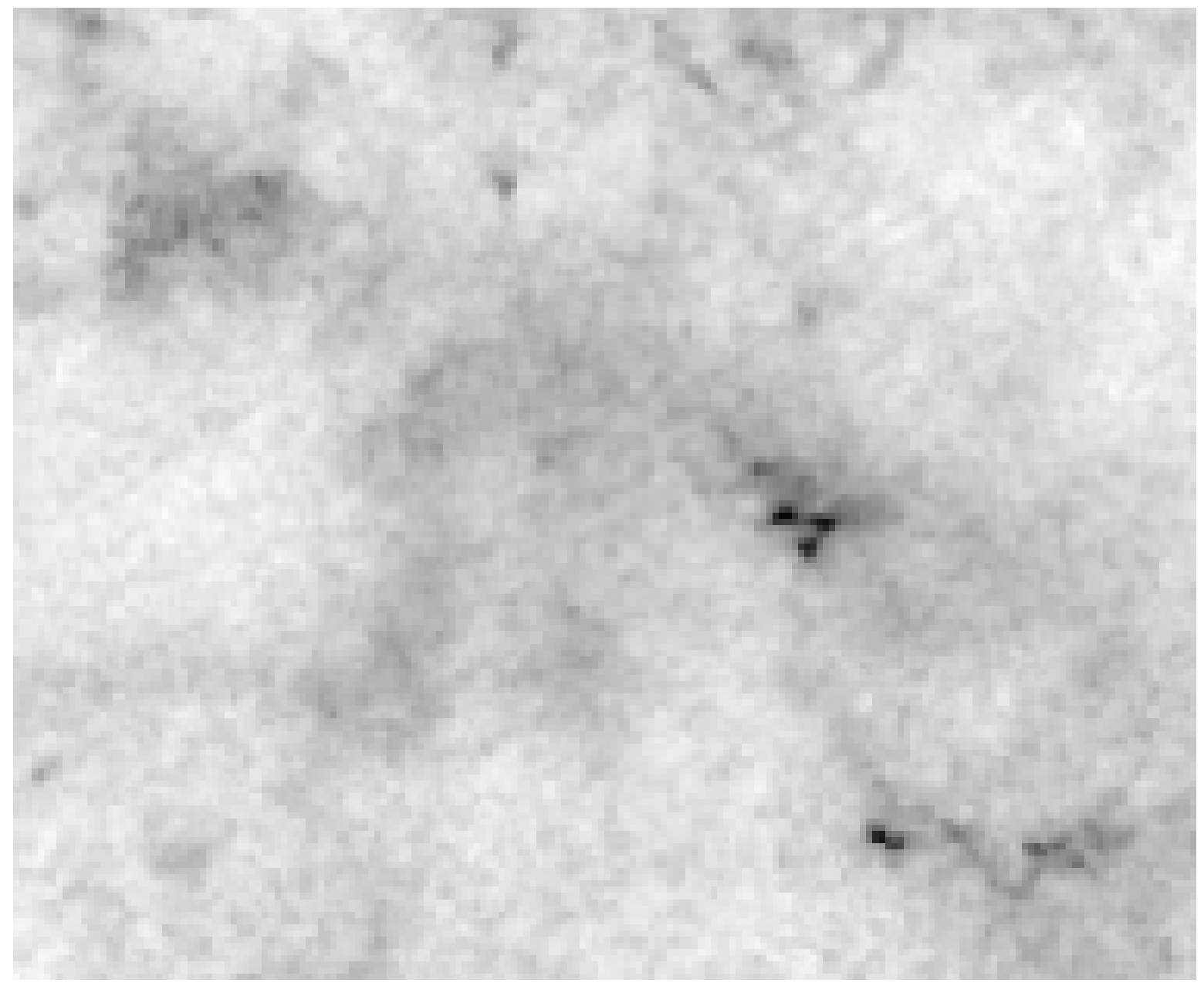}} at 180.5 -6.5
\axis left label {$b$\,[$^\circ$]}
ticks in long numbered from -11 to -2 by 2
      short unlabeled from -11 to -2 by 1 /
\axis right label {}
ticks in long unlabeled from -11 to -2 by 2
      short unlabeled from -11 to -2 by 1 /
\axis bottom label {$l$\,[$^\circ$]}
ticks in long numbered from 176 to 186 by 2
      short unlabeled from 175 to 186 by 1 /
\axis top label {}
ticks in long unlabeled from 176 to 186 by 2
      short unlabeled from 175 to 186 by 1 /
\put {\tiny $+$} at 176.579  -7.676	 
\put {\tiny $+$} at 181.507  -3.89 	 
\put {\tiny $+$} at 183.85   -10.88	 
\put {\tiny $+$} at 184.83   -3.817	 
\put {\tiny $+$} at 184.97   -10.479	 
\put {\tiny $\circ$} at 175.65  -3.65  
\put {\tiny $\circ$} at 176.57  -3.07  
\put {\tiny $\circ$} at 176.90  -3.59  
\put {\tiny $\circ$} at 179.07 -10.65  
\put {\tiny $\circ$} at 179.14 -10.53  
\put {\tiny $\circ$} at 179.18 -10.70  
\put {\tiny $\circ$} at 179.76  -3.81  
\put {\tiny $\circ$} at 180.68  -3.54  
\put {\tiny $\circ$} at 180.79 -10.86  
\put {\tiny $\circ$} at 181.39  -3.41  
\put {\tiny $\circ$} at 183.72  -3.67  
\setcoordinatesystem units <-6.3636mm,6.3636mm> point at 12.5715 0
\setplotarea x from 186 to 175 , y from -11 to -2
\put {\includegraphics[width=7cm]{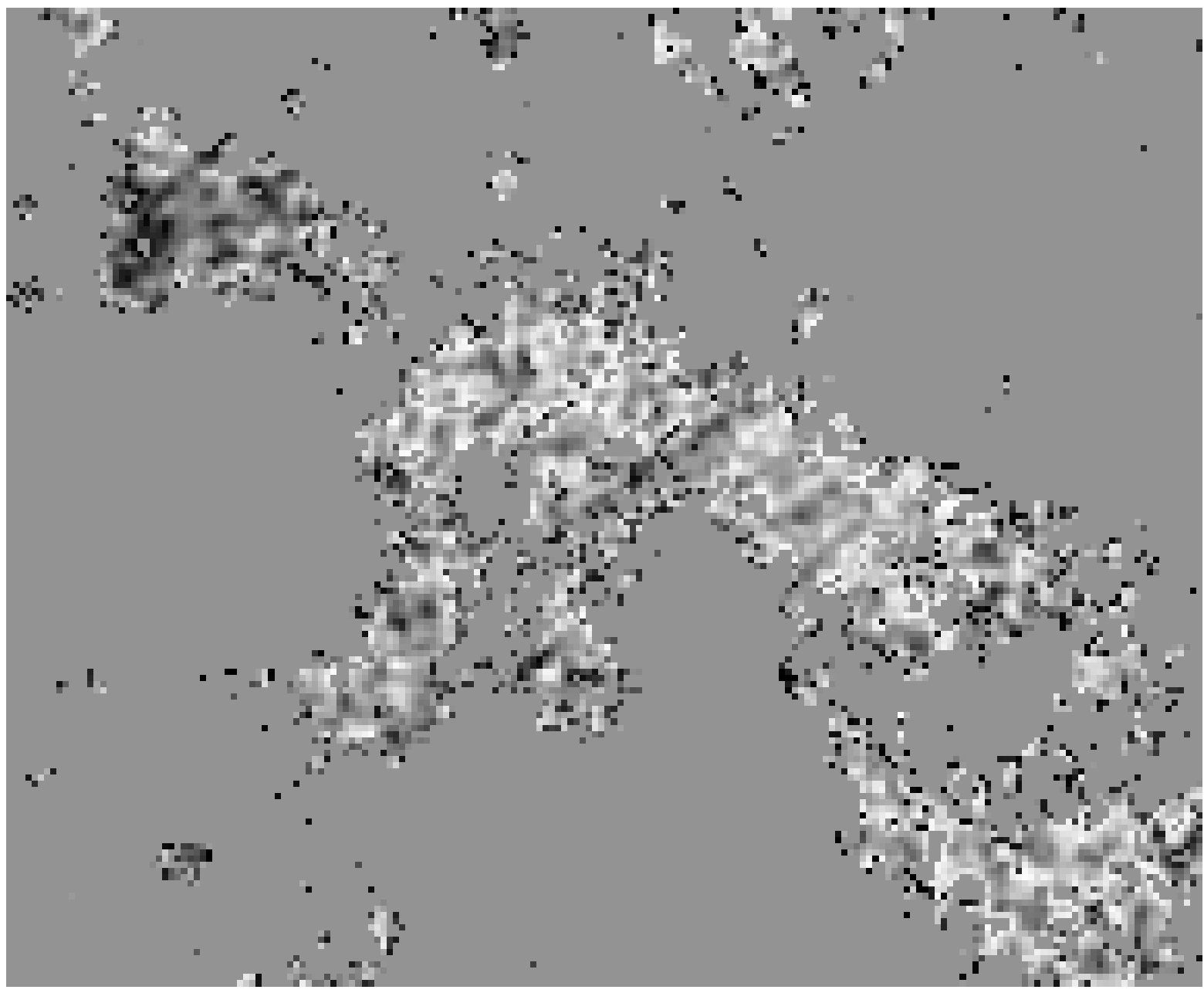}} at 180.5 -6.5
\axis left label {}
ticks in long unlabeled from -11 to -2 by 2
      short unlabeled from -11 to -2 by 1 /
\axis right label {}
ticks in long numbered from -11 to -2 by 2
      short unlabeled from -11 to -2 by 1 /
\axis bottom label {$l$\,[$^\circ$]}
ticks in long numbered from 176 to 186 by 2
      short unlabeled from 175 to 186 by 1 /
\axis top label {}
ticks in long unlabeled from 176 to 186 by 2
      short unlabeled from 175 to 186 by 1 /
\put {\tiny $+$} at 176.579  -7.676	 
\put {\tiny $+$} at 181.507  -3.89 	 
\put {\tiny $+$} at 183.85   -10.88	 
\put {\tiny $+$} at 184.83   -3.817	 
\put {\tiny $+$} at 184.97   -10.479	 
\put {\tiny $\circ$} at 175.65  -3.65  
\put {\tiny $\circ$} at 176.57  -3.07  
\put {\tiny $\circ$} at 176.90  -3.59  
\put {\tiny $\circ$} at 179.07 -10.65  
\put {\tiny $\circ$} at 179.14 -10.53  
\put {\tiny $\circ$} at 179.18 -10.70  
\put {\tiny $\circ$} at 179.76  -3.81  
\put {\tiny $\circ$} at 180.68  -3.54  
\put {\tiny $\circ$} at 180.79 -10.86  
\put {\tiny $\circ$} at 181.39  -3.41  
\put {\tiny $\circ$} at 183.72  -3.67  
\endpicture 
\caption{\label{map_aurigae2} As Fig.\,\ref{map_aurigae2} but for the region
Auriga\,2. Extinction values are scaled linearly from -0.33 to 6\,mag of 
optical extinction.} 
\end{figure*}

\begin{figure*}
\beginpicture
\setcoordinatesystem units <-6.3636mm,6.3636mm> point at 0 0
\setplotarea x from 127 to 116 , y from -4 to 10
\put {\includegraphics[width=7cm]{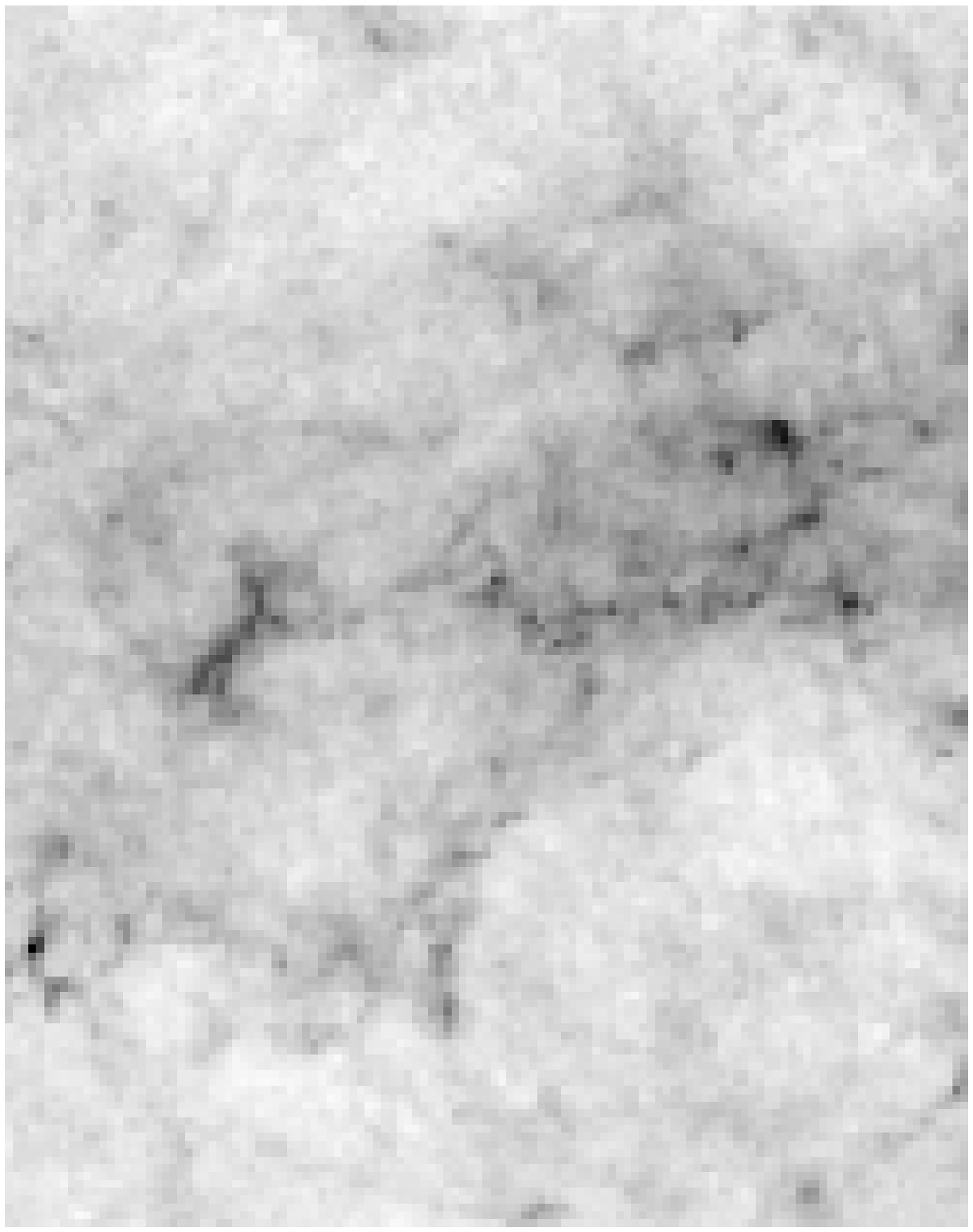}} at 121.5 3
\axis left label {$b$\,[$^\circ$]}
ticks in long numbered from -4 to 10 by 2
      short unlabeled from -4 to 10 by 1 /
\axis right label {}
ticks in long unlabeled from -4 to 10 by 2
      short unlabeled from -4 to 10 by 1 /
\axis bottom label {$l$\,[$^\circ$]}
ticks in long numbered from 116 to 127 by 2
      short unlabeled from 116 to 127 by 1 /
\axis top label {}
ticks in long unlabeled from 116 to 127 by 2
      short unlabeled from 116 to 127 by 1 /
\put {\tiny $+$} at 116.47   -2.994	 
\put {\tiny $+$} at 116.576  -1.533	 
\put {\tiny $+$} at 116.847  8.639 	 
\put {\tiny $+$} at 116.872  4.027 	 
\put {\tiny $+$} at 117.317  3.137 	 
\put {\tiny $+$} at 117.84   -2.723	 
\put {\tiny $+$} at 118.586  -1.09 	 
\put {\tiny $+$} at 118.969  -3.314	 
\put {\tiny $+$} at 119.646  3.187 	 
\put {\tiny $+$} at 120.074  1.03  	 
\put {\tiny $+$} at 120.257  1.289 	 
\put {\tiny $+$} at 120.364  -0.345	 
\put {\tiny $+$} at 120.944  6.119 
\put {\tiny $+$} at 121.581  -1.244	 
\put {\tiny $+$} at 121.931  2.128 	 
\put {\tiny $+$} at 122.864  3.097 	 
\put {\tiny $+$} at 123.046  1.78  	 
\put {\tiny $+$} at 123.129  -3.021	 
\put {\tiny $+$} at 123.179  2.923 	 
\put {\tiny $+$} at 123.591  5.605 
\put {\tiny $+$} at 123.624  -0.749
\put {\tiny $+$} at 126.049  -1.414	 
\put {\tiny $+$} at 126.131  0.373 	 
\put {\tiny $+$} at 126.317  -2.341	 
\put {\tiny $+$} at 126.326  1.024 	 
\put {\tiny $\circ$} at 116.12  -0.13  
\put {\tiny $\circ$} at 116.16  -0.51  
\put {\tiny $\circ$} at 116.21  +0.33  
\put {\tiny $\circ$} at 116.21  -0.35  
\put {\tiny $\circ$} at 116.27  -0.58  
\put {\tiny $\circ$} at 116.33  +0.66  
\put {\tiny $\circ$} at 116.44  -0.78  
\put {\tiny $\circ$} at 116.56  -0.57  
\put {\tiny $\circ$} at 116.59  -1.00  
\put {\tiny $\circ$} at 116.73  -1.28  
\put {\tiny $\circ$} at 117.05  +2.30  
\put {\tiny $\circ$} at 117.05  +2.31  
\put {\tiny $\circ$} at 117.16  +6.48  
\put {\tiny $\circ$} at 117.17  +5.66  
\put {\tiny $\circ$} at 117.20  +5.85  
\put {\tiny $\circ$} at 117.48  -3.76  
\put {\tiny $\circ$} at 117.56  +6.15  
\put {\tiny $\circ$} at 117.63  +1.22  
\put {\tiny $\circ$} at 117.63  +2.29  
\put {\tiny $\circ$} at 117.72  -1.01  
\put {\tiny $\circ$} at 117.79  -2.02  
\put {\tiny $\circ$} at 117.98  -1.26  
\put {\tiny $\circ$} at 118.15  +0.20  
\put {\tiny $\circ$} at 118.22  +5.00  
\put {\tiny $\circ$} at 118.37  +3.17  
\put {\tiny $\circ$} at 118.44  +4.73  
\put {\tiny $\circ$} at 118.55  -2.61  
\put {\tiny $\circ$} at 118.55  +5.15  
\put {\tiny $\circ$} at 118.60  +6.11  
\put {\tiny $\circ$} at 118.62  -1.32  
\put {\tiny $\circ$} at 118.63  +6.17  
\put {\tiny $\circ$} at 118.85  -1.64  
\put {\tiny $\circ$} at 118.98  -3.35  
\put {\tiny $\circ$} at 119.44  -0.92  
\put {\tiny $\circ$} at 119.51  +3.19  
\put {\tiny $\circ$} at 119.57  +5.62  
\put {\tiny $\circ$} at 119.71  -2.31  
\put {\tiny $\circ$} at 119.76  +1.71  
\put {\tiny $\circ$} at 119.80  -1.36  
\put {\tiny $\circ$} at 119.92  -0.10  
\put {\tiny $\circ$} at 120.11  -0.34  
\put {\tiny $\circ$} at 120.27  -2.54  
\put {\tiny $\circ$} at 120.56  -1.25  
\put {\tiny $\circ$} at 120.67  +0.58  
\put {\tiny $\circ$} at 120.73  +0.38  
\put {\tiny $\circ$} at 120.75  -0.93  
\put {\tiny $\circ$} at 120.87  +0.50  
\put {\tiny $\circ$} at 120.99  +8.62  
\put {\tiny $\circ$} at 121.01  +2.31  
\put {\tiny $\circ$} at 121.30  +0.66  
\put {\tiny $\circ$} at 121.50  -1.77  
\put {\tiny $\circ$} at 121.55  -0.89  
\put {\tiny $\circ$} at 121.64  -1.99  
\put {\tiny $\circ$} at 121.95  +1.20  
\put {\tiny $\circ$} at 121.97  -2.71  
\put {\tiny $\circ$} at 122.02  -1.07  
\put {\tiny $\circ$} at 122.09  +1.32  
\put {\tiny $\circ$} at 122.25  +1.53  
\put {\tiny $\circ$} at 122.65  +4.36  
\put {\tiny $\circ$} at 122.78  +1.26  
\put {\tiny $\circ$} at 122.80  +1.20  
\put {\tiny $\circ$} at 123.14  -1.29  
\put {\tiny $\circ$} at 123.20  +1.59  
\put {\tiny $\circ$} at 123.56  -3.43  
\put {\tiny $\circ$} at 123.81  -1.78  
\put {\tiny $\circ$} at 123.98  +1.10  
\put {\tiny $\circ$} at 124.27  -0.04  
\put {\tiny $\circ$} at 124.65  +2.83  
\put {\tiny $\circ$} at 124.69  -0.59  
\put {\tiny $\circ$} at 124.87  -3.14  
\put {\tiny $\circ$} at 124.90  +0.32  
\put {\tiny $\circ$} at 124.94  -1.22  
\put {\tiny $\circ$} at 125.35  -0.44  
\put {\tiny $\circ$} at 125.89  -2.60  
\put {\tiny $\circ$} at 126.06  -3.91  
\setcoordinatesystem units <-6.3636mm,6.3636mm> point at 12.5715 0
\setplotarea x from 127 to 116 , y from -4 to 10
\put {\includegraphics[width=7cm]{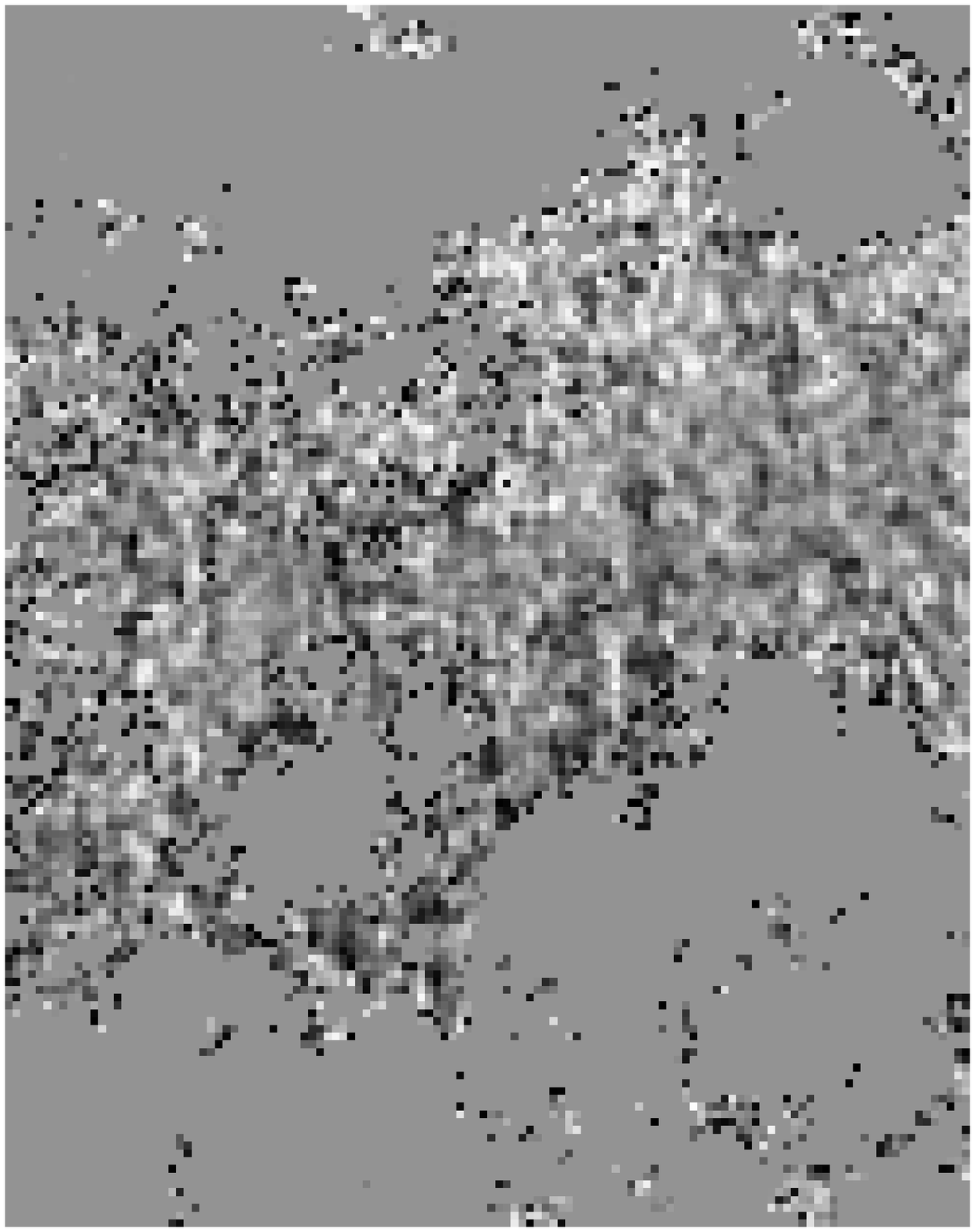}} at 121.5 3
\axis left label {}
ticks in long unlabeled from -4 to 10 by 2
      short unlabeled from -4 to 10 by 1 /
\axis right label {}
ticks in long numbered from -4 to 10 by 2
      short unlabeled from -4 to 10 by 1 /
\axis bottom label {$l$\,[$^\circ$]}
ticks in long numbered from 116 to 127 by 2
      short unlabeled from 116 to 127 by 1 /
\axis top label {}
ticks in long unlabeled from 116 to 127 by 2
      short unlabeled from 116 to 127 by 1 /
\put {\tiny $+$} at 116.47   -2.994	 
\put {\tiny $+$} at 116.576  -1.533	 
\put {\tiny $+$} at 116.847  8.639 	 
\put {\tiny $+$} at 116.872  4.027 	 
\put {\tiny $+$} at 117.317  3.137 	 
\put {\tiny $+$} at 117.84   -2.723	 
\put {\tiny $+$} at 118.586  -1.09 	 
\put {\tiny $+$} at 118.969  -3.314	 
\put {\tiny $+$} at 119.646  3.187 	 
\put {\tiny $+$} at 120.074  1.03  	 
\put {\tiny $+$} at 120.257  1.289 	 
\put {\tiny $+$} at 120.364  -0.345	 
\put {\tiny $+$} at 120.944  6.119 
\put {\tiny $+$} at 121.581  -1.244	 
\put {\tiny $+$} at 121.931  2.128 	 
\put {\tiny $+$} at 122.864  3.097 	 
\put {\tiny $+$} at 123.046  1.78  	 
\put {\tiny $+$} at 123.129  -3.021	 
\put {\tiny $+$} at 123.179  2.923 	 
\put {\tiny $+$} at 123.591  5.605 
\put {\tiny $+$} at 123.624  -0.749
\put {\tiny $+$} at 126.049  -1.414	 
\put {\tiny $+$} at 126.131  0.373 	 
\put {\tiny $+$} at 126.317  -2.341	 
\put {\tiny $+$} at 126.326  1.024 	 
\put {\tiny $\circ$} at 116.12  -0.13  
\put {\tiny $\circ$} at 116.16  -0.51  
\put {\tiny $\circ$} at 116.21  +0.33  
\put {\tiny $\circ$} at 116.21  -0.35  
\put {\tiny $\circ$} at 116.27  -0.58  
\put {\tiny $\circ$} at 116.33  +0.66  
\put {\tiny $\circ$} at 116.44  -0.78  
\put {\tiny $\circ$} at 116.56  -0.57  
\put {\tiny $\circ$} at 116.59  -1.00  
\put {\tiny $\circ$} at 116.73  -1.28  
\put {\tiny $\circ$} at 117.05  +2.30  
\put {\tiny $\circ$} at 117.05  +2.31  
\put {\tiny $\circ$} at 117.16  +6.48  
\put {\tiny $\circ$} at 117.17  +5.66  
\put {\tiny $\circ$} at 117.20  +5.85  
\put {\tiny $\circ$} at 117.48  -3.76  
\put {\tiny $\circ$} at 117.56  +6.15  
\put {\tiny $\circ$} at 117.63  +1.22  
\put {\tiny $\circ$} at 117.63  +2.29  
\put {\tiny $\circ$} at 117.72  -1.01  
\put {\tiny $\circ$} at 117.79  -2.02  
\put {\tiny $\circ$} at 117.98  -1.26  
\put {\tiny $\circ$} at 118.15  +0.20  
\put {\tiny $\circ$} at 118.22  +5.00  
\put {\tiny $\circ$} at 118.37  +3.17  
\put {\tiny $\circ$} at 118.44  +4.73  
\put {\tiny $\circ$} at 118.55  -2.61  
\put {\tiny $\circ$} at 118.55  +5.15  
\put {\tiny $\circ$} at 118.60  +6.11  
\put {\tiny $\circ$} at 118.62  -1.32  
\put {\tiny $\circ$} at 118.63  +6.17  
\put {\tiny $\circ$} at 118.85  -1.64  
\put {\tiny $\circ$} at 118.98  -3.35  
\put {\tiny $\circ$} at 119.44  -0.92  
\put {\tiny $\circ$} at 119.51  +3.19  
\put {\tiny $\circ$} at 119.57  +5.62  
\put {\tiny $\circ$} at 119.71  -2.31  
\put {\tiny $\circ$} at 119.76  +1.71  
\put {\tiny $\circ$} at 119.80  -1.36  
\put {\tiny $\circ$} at 119.92  -0.10  
\put {\tiny $\circ$} at 120.11  -0.34  
\put {\tiny $\circ$} at 120.27  -2.54  
\put {\tiny $\circ$} at 120.56  -1.25  
\put {\tiny $\circ$} at 120.67  +0.58  
\put {\tiny $\circ$} at 120.73  +0.38  
\put {\tiny $\circ$} at 120.75  -0.93  
\put {\tiny $\circ$} at 120.87  +0.50  
\put {\tiny $\circ$} at 120.99  +8.62  
\put {\tiny $\circ$} at 121.01  +2.31  
\put {\tiny $\circ$} at 121.30  +0.66  
\put {\tiny $\circ$} at 121.50  -1.77  
\put {\tiny $\circ$} at 121.55  -0.89  
\put {\tiny $\circ$} at 121.64  -1.99  
\put {\tiny $\circ$} at 121.95  +1.20  
\put {\tiny $\circ$} at 121.97  -2.71  
\put {\tiny $\circ$} at 122.02  -1.07  
\put {\tiny $\circ$} at 122.09  +1.32  
\put {\tiny $\circ$} at 122.25  +1.53  
\put {\tiny $\circ$} at 122.65  +4.36  
\put {\tiny $\circ$} at 122.78  +1.26  
\put {\tiny $\circ$} at 122.80  +1.20  
\put {\tiny $\circ$} at 123.14  -1.29  
\put {\tiny $\circ$} at 123.20  +1.59  
\put {\tiny $\circ$} at 123.56  -3.43  
\put {\tiny $\circ$} at 123.81  -1.78  
\put {\tiny $\circ$} at 123.98  +1.10  
\put {\tiny $\circ$} at 124.27  -0.04  
\put {\tiny $\circ$} at 124.65  +2.83  
\put {\tiny $\circ$} at 124.69  -0.59  
\put {\tiny $\circ$} at 124.87  -3.14  
\put {\tiny $\circ$} at 124.90  +0.32  
\put {\tiny $\circ$} at 124.94  -1.22  
\put {\tiny $\circ$} at 125.35  -0.44  
\put {\tiny $\circ$} at 125.89  -2.60  
\put {\tiny $\circ$} at 126.06  -3.91  
\endpicture 
\caption{\label{map_cassiopea} As Fig.\,\ref{map_aurigae2} but for the region
Cassiopeia. Extinction values are scaled linearly from -0.33 to 7\,mag of
optical extinction.} 
\end{figure*}

\begin{figure*}
\beginpicture
\setcoordinatesystem units <-3.3333mm,3.3333mm> point at 0 0
\setplotarea x from 148 to 127 , y from 3 to 15
\put {\includegraphics[width=7cm]{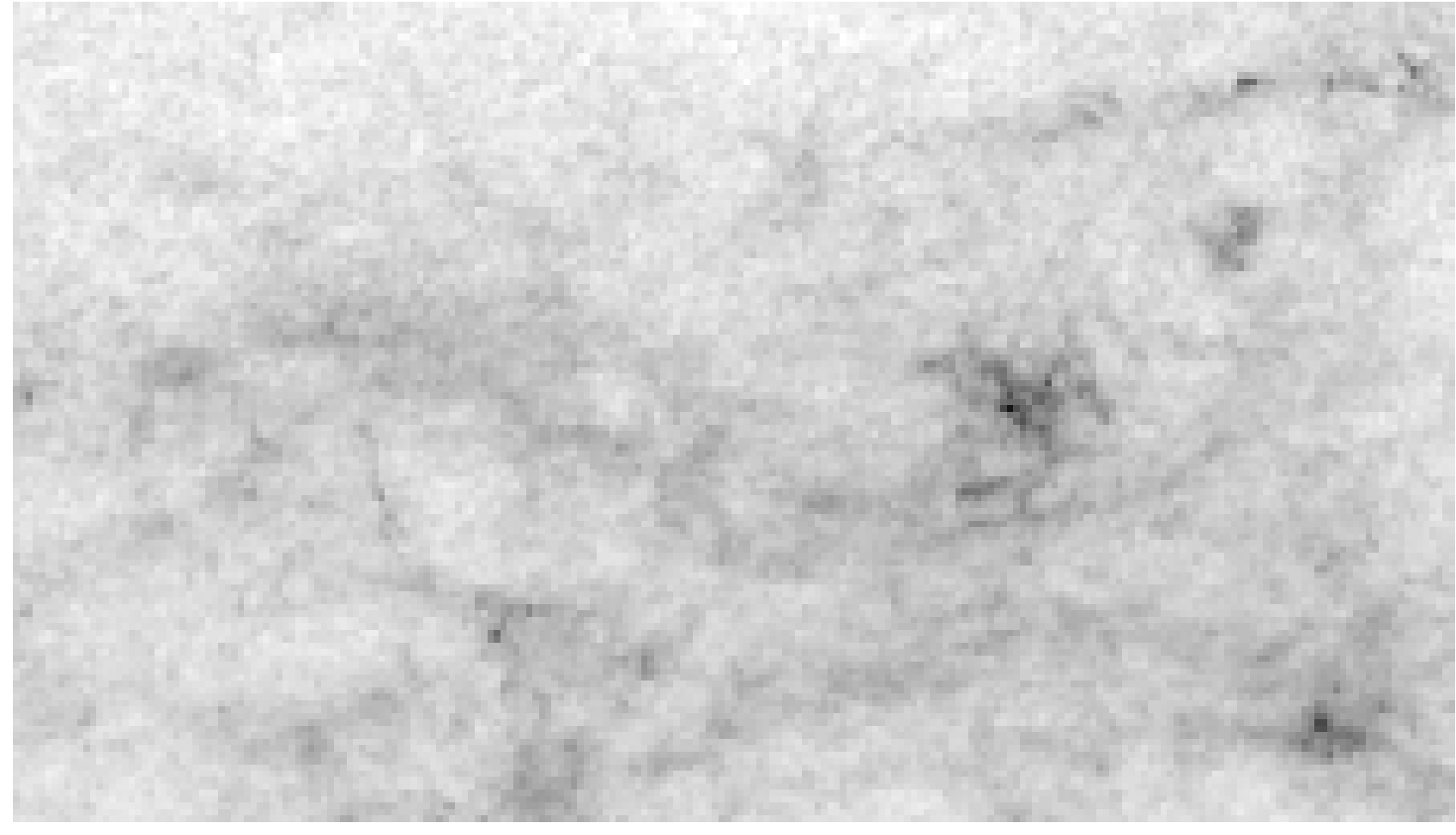}} at 137.5 9
\axis left label {$b$\,[$^\circ$]}
ticks in long numbered from 4 to 15 by 4
      short unlabeled from 3 to 15 by 1 /
\axis right label {}
ticks in long unlabeled from 4 to 15 by 4
      short unlabeled from 3 to 15 by 1 /
\axis bottom label {$l$\,[$^\circ$]}
ticks in long numbered from 128 to 148 by 4
      short unlabeled from 127 to 148 by 1 /
\axis top label {}
ticks in long unlabeled from 128 to 148 by 4
      short unlabeled from 127 to 148 by 1 /
\put {\tiny $+$} at 127.356  13.218
\put {\tiny $+$} at 127.595  3.402 	 
\put {\tiny $+$} at 127.83   3.508 	 
\put {\tiny $+$} at 128.811  8.652 
\put {\tiny $+$} at 129.426  5.813 
\put {\tiny $+$} at 129.914  5.888 
\put {\tiny $+$} at 130.122  11.544
\put {\tiny $+$} at 130.324  10.742
\put {\tiny $+$} at 130.374  11.061	 
\put {\tiny $+$} at 133.275  8.825 	 
\put {\tiny $+$} at 133.482  9.01  	 
\put {\tiny $+$} at 133.851  8.637 	 
\put {\tiny $+$} at 136.547  4.129 	 
\put {\tiny $+$} at 137.282  5.363 	 
\put {\tiny $+$} at 138.624  8.896 
\put {\tiny $+$} at 142.966  3.289 	 
\put {\tiny $+$} at 144.783  13.645
\put {\tiny $+$} at 146.265  3.122 	 
\put {\tiny $\circ$} at 127.28  +9.40  
\put {\tiny $\circ$} at 127.32 +13.31  
\put {\tiny $\circ$} at 129.17  +6.00  
\put {\tiny $\circ$} at 130.08 +11.10  
\put {\tiny $\circ$} at 130.11 +11.11  
\put {\tiny $\circ$} at 131.91  +4.60  
\put {\tiny $\circ$} at 138.01  +4.50  
\put {\tiny $\circ$} at 138.17 +10.58  
\put {\tiny $\circ$} at 138.58  +8.91  
\put {\tiny $\circ$} at 142.52  +6.22  
\put {\tiny $\circ$} at 143.68 +11.24  
\put {\tiny $\circ$} at 143.68  +7.66  
\put {\tiny $\circ$} at 143.94  +3.57  
\put {\tiny $\circ$} at 145.12  +3.68  
\put {\tiny $\circ$} at 147.48  +5.70  
\setcoordinatesystem units <-3.3333mm,3.3333mm> point at 24.0002 0
\setplotarea x from 148 to 127 , y from 3 to 15
\put {\includegraphics[width=7cm]{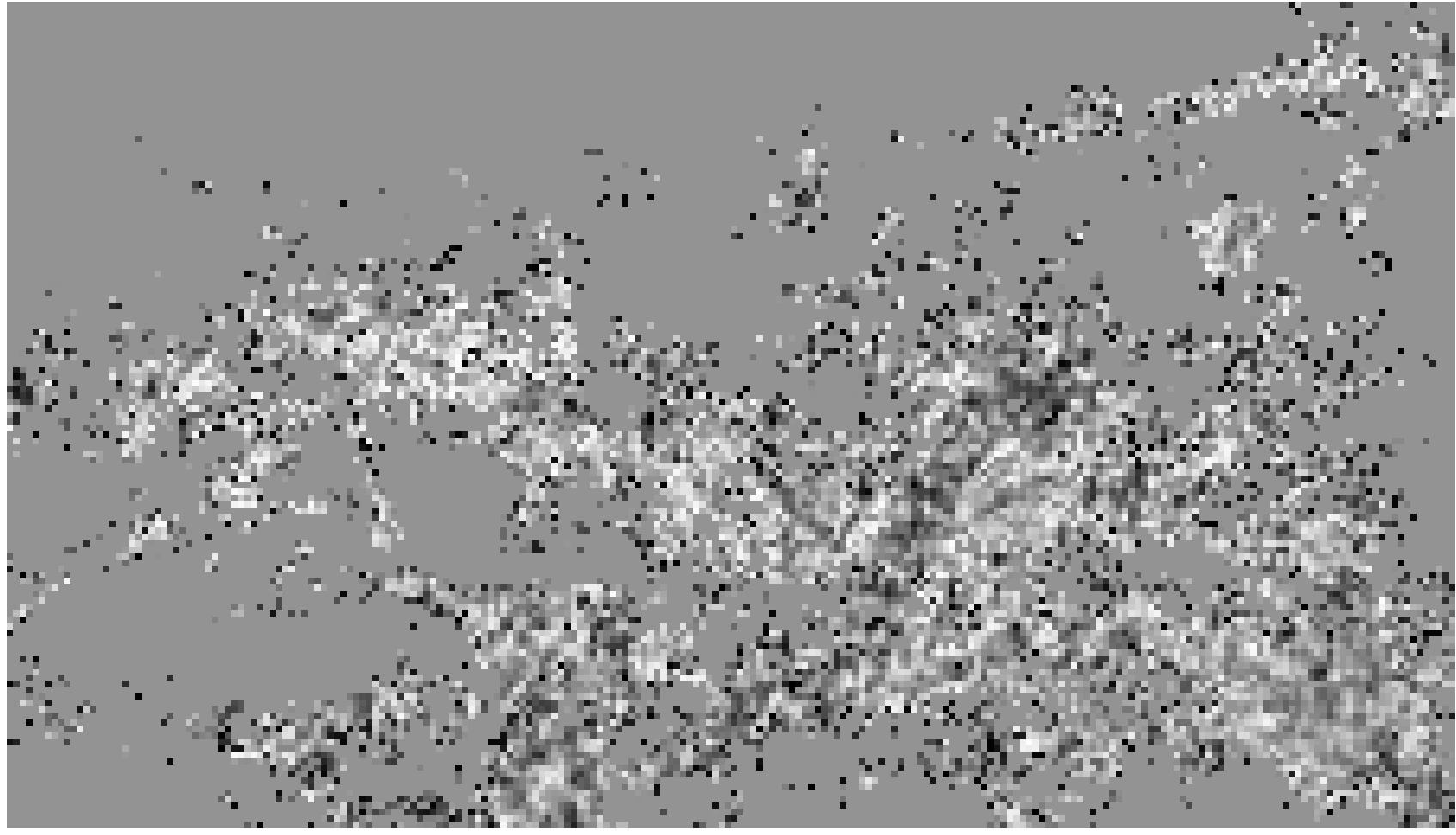}} at 137.5 9
\axis left label {}
ticks in long unlabeled from 4 to 15 by 4
      short unlabeled from 3 to 15 by 1 /
\axis right label {}
ticks in long numbered from 4 to 15 by 4
      short unlabeled from 3 to 15 by 1 /
\axis bottom label {$l$\,[$^\circ$]}
ticks in long numbered from 128 to 148 by 4
      short unlabeled from 127 to 148 by 1 /
\axis top label {}
ticks in long unlabeled from 128 to 148 by 4
      short unlabeled from 127 to 148 by 1 /
\put {\tiny $+$} at 127.356  13.218
\put {\tiny $+$} at 127.595  3.402 	 
\put {\tiny $+$} at 127.83   3.508 	 
\put {\tiny $+$} at 128.811  8.652 
\put {\tiny $+$} at 129.426  5.813 
\put {\tiny $+$} at 129.914  5.888 
\put {\tiny $+$} at 130.122  11.544
\put {\tiny $+$} at 130.324  10.742
\put {\tiny $+$} at 130.374  11.061	 
\put {\tiny $+$} at 133.275  8.825 	 
\put {\tiny $+$} at 133.482  9.01  	 
\put {\tiny $+$} at 133.851  8.637 	 
\put {\tiny $+$} at 136.547  4.129 	 
\put {\tiny $+$} at 137.282  5.363 	 
\put {\tiny $+$} at 138.624  8.896 
\put {\tiny $+$} at 142.966  3.289 	 
\put {\tiny $+$} at 144.783  13.645
\put {\tiny $+$} at 146.265  3.122 	 
\put {\tiny $\circ$} at 127.28  +9.40  
\put {\tiny $\circ$} at 127.32 +13.31  
\put {\tiny $\circ$} at 129.17  +6.00  
\put {\tiny $\circ$} at 130.08 +11.10  
\put {\tiny $\circ$} at 130.11 +11.11  
\put {\tiny $\circ$} at 131.91  +4.60  
\put {\tiny $\circ$} at 138.01  +4.50  
\put {\tiny $\circ$} at 138.17 +10.58  
\put {\tiny $\circ$} at 138.58  +8.91  
\put {\tiny $\circ$} at 142.52  +6.22  
\put {\tiny $\circ$} at 143.68 +11.24  
\put {\tiny $\circ$} at 143.68  +7.66  
\put {\tiny $\circ$} at 143.94  +3.57  
\put {\tiny $\circ$} at 145.12  +3.68  
\put {\tiny $\circ$} at 147.48  +5.70  
\endpicture 
\caption{\label{map_chameleopardalis1} As Fig.\,\ref{map_aurigae2} but for the
region Camelopardalis\,1. Extinction values are scaled linearly from -0.33
to 7\,mag of optical extinction.} 
\end{figure*}

\begin{figure*}
\beginpicture
\setcoordinatesystem units <-4.6667mm,4.6667mm> point at 0 0
\setplotarea x from 152 to 137 , y from -5 to 4
\put {\includegraphics[width=7cm]{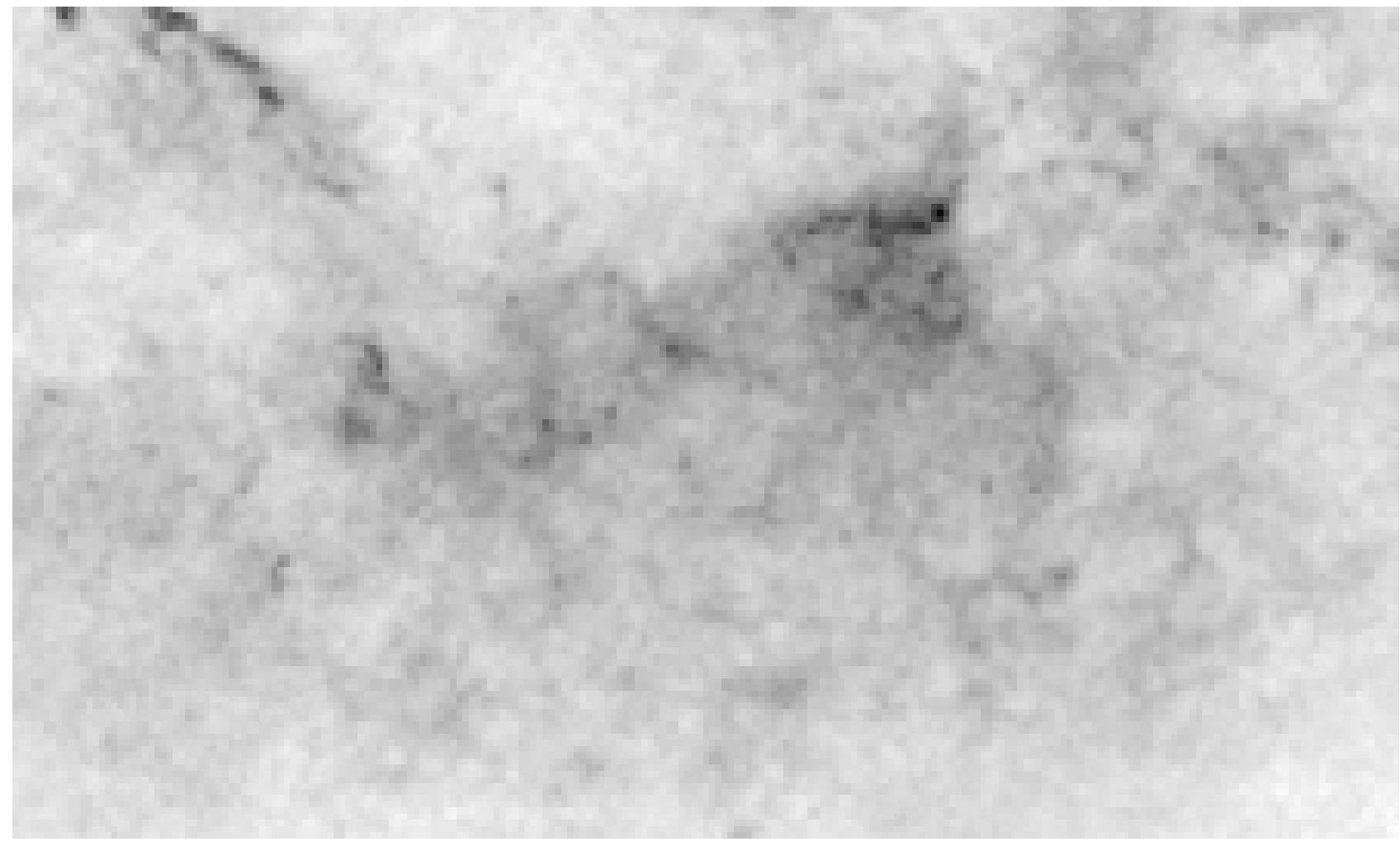}} at 144.5 -0.5
\axis left label {$b$\,[$^\circ$]}
ticks in long numbered from -5 to 4 by 4
      short unlabeled from -5 to 4 by 1 /
\axis right label {}
ticks in long unlabeled from -5 to 4 by 4
      short unlabeled from -5 to 4 by 1 /
\axis bottom label {$l$\,[$^\circ$]}
ticks in long numbered from 138 to 152 by 4
      short unlabeled from 137 to 152 by 1 /
\axis top label {}
ticks in long unlabeled from 138 to 152 by 4
      short unlabeled from 137 to 152 by 1 /
\put {\tiny $+$} at 137.03   1.096 	 
\put {\tiny $+$} at 137.275  -3.47 	 
\put {\tiny $+$} at 137.766  -2.903	 
\put {\tiny $+$} at 140.411  2.572 
\put {\tiny $+$} at 140.545  2.99  	 
\put {\tiny $+$} at 142.284  2.498 
\put {\tiny $+$} at 142.966  3.289 	 
\put {\tiny $+$} at 146.265  3.122 	 
\put {\tiny $+$} at 147.939  2.182 
\put {\tiny $+$} at 147.949  0.346 	 
\put {\tiny $+$} at 148.122  0.288 	 
\put {\tiny $+$} at 148.572  0.363 	 
\put {\tiny $+$} at 150.113  -0.551	 
\put {\tiny $+$} at 150.551  -0.213
\put {\tiny $+$} at 150.683  -0.594	 
\put {\tiny $+$} at 150.79   -0.583
\put {\tiny $+$} at 151.144  -0.651	 
\put {\tiny $\circ$} at 137.07  +3.00  
\put {\tiny $\circ$} at 137.25  +1.12  
\put {\tiny $\circ$} at 137.38  -3.97  
\put {\tiny $\circ$} at 137.54  +1.28  
\put {\tiny $\circ$} at 137.76  +1.50  
\put {\tiny $\circ$} at 137.78  -1.06  
\put {\tiny $\circ$} at 137.87  -1.75  
\put {\tiny $\circ$} at 138.07  -4.74  
\put {\tiny $\circ$} at 138.10  +1.35  
\put {\tiny $\circ$} at 138.15  +1.69  
\put {\tiny $\circ$} at 138.16  +1.23  
\put {\tiny $\circ$} at 138.26  +1.46  
\put {\tiny $\circ$} at 138.30  +1.56  
\put {\tiny $\circ$} at 138.50  +1.64  
\put {\tiny $\circ$} at 139.43  +0.22  
\put {\tiny $\circ$} at 139.91  +0.20  
\put {\tiny $\circ$} at 140.09  +2.09  
\put {\tiny $\circ$} at 140.93  +0.91  
\put {\tiny $\circ$} at 141.08  -1.06  
\put {\tiny $\circ$} at 142.00  +1.82  
\put {\tiny $\circ$} at 142.51  +1.99  
\put {\tiny $\circ$} at 143.05  -3.97  
\put {\tiny $\circ$} at 143.35  -0.07  
\put {\tiny $\circ$} at 143.74  -4.27  
\put {\tiny $\circ$} at 143.82  -1.57  
\put {\tiny $\circ$} at 144.19  -4.76  
\put {\tiny $\circ$} at 145.10  -3.97  
\put {\tiny $\circ$} at 145.84  -2.98  
\put {\tiny $\circ$} at 146.04  -2.78  
\put {\tiny $\circ$} at 146.92  -3.71  
\put {\tiny $\circ$} at 147.07  -0.51  
\put {\tiny $\circ$} at 148.09  -1.29  
\put {\tiny $\circ$} at 149.08  -1.99  
\put {\tiny $\circ$} at 149.74  -0.20  
\put {\tiny $\circ$} at 149.77  -1.02  
\put {\tiny $\circ$} at 149.86  +0.14  
\put {\tiny $\circ$} at 150.59  -0.94  
\put {\tiny $\circ$} at 150.86  -1.12  
\put {\tiny $\circ$} at 150.98  -0.47  
\put {\tiny $\circ$} at 151.18  +2.12  
\put {\tiny $\circ$} at 151.29  +1.97  
\setcoordinatesystem units <-4.6667mm,4.6667mm> point at 17.1427 0
\setplotarea x from 152 to 137 , y from -5 to 4
\put {\includegraphics[width=7cm]{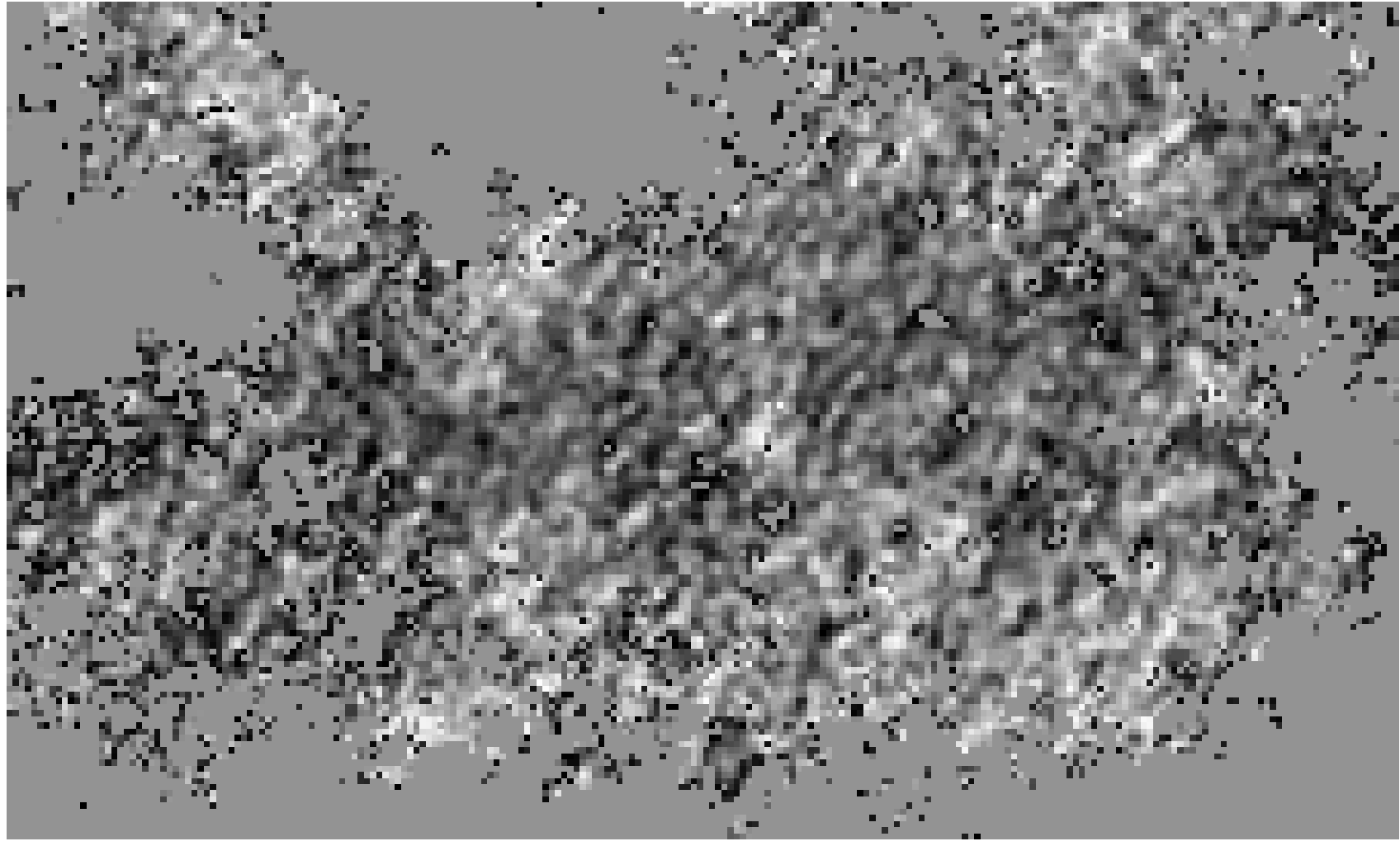}} at 144.5 -0.5
\axis left label {}
ticks in long unlabeled from -5 to 4 by 4
      short unlabeled from -5 to 4 by 1 /
\axis right label {}
ticks in long numbered from -5 to 4 by 4
      short unlabeled from -5 to 4 by 1 /
\axis bottom label {$l$\,[$^\circ$]}
ticks in long numbered from 138 to 152 by 4
      short unlabeled from 137 to 152 by 1 /
\axis top label {}
ticks in long unlabeled from 138 to 152 by 4
      short unlabeled from 137 to 152 by 1 /
\put {\tiny $+$} at 137.03   1.096 	 
\put {\tiny $+$} at 137.275  -3.47 	 
\put {\tiny $+$} at 137.766  -2.903	 
\put {\tiny $+$} at 140.411  2.572 
\put {\tiny $+$} at 140.545  2.99  	 
\put {\tiny $+$} at 142.284  2.498 
\put {\tiny $+$} at 142.966  3.289 	 
\put {\tiny $+$} at 146.265  3.122 	 
\put {\tiny $+$} at 147.939  2.182 
\put {\tiny $+$} at 147.949  0.346 	 
\put {\tiny $+$} at 148.122  0.288 	 
\put {\tiny $+$} at 148.572  0.363 	 
\put {\tiny $+$} at 150.113  -0.551	 
\put {\tiny $+$} at 150.551  -0.213
\put {\tiny $+$} at 150.683  -0.594	 
\put {\tiny $+$} at 150.79   -0.583
\put {\tiny $+$} at 151.144  -0.651	 
\put {\tiny $\circ$} at 137.07  +3.00  
\put {\tiny $\circ$} at 137.25  +1.12  
\put {\tiny $\circ$} at 137.38  -3.97  
\put {\tiny $\circ$} at 137.54  +1.28  
\put {\tiny $\circ$} at 137.76  +1.50  
\put {\tiny $\circ$} at 137.78  -1.06  
\put {\tiny $\circ$} at 137.87  -1.75  
\put {\tiny $\circ$} at 138.07  -4.74  
\put {\tiny $\circ$} at 138.10  +1.35  
\put {\tiny $\circ$} at 138.15  +1.69  
\put {\tiny $\circ$} at 138.16  +1.23  
\put {\tiny $\circ$} at 138.26  +1.46  
\put {\tiny $\circ$} at 138.30  +1.56  
\put {\tiny $\circ$} at 138.50  +1.64  
\put {\tiny $\circ$} at 139.43  +0.22  
\put {\tiny $\circ$} at 139.91  +0.20  
\put {\tiny $\circ$} at 140.09  +2.09  
\put {\tiny $\circ$} at 140.93  +0.91  
\put {\tiny $\circ$} at 141.08  -1.06  
\put {\tiny $\circ$} at 142.00  +1.82  
\put {\tiny $\circ$} at 142.51  +1.99  
\put {\tiny $\circ$} at 143.05  -3.97  
\put {\tiny $\circ$} at 143.35  -0.07  
\put {\tiny $\circ$} at 143.74  -4.27  
\put {\tiny $\circ$} at 143.82  -1.57  
\put {\tiny $\circ$} at 144.19  -4.76  
\put {\tiny $\circ$} at 145.10  -3.97  
\put {\tiny $\circ$} at 145.84  -2.98  
\put {\tiny $\circ$} at 146.04  -2.78  
\put {\tiny $\circ$} at 146.92  -3.71  
\put {\tiny $\circ$} at 147.07  -0.51  
\put {\tiny $\circ$} at 148.09  -1.29  
\put {\tiny $\circ$} at 149.08  -1.99  
\put {\tiny $\circ$} at 149.74  -0.20  
\put {\tiny $\circ$} at 149.77  -1.02  
\put {\tiny $\circ$} at 149.86  +0.14  
\put {\tiny $\circ$} at 150.59  -0.94  
\put {\tiny $\circ$} at 150.86  -1.12  
\put {\tiny $\circ$} at 150.98  -0.47  
\put {\tiny $\circ$} at 151.18  +2.12  
\put {\tiny $\circ$} at 151.29  +1.97  
\endpicture 
\caption{\label{map_chameleopardalis2} As Fig.\,\ref{map_aurigae2} but for the
region Camelopardalis\,2. Extinction values are scaled linearly from -0.33
to 8\,mag of optical extinction.} 
\end{figure*}

\begin{figure*}
\beginpicture
\setcoordinatesystem units <-7.7778mm,7.7778mm> point at 0 0
\setplotarea x from 161 to 152 , y from -4 to 8
\put {\includegraphics[width=7cm]{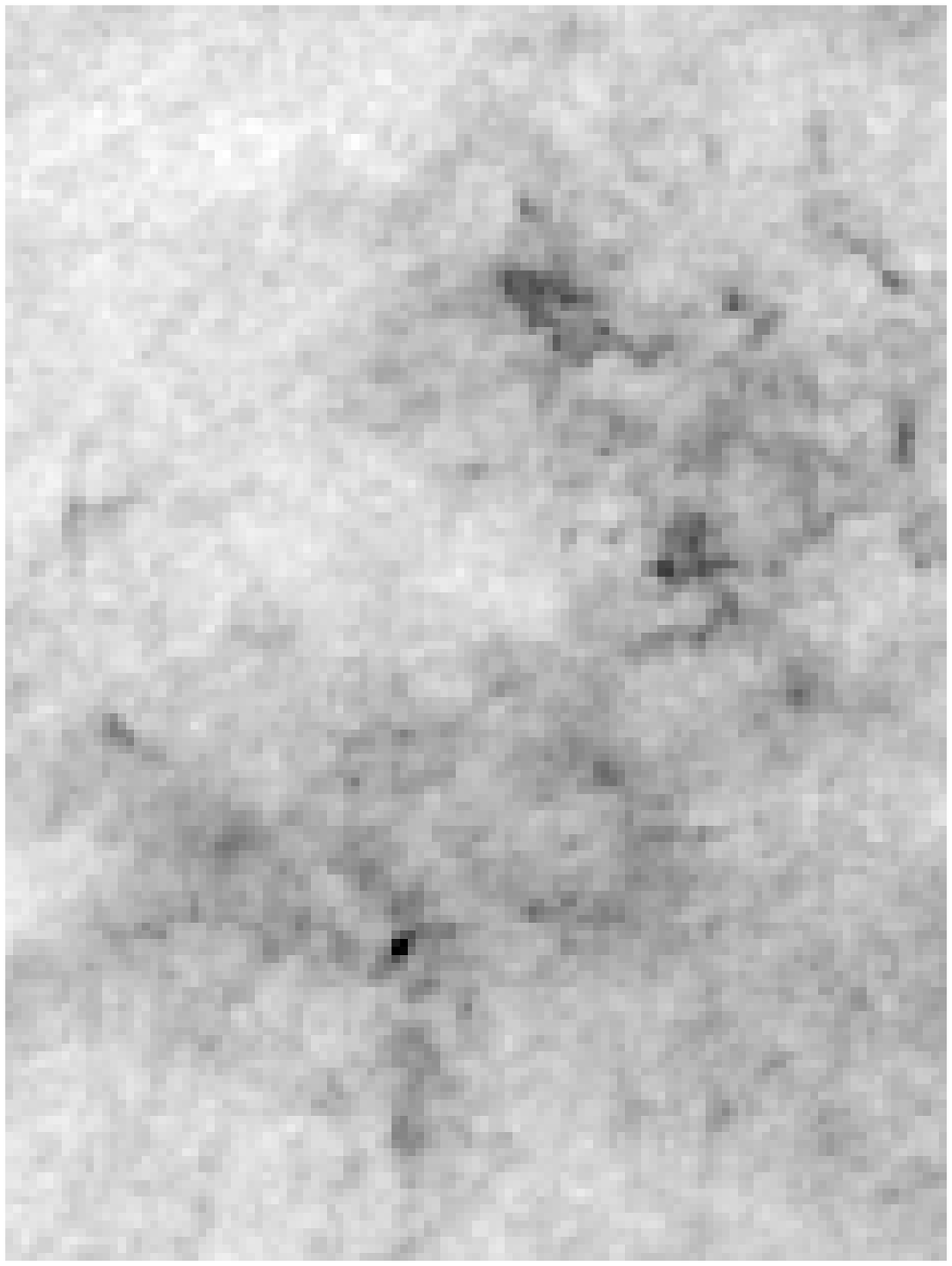}} at 156.5 2
\axis left label {$b$\,[$^\circ$]}
ticks in long numbered from -4 to 8 by 2
      short unlabeled from -4 to 8 by 1 /
\axis right label {}
ticks in long unlabeled from -4 to 8 by 2
      short unlabeled from -4 to 8 by 1 /
\axis bottom label {$l$\,[$^\circ$]}
ticks in long numbered from 152 to 161 by 2
      short unlabeled from 152 to 161 by 1 /
\axis top label {}
ticks in long unlabeled from 152 to 161 by 2
      short unlabeled from 152 to 161 by 1 /
\put {\tiny $+$} at 152.408  1.484 	 
\put {\tiny $+$} at 153.409  -1.877	 
\put {\tiny $+$} at 156.454  5.765 	 
\put {\tiny $+$} at 156.649  5.359 	 
\put {\tiny $+$} at 156.888  -2.157	 
\put {\tiny $+$} at 156.934  0.968 	 
\put {\tiny $+$} at 157.905  5.126 	 
\put {\tiny $+$} at 159.144  3.279 	 
\put {\tiny $+$} at 159.357  2.58  	 
\put {\tiny $+$} at 160.131  0.964 	 
\put {\tiny $\circ$} at 152.06  +0.26  
\put {\tiny $\circ$} at 152.29  +6.40  
\put {\tiny $\circ$} at 152.34  -0.29  
\put {\tiny $\circ$} at 152.67  -1.50  
\put {\tiny $\circ$} at 153.36  +0.18  
\put {\tiny $\circ$} at 154.35  +2.61  
\put {\tiny $\circ$} at 154.49  -3.41  
\put {\tiny $\circ$} at 154.65  +2.44  
\put {\tiny $\circ$} at 154.88  +2.50  
\put {\tiny $\circ$} at 155.10  +5.86  
\put {\tiny $\circ$} at 155.34  +2.58  
\put {\tiny $\circ$} at 155.36  +2.61  
\put {\tiny $\circ$} at 157.08  -3.64  
\put {\tiny $\circ$} at 158.44  +4.85  
\put {\tiny $\circ$} at 158.60  -1.58  
\put {\tiny $\circ$} at 159.30  -2.89  
\put {\tiny $\circ$} at 160.45  -2.46  
\setcoordinatesystem units <-7.7778mm,7.7778mm> point at 10.2857 0
\setplotarea x from 161 to 152 , y from -4 to 8
\put {\includegraphics[width=7cm]{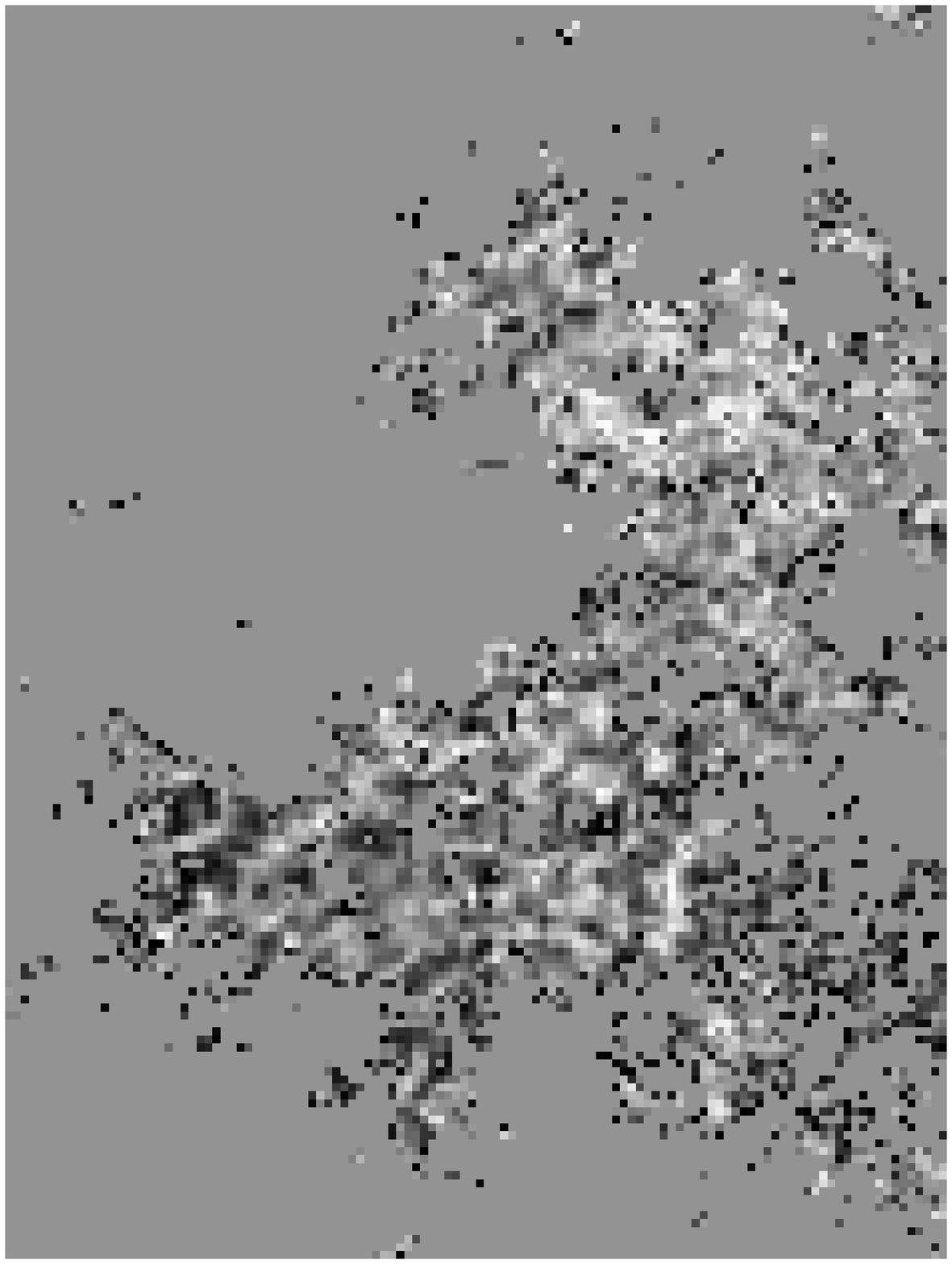}} at 156.5 2
\axis left label {}
ticks in long unlabeled from -4 to 8 by 2
      short unlabeled from -4 to 8 by 1 /
\axis right label {}
ticks in long numbered from -4 to 8 by 2
      short unlabeled from -4 to 8 by 1 /
\axis bottom label {$l$\,[$^\circ$]}
ticks in long numbered from 152 to 161 by 2
      short unlabeled from 152 to 161 by 1 /
\axis top label {}
ticks in long unlabeled from 152 to 161 by 2
      short unlabeled from 152 to 161 by 1 /
\put {\tiny $+$} at 152.408  1.484 	 
\put {\tiny $+$} at 153.409  -1.877	 
\put {\tiny $+$} at 156.454  5.765 	 
\put {\tiny $+$} at 156.649  5.359 	 
\put {\tiny $+$} at 156.888  -2.157	 
\put {\tiny $+$} at 156.934  0.968 	 
\put {\tiny $+$} at 157.905  5.126 	 
\put {\tiny $+$} at 159.144  3.279 	 
\put {\tiny $+$} at 159.357  2.58  	 
\put {\tiny $+$} at 160.131  0.964 	 
\put {\tiny $\circ$} at 152.06  +0.26  
\put {\tiny $\circ$} at 152.29  +6.40  
\put {\tiny $\circ$} at 152.34  -0.29  
\put {\tiny $\circ$} at 152.67  -1.50  
\put {\tiny $\circ$} at 153.36  +0.18  
\put {\tiny $\circ$} at 154.35  +2.61  
\put {\tiny $\circ$} at 154.49  -3.41  
\put {\tiny $\circ$} at 154.65  +2.44  
\put {\tiny $\circ$} at 154.88  +2.50  
\put {\tiny $\circ$} at 155.10  +5.86  
\put {\tiny $\circ$} at 155.34  +2.58  
\put {\tiny $\circ$} at 155.36  +2.61  
\put {\tiny $\circ$} at 157.08  -3.64  
\put {\tiny $\circ$} at 158.44  +4.85  
\put {\tiny $\circ$} at 158.60  -1.58  
\put {\tiny $\circ$} at 159.30  -2.89  
\put {\tiny $\circ$} at 160.45  -2.46  
\endpicture 
\caption{\label{map_chameleopardalis3} As Fig.\,\ref{map_aurigae2} but for the
region Camelopardalis\,3. Extinction values are scaled linearly from -0.33
to 5\,mag of optical extinction.} 
\end{figure*}

\begin{figure*}
\beginpicture
\setcoordinatesystem units <-5.3846mm,5.3846mm> point at 0 0
\setplotarea x from 201 to 188 , y from -18 to -6
\put {\includegraphics[width=7cm]{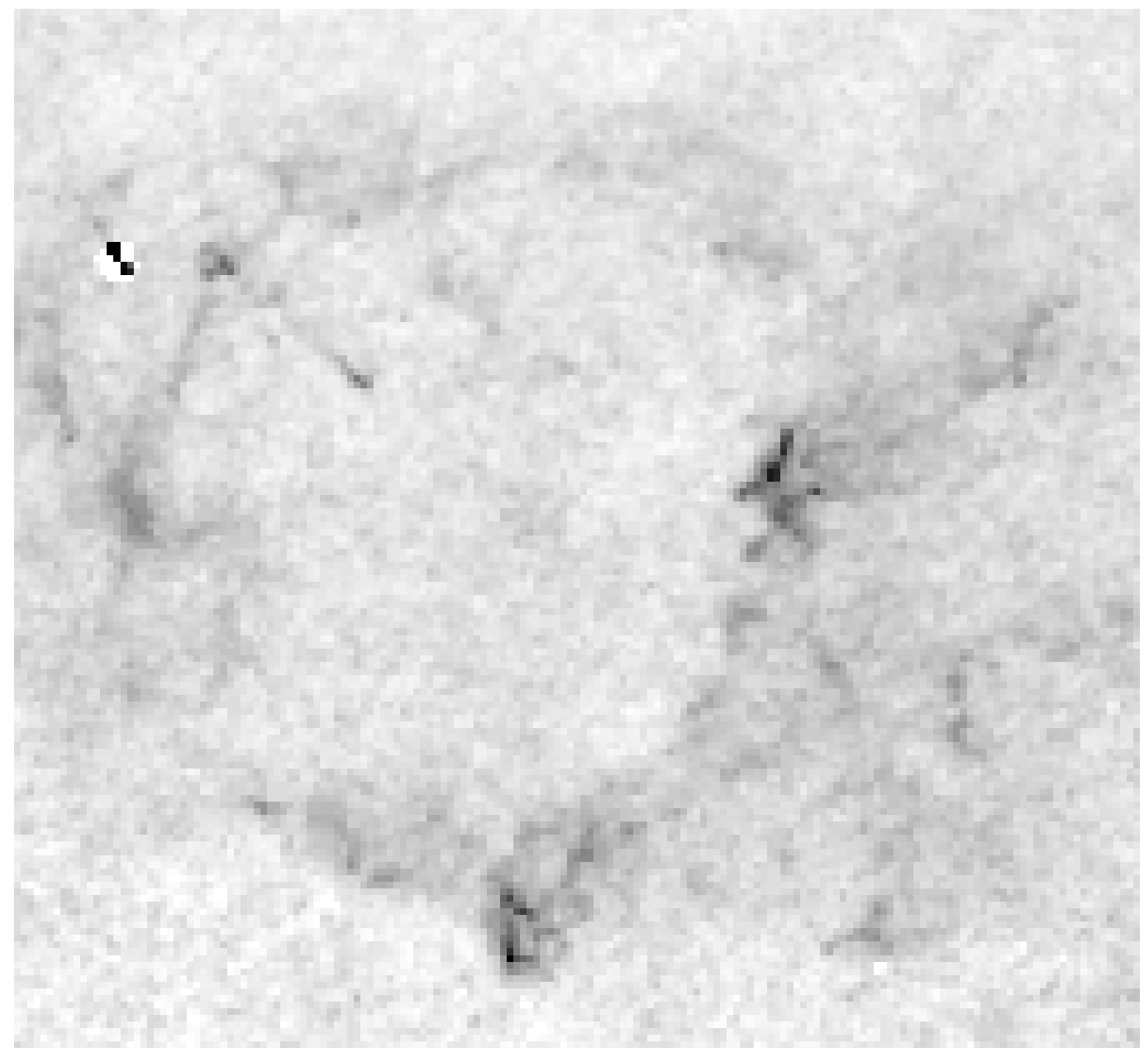}} at 194.5 -12
\axis left label {$b$\,[$^\circ$]}
ticks in long numbered from -18 to -6 by 4
      short unlabeled from -18 to -6 by 1 /
\axis right label {}
ticks in long unlabeled from -18 to -6 by 4
      short unlabeled from -18 to -6 by 1 /
\axis bottom label {$l$\,[$^\circ$]}
ticks in long numbered from 188 to 201 by 4
      short unlabeled from 188 to 201 by 1 /
\axis top label {}
ticks in long unlabeled from 188 to 201 by 4
      short unlabeled from 188 to 201 by 1 /
\put {\tiny $+$} at 188.062  -9.841	 
\put {\tiny $+$} at 188.403  -8.792	 
\put {\tiny $+$} at 188.665  -8.217	 
\put {\tiny $+$} at 191.326  -11.41	 
\put {\tiny $+$} at 192.346  -13.153	 
\put {\tiny $+$} at 192.415  -16.666	 
\put {\tiny $+$} at 194.359  -10.155	 
\put {\tiny $+$} at 195.127  -11.969	 
\put {\tiny $+$} at 197.626  -16.548	 
\put {\tiny $\circ$} at 188.86 -10.85  
\put {\tiny $\circ$} at 192.15 -13.71  
\put {\tiny $\circ$} at 192.45 -11.31  
\put {\tiny $\circ$} at 192.62 -11.63  
\put {\tiny $\circ$} at 195.10 -15.32  
\put {\tiny $\circ$} at 195.27 -12.10  
\put {\tiny $\circ$} at 195.93 -16.29  
\put {\tiny $\circ$} at 196.59 -15.10  
\put {\tiny $\circ$} at 196.82 -10.56  
\put {\tiny $\circ$} at 198.97 -10.41  
\put {\tiny $\circ$} at 199.80  -8.05  
\put {\tiny $\circ$} at 199.86 -16.48  
\setcoordinatesystem units <-5.3846mm,5.3846mm> point at 14.8572 0
\setplotarea x from 201 to 188 , y from -18 to -6
\put {\includegraphics[width=7cm]{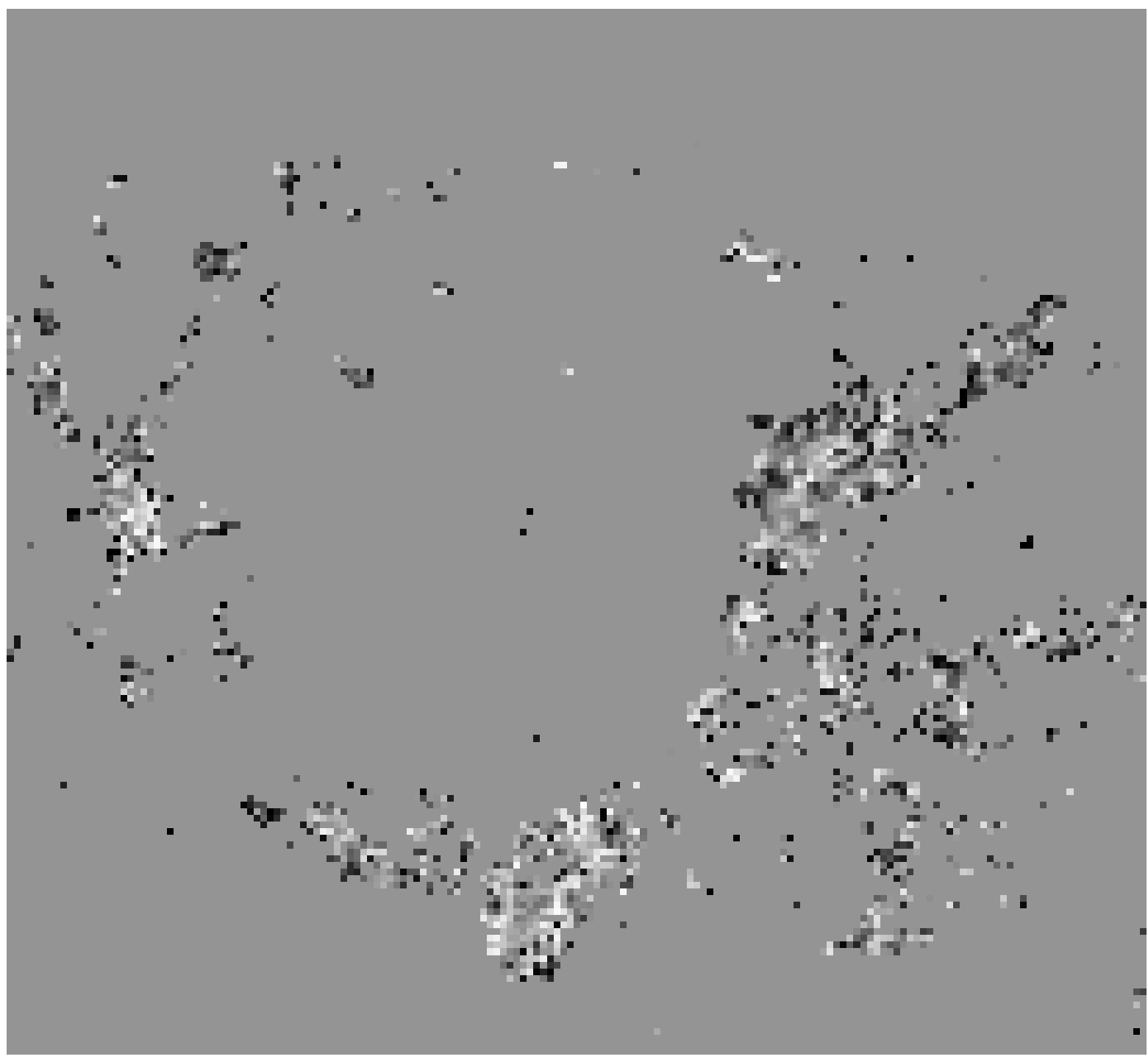}} at 194.5 -12
\axis left label {}
ticks in long unlabeled from -18 to -6 by 4
      short unlabeled from -18 to -6 by 1 /
\axis right label {}
ticks in long numbered from -18 to -6 by 4
      short unlabeled from -18 to -6 by 1 /
\axis bottom label {$l$\,[$^\circ$]}
ticks in long numbered from 188 to 201 by 4
      short unlabeled from 188 to 201 by 1 /
\axis top label {}
ticks in long unlabeled from 188 to 201 by 4
      short unlabeled from 188 to 201 by 1 /
\put {\tiny $+$} at 188.062  -9.841	 
\put {\tiny $+$} at 188.403  -8.792	 
\put {\tiny $+$} at 188.665  -8.217	 
\put {\tiny $+$} at 191.326  -11.41	 
\put {\tiny $+$} at 192.346  -13.153	 
\put {\tiny $+$} at 192.415  -16.666	 
\put {\tiny $+$} at 194.359  -10.155	 
\put {\tiny $+$} at 195.127  -11.969	 
\put {\tiny $+$} at 197.626  -16.548	 
\put {\tiny $\circ$} at 188.86 -10.85  
\put {\tiny $\circ$} at 192.15 -13.71  
\put {\tiny $\circ$} at 192.45 -11.31  
\put {\tiny $\circ$} at 192.62 -11.63  
\put {\tiny $\circ$} at 195.10 -15.32  
\put {\tiny $\circ$} at 195.27 -12.10  
\put {\tiny $\circ$} at 195.93 -16.29  
\put {\tiny $\circ$} at 196.59 -15.10  
\put {\tiny $\circ$} at 196.82 -10.56  
\put {\tiny $\circ$} at 198.97 -10.41  
\put {\tiny $\circ$} at 199.80  -8.05  
\put {\tiny $\circ$} at 199.86 -16.48  
\endpicture 
\caption{\label{map_lori} As Fig.\,\ref{map_aurigae2} but for the region
$\lambda$-Ori. Extinction values are scaled linearly from -0.33 to 6\,mag of 
optical extinction.} 
\end{figure*}

\begin{figure*}
\beginpicture
\setcoordinatesystem units <-4.6667mm,4.6667mm> point at 0 0
\setplotarea x from 227 to 212 , y from -13 to 1
\put {\includegraphics[width=7cm]{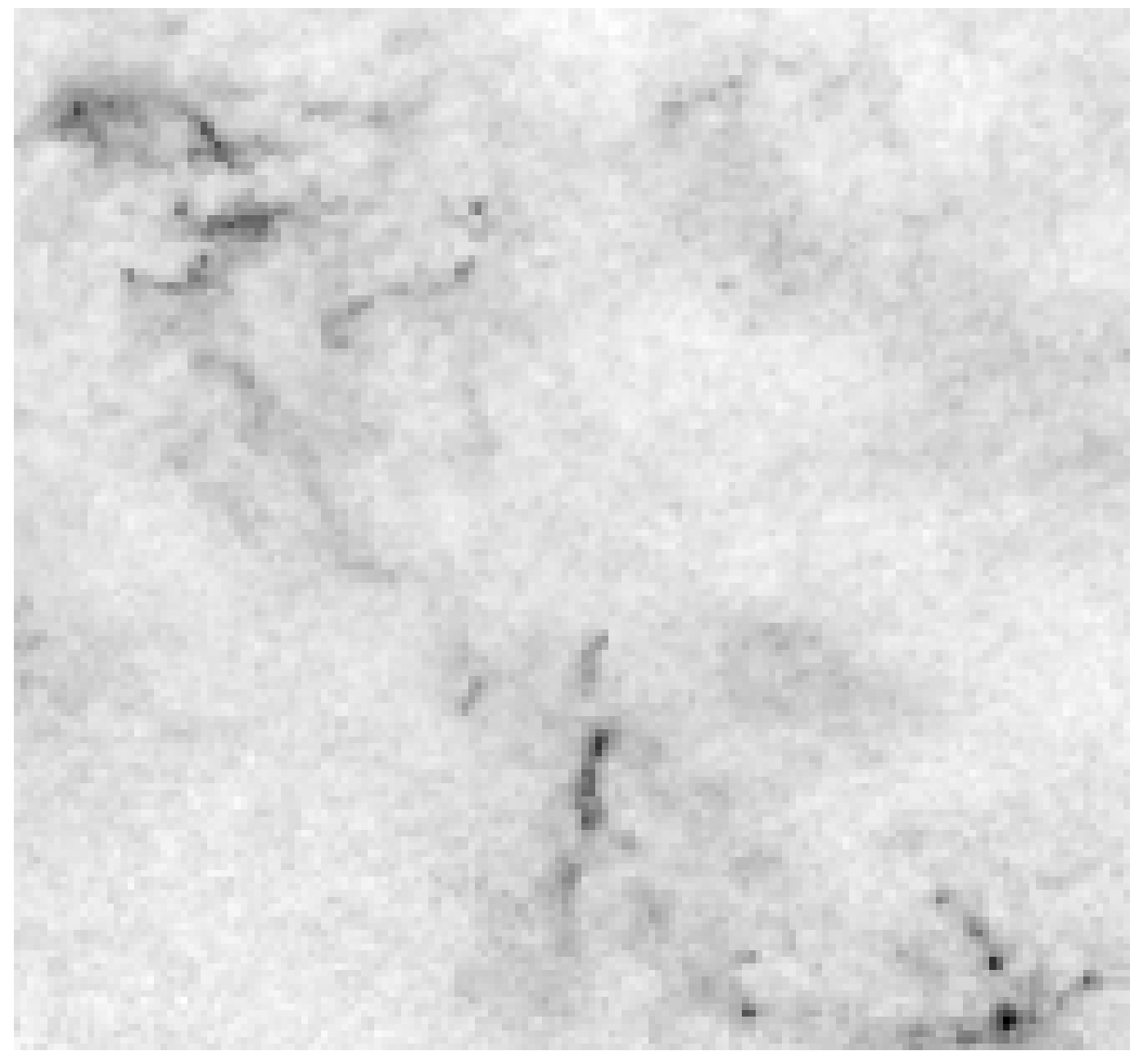}} at 219.5 -6
\axis left label {$b$\,[$^\circ$]}
ticks in long numbered from -12 to 1 by 4
      short unlabeled from -13 to 1 by 1 /
\axis right label {}
ticks in long unlabeled from -12 to 1 by 4
      short unlabeled from -13 to 1 by 1 /
\axis bottom label {$l$\,[$^\circ$]}
ticks in long numbered from 212 to 227 by 4
      short unlabeled from 212 to 227 by 1 /
\axis top label {}
ticks in long unlabeled from 212 to 227 by 4
      short unlabeled from 212 to 227 by 1 /
\put {\tiny $+$} at 212.426  0.244 	 
\put {\tiny $+$} at 212.453  -0.856	 
\put {\tiny $+$} at 212.647  -2.523	 
\put {\tiny $+$} at 213.181  -3.289	 
\put {\tiny $+$} at 213.305  0.302 	 
\put {\tiny $+$} at 213.892  -4.311	 
\put {\tiny $+$} at 213.913  -12.092	 
\put {\tiny $+$} at 214.413  -5.518	 
\put {\tiny $+$} at 215.42   -12.823	 
\put {\tiny $+$} at 215.706  -6.157	 
\put {\tiny $+$} at 215.711  -10.065	 
\put {\tiny $+$} at 217.034  -11.789	 
\put {\tiny $+$} at 217.157  -12.517	 
\put {\tiny $+$} at 217.674  -2.696	 
\put {\tiny $+$} at 217.748  -6.646	 
\put {\tiny $+$} at 218.157  -0.626	 
\put {\tiny $+$} at 219.147  -10.556	 
\put {\tiny $+$} at 219.337  -9.942	 
\put {\tiny $+$} at 219.488  -1.724	 
\put {\tiny $+$} at 220.07   -0.112	 
\put {\tiny $+$} at 220.312  -0.084	 
\put {\tiny $+$} at 220.352  -7.708	 
\put {\tiny $+$} at 220.878  -2.45 	 
\put {\tiny $+$} at 221.475  -2.518	 
\put {\tiny $+$} at 221.927  -2.399	 
\put {\tiny $+$} at 221.933  -2.57 	 
\put {\tiny $+$} at 222.13   -6.107	 
\put {\tiny $+$} at 222.184  -6.074	 
\put {\tiny $+$} at 222.491  -1.699	 
\put {\tiny $+$} at 222.643  -0.507	 
\put {\tiny $+$} at 223.116  -2.758	 
\put {\tiny $+$} at 223.231  -4.078	 
\put {\tiny $+$} at 223.287  -0.481	 
\put {\tiny $+$} at 224.012  -1.811	 
\put {\tiny $+$} at 224.013  -9.664	 
\put {\tiny $+$} at 224.208  0.324 	 
\put {\tiny $+$} at 224.415  -2.731	 
\put {\tiny $+$} at 224.521  -3.657	 
\put {\tiny $+$} at 224.64   -10.265	 
\put {\tiny $+$} at 224.663  -2.506	 
\put {\tiny $+$} at 224.792  -1.73 	 
\put {\tiny $+$} at 224.87   -5.763	 
\put {\tiny $+$} at 224.91   -2.128	 
\put {\tiny $+$} at 225.102  -2.599	 
\put {\tiny $+$} at 225.387  -4.2  	 
\put {\tiny $+$} at 225.469  -5.515	 
\put {\tiny $+$} at 225.81   -2.444	 
\put {\tiny $+$} at 225.906  -2.028	 
\put {\tiny $+$} at 226.133  -2.318	 
\put {\tiny $+$} at 226.315  -2.168	 
\put {\tiny $\circ$} at 212.18  -3.47  
\put {\tiny $\circ$} at 212.30  -0.39  
\put {\tiny $\circ$} at 212.56  +0.29  
\put {\tiny $\circ$} at 212.63  -1.73  
\put {\tiny $\circ$} at 212.82  -2.11  
\put {\tiny $\circ$} at 212.88  -2.09  
\put {\tiny $\circ$} at 213.34 -12.60  
\put {\tiny $\circ$} at 213.70 -12.59  
\put {\tiny $\circ$} at 213.70 -12.60  
\put {\tiny $\circ$} at 213.88 -11.84  
\put {\tiny $\circ$} at 214.13 -11.42  
\put {\tiny $\circ$} at 214.35  -6.51  
\put {\tiny $\circ$} at 214.49  -1.81  
\put {\tiny $\circ$} at 214.54  -0.83  
\put {\tiny $\circ$} at 214.60  -7.41  
\put {\tiny $\circ$} at 214.73  -0.12  
\put {\tiny $\circ$} at 215.26  -9.39  
\put {\tiny $\circ$} at 215.31  -2.30  
\put {\tiny $\circ$} at 215.31  -6.43  
\put {\tiny $\circ$} at 215.99 -10.11  
\put {\tiny $\circ$} at 216.64  -8.23  
\put {\tiny $\circ$} at 217.11  -1.45  
\put {\tiny $\circ$} at 217.20  -5.88  
\put {\tiny $\circ$} at 217.30  -0.06  
\put {\tiny $\circ$} at 217.33  -1.36  
\put {\tiny $\circ$} at 217.38  -0.08  
\put {\tiny $\circ$} at 217.44  +0.34  
\put {\tiny $\circ$} at 217.49  -0.02  
\put {\tiny $\circ$} at 217.63  -0.18  
\put {\tiny $\circ$} at 217.77  -0.69  
\put {\tiny $\circ$} at 218.02  -0.32  
\put {\tiny $\circ$} at 218.20  -0.39  
\put {\tiny $\circ$} at 218.79  -9.87  
\put {\tiny $\circ$} at 218.80  +0.34  
\put {\tiny $\circ$} at 218.86  -1.41  
\put {\tiny $\circ$} at 219.09  -1.54  
\put {\tiny $\circ$} at 219.10  -1.60  
\put {\tiny $\circ$} at 219.27  -8.95  
\put {\tiny $\circ$} at 219.29  -3.12  
\put {\tiny $\circ$} at 219.47 -10.56  
\put {\tiny $\circ$} at 219.69  -2.60  
\put {\tiny $\circ$} at 219.82  -0.01  
\put {\tiny $\circ$} at 219.88  -2.22  
\put {\tiny $\circ$} at 220.79  -1.71  
\put {\tiny $\circ$} at 221.01  -2.51  
\put {\tiny $\circ$} at 221.68  -1.31  
\put {\tiny $\circ$} at 221.86  -2.03  
\put {\tiny $\circ$} at 221.96  -1.99  
\put {\tiny $\circ$} at 222.05  -5.34  
\put {\tiny $\circ$} at 222.60  +0.35  
\put {\tiny $\circ$} at 223.44  +0.01  
\put {\tiny $\circ$} at 223.49  -7.93  
\put {\tiny $\circ$} at 223.60  -1.18  
\put {\tiny $\circ$} at 223.99  -9.69  
\put {\tiny $\circ$} at 224.00  -1.35  
\put {\tiny $\circ$} at 224.27  -1.17  
\put {\tiny $\circ$} at 224.36  -1.44  
\put {\tiny $\circ$} at 224.43  -2.25  
\put {\tiny $\circ$} at 224.44  -2.36  
\put {\tiny $\circ$} at 224.50  -2.41  
\put {\tiny $\circ$} at 224.57  -2.49  
\put {\tiny $\circ$} at 224.61  -1.00  
\put {\tiny $\circ$} at 224.69  +0.38  
\put {\tiny $\circ$} at 224.71  -1.33  
\put {\tiny $\circ$} at 225.09  -1.99  
\put {\tiny $\circ$} at 225.30  -5.02  
\put {\tiny $\circ$} at 225.34  -5.57  
\put {\tiny $\circ$} at 225.42  -4.62  
\put {\tiny $\circ$} at 225.47  -2.57  
\put {\tiny $\circ$} at 226.19  -3.92  
\setcoordinatesystem units <-4.6667mm,4.6667mm> point at 17.1427 0
\setplotarea x from 227 to 212 , y from -13 to 1
\put {\includegraphics[width=7cm]{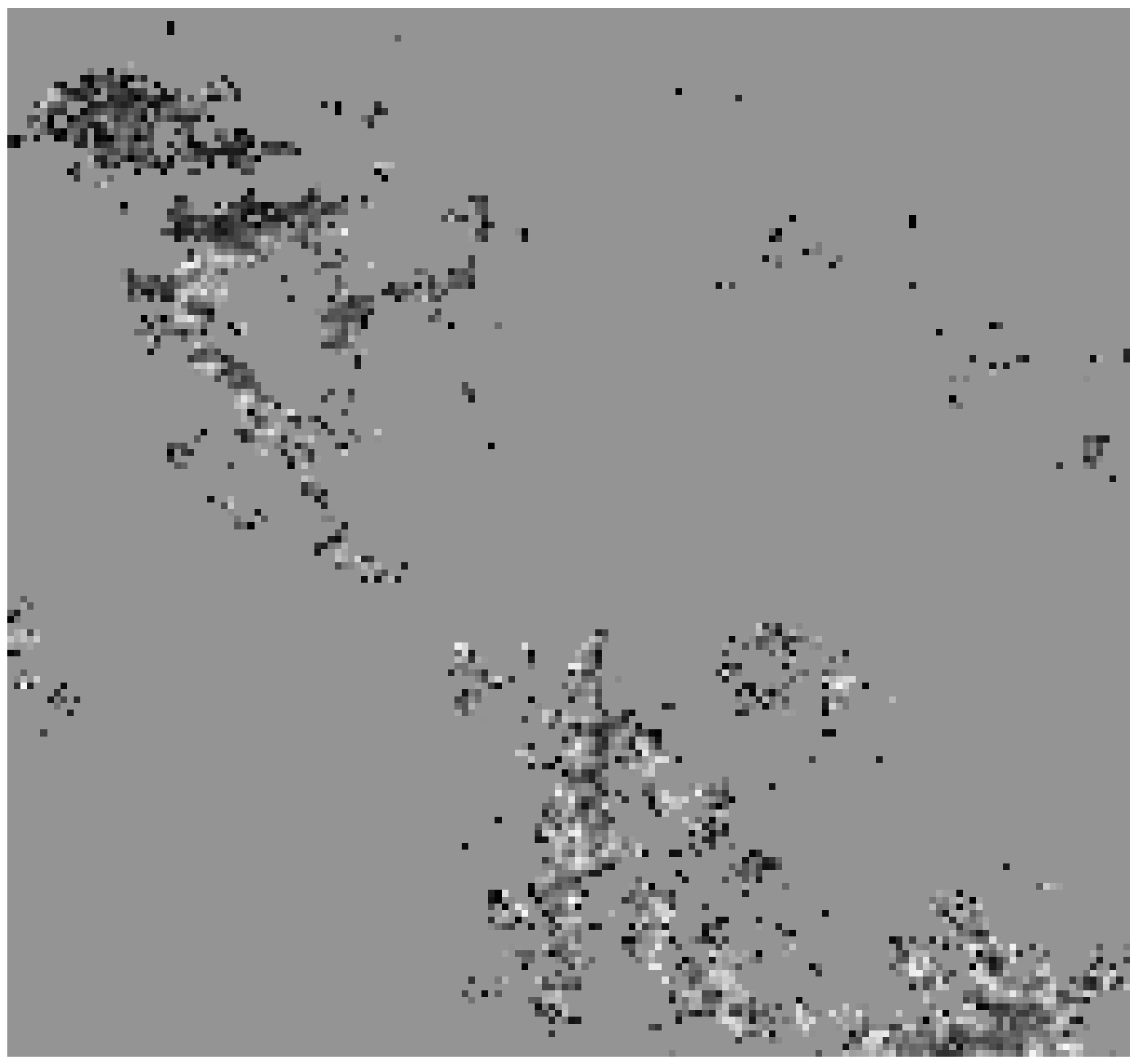}} at 219.5 -6
\axis left label {}
ticks in long unlabeled from -12 to 1 by 4
      short unlabeled from -13 to 1 by 1 /
\axis right label {}
ticks in long numbered from -12 to 1 by 4
      short unlabeled from -13 to 1 by 1 /
\axis bottom label {$l$\,[$^\circ$]}
ticks in long numbered from 212 to 227 by 4
      short unlabeled from 212 to 227 by 1 /
\axis top label {}
ticks in long unlabeled from 212 to 227 by 4
      short unlabeled from 212 to 227 by 1 /
\put {\tiny $+$} at 212.426  0.244 	 
\put {\tiny $+$} at 212.453  -0.856	 
\put {\tiny $+$} at 212.647  -2.523	 
\put {\tiny $+$} at 213.181  -3.289	 
\put {\tiny $+$} at 213.305  0.302 	 
\put {\tiny $+$} at 213.892  -4.311	 
\put {\tiny $+$} at 213.913  -12.092	 
\put {\tiny $+$} at 214.413  -5.518	 
\put {\tiny $+$} at 215.42   -12.823	 
\put {\tiny $+$} at 215.706  -6.157	 
\put {\tiny $+$} at 215.711  -10.065	 
\put {\tiny $+$} at 217.034  -11.789	 
\put {\tiny $+$} at 217.157  -12.517	 
\put {\tiny $+$} at 217.674  -2.696	 
\put {\tiny $+$} at 217.748  -6.646	 
\put {\tiny $+$} at 218.157  -0.626	 
\put {\tiny $+$} at 219.147  -10.556	 
\put {\tiny $+$} at 219.337  -9.942	 
\put {\tiny $+$} at 219.488  -1.724	 
\put {\tiny $+$} at 220.07   -0.112	 
\put {\tiny $+$} at 220.312  -0.084	 
\put {\tiny $+$} at 220.352  -7.708	 
\put {\tiny $+$} at 220.878  -2.45 	 
\put {\tiny $+$} at 221.475  -2.518	 
\put {\tiny $+$} at 221.927  -2.399	 
\put {\tiny $+$} at 221.933  -2.57 	 
\put {\tiny $+$} at 222.13   -6.107	 
\put {\tiny $+$} at 222.184  -6.074	 
\put {\tiny $+$} at 222.491  -1.699	 
\put {\tiny $+$} at 222.643  -0.507	 
\put {\tiny $+$} at 223.116  -2.758	 
\put {\tiny $+$} at 223.231  -4.078	 
\put {\tiny $+$} at 223.287  -0.481	 
\put {\tiny $+$} at 224.012  -1.811	 
\put {\tiny $+$} at 224.013  -9.664	 
\put {\tiny $+$} at 224.208  0.324 	 
\put {\tiny $+$} at 224.415  -2.731	 
\put {\tiny $+$} at 224.521  -3.657	 
\put {\tiny $+$} at 224.64   -10.265	 
\put {\tiny $+$} at 224.663  -2.506	 
\put {\tiny $+$} at 224.792  -1.73 	 
\put {\tiny $+$} at 224.87   -5.763	 
\put {\tiny $+$} at 224.91   -2.128	 
\put {\tiny $+$} at 225.102  -2.599	 
\put {\tiny $+$} at 225.387  -4.2  	 
\put {\tiny $+$} at 225.469  -5.515	 
\put {\tiny $+$} at 225.81   -2.444	 
\put {\tiny $+$} at 225.906  -2.028	 
\put {\tiny $+$} at 226.133  -2.318	 
\put {\tiny $+$} at 226.315  -2.168	 
\put {\tiny $\circ$} at 212.18  -3.47  
\put {\tiny $\circ$} at 212.30  -0.39  
\put {\tiny $\circ$} at 212.56  +0.29  
\put {\tiny $\circ$} at 212.63  -1.73  
\put {\tiny $\circ$} at 212.82  -2.11  
\put {\tiny $\circ$} at 212.88  -2.09  
\put {\tiny $\circ$} at 213.34 -12.60  
\put {\tiny $\circ$} at 213.70 -12.59  
\put {\tiny $\circ$} at 213.70 -12.60  
\put {\tiny $\circ$} at 213.88 -11.84  
\put {\tiny $\circ$} at 214.13 -11.42  
\put {\tiny $\circ$} at 214.35  -6.51  
\put {\tiny $\circ$} at 214.49  -1.81  
\put {\tiny $\circ$} at 214.54  -0.83  
\put {\tiny $\circ$} at 214.60  -7.41  
\put {\tiny $\circ$} at 214.73  -0.12  
\put {\tiny $\circ$} at 215.26  -9.39  
\put {\tiny $\circ$} at 215.31  -2.30  
\put {\tiny $\circ$} at 215.31  -6.43  
\put {\tiny $\circ$} at 215.99 -10.11  
\put {\tiny $\circ$} at 216.64  -8.23  
\put {\tiny $\circ$} at 217.11  -1.45  
\put {\tiny $\circ$} at 217.20  -5.88  
\put {\tiny $\circ$} at 217.30  -0.06  
\put {\tiny $\circ$} at 217.33  -1.36  
\put {\tiny $\circ$} at 217.38  -0.08  
\put {\tiny $\circ$} at 217.44  +0.34  
\put {\tiny $\circ$} at 217.49  -0.02  
\put {\tiny $\circ$} at 217.63  -0.18  
\put {\tiny $\circ$} at 217.77  -0.69  
\put {\tiny $\circ$} at 218.02  -0.32  
\put {\tiny $\circ$} at 218.20  -0.39  
\put {\tiny $\circ$} at 218.79  -9.87  
\put {\tiny $\circ$} at 218.80  +0.34  
\put {\tiny $\circ$} at 218.86  -1.41  
\put {\tiny $\circ$} at 219.09  -1.54  
\put {\tiny $\circ$} at 219.10  -1.60  
\put {\tiny $\circ$} at 219.27  -8.95  
\put {\tiny $\circ$} at 219.29  -3.12  
\put {\tiny $\circ$} at 219.47 -10.56  
\put {\tiny $\circ$} at 219.69  -2.60  
\put {\tiny $\circ$} at 219.82  -0.01  
\put {\tiny $\circ$} at 219.88  -2.22  
\put {\tiny $\circ$} at 220.79  -1.71  
\put {\tiny $\circ$} at 221.01  -2.51  
\put {\tiny $\circ$} at 221.68  -1.31  
\put {\tiny $\circ$} at 221.86  -2.03  
\put {\tiny $\circ$} at 221.96  -1.99  
\put {\tiny $\circ$} at 222.05  -5.34  
\put {\tiny $\circ$} at 222.60  +0.35  
\put {\tiny $\circ$} at 223.44  +0.01  
\put {\tiny $\circ$} at 223.49  -7.93  
\put {\tiny $\circ$} at 223.60  -1.18  
\put {\tiny $\circ$} at 223.99  -9.69  
\put {\tiny $\circ$} at 224.00  -1.35  
\put {\tiny $\circ$} at 224.27  -1.17  
\put {\tiny $\circ$} at 224.36  -1.44  
\put {\tiny $\circ$} at 224.43  -2.25  
\put {\tiny $\circ$} at 224.44  -2.36  
\put {\tiny $\circ$} at 224.50  -2.41  
\put {\tiny $\circ$} at 224.57  -2.49  
\put {\tiny $\circ$} at 224.61  -1.00  
\put {\tiny $\circ$} at 224.69  +0.38  
\put {\tiny $\circ$} at 224.71  -1.33  
\put {\tiny $\circ$} at 225.09  -1.99  
\put {\tiny $\circ$} at 225.30  -5.02  
\put {\tiny $\circ$} at 225.34  -5.57  
\put {\tiny $\circ$} at 225.42  -4.62  
\put {\tiny $\circ$} at 225.47  -2.57  
\put {\tiny $\circ$} at 226.19  -3.92  
\endpicture 
\caption{\label{map_monocerus} As Fig.\,\ref{map_aurigae2} but for the
region Monoceros. Extinction values are scaled linearly from -0.33 to 7\,mag of
optical extinction.} 
\end{figure*}

\begin{figure*}
\beginpicture
\setcoordinatesystem units <-7mm,7mm> point at 0 0
\setplotarea x from 211 to 201 , y from -17 to -8
\put {\includegraphics[width=7cm]{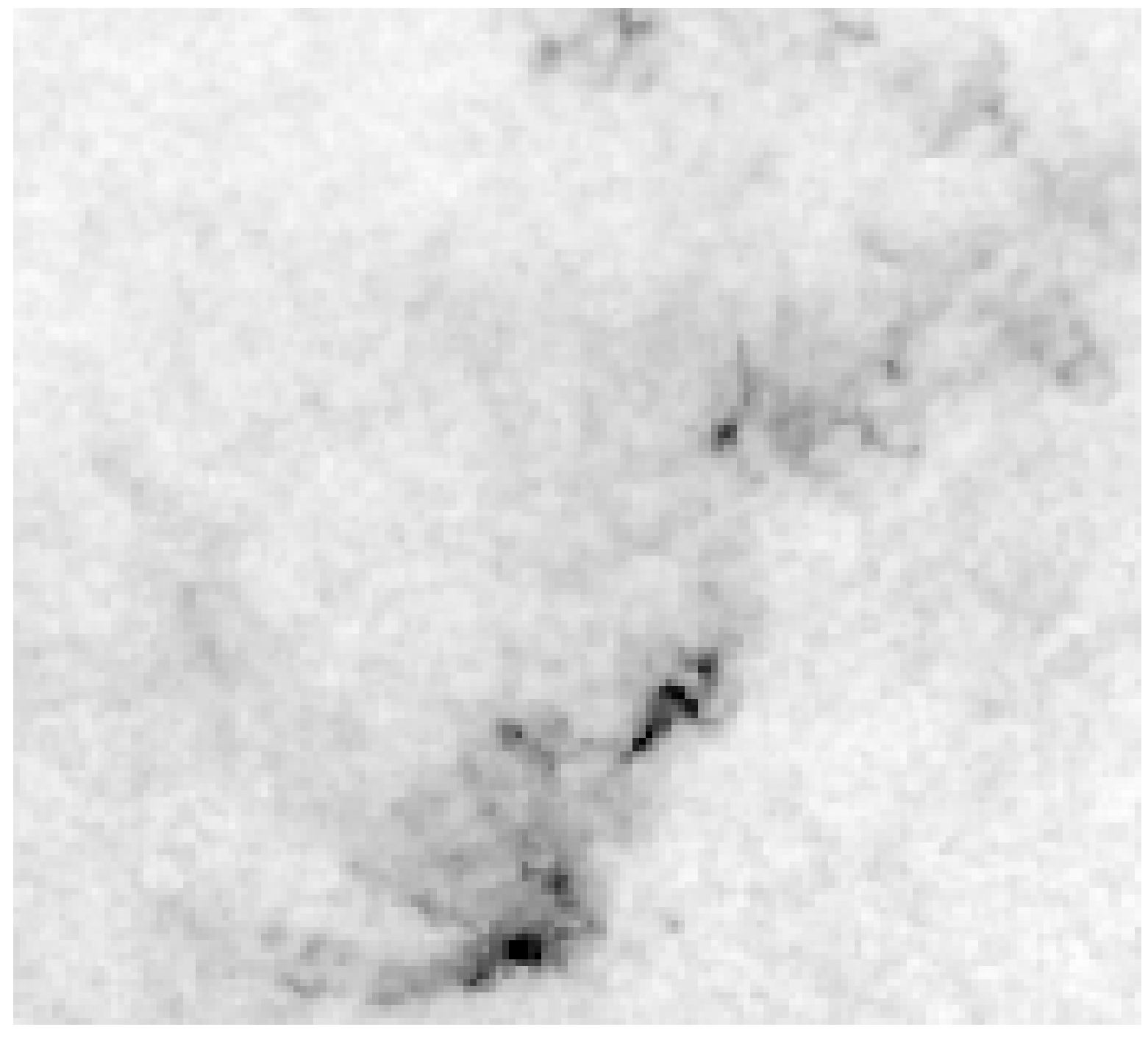}} at 206 -12.5
\axis left label {$b$\,[$^\circ$]}
ticks in long numbered from -17 to -8 by 2
      short unlabeled from -17 to -8 by 1 /
\axis right label {}
ticks in long unlabeled from -17 to -8 by 2
      short unlabeled from -17 to -8 by 1 /
\axis bottom label {$l$\,[$^\circ$]}
ticks in long numbered from 202 to 211 by 2
      short unlabeled from 201 to 211 by 1 /
\axis top label {}
ticks in long unlabeled from 202 to 211 by 2
      short unlabeled from 201 to 211 by 1 /
\put {\tiny $\circ$} at 201.96 -12.11  
\put {\tiny $\circ$} at 205.18 -14.13  
\put {\tiny $\circ$} at 205.31 -14.30  
\put {\tiny $\circ$} at 205.34 -14.32  
\put {\tiny $\circ$} at 205.87 -12.61  
\put {\tiny $\circ$} at 206.48 -16.33  
\put {\tiny $\circ$} at 206.85 -16.54  
\setcoordinatesystem units <-7mm,7mm> point at 11.4286 0
\setplotarea x from 211 to 201 , y from -17 to -8
\put {\includegraphics[width=7cm]{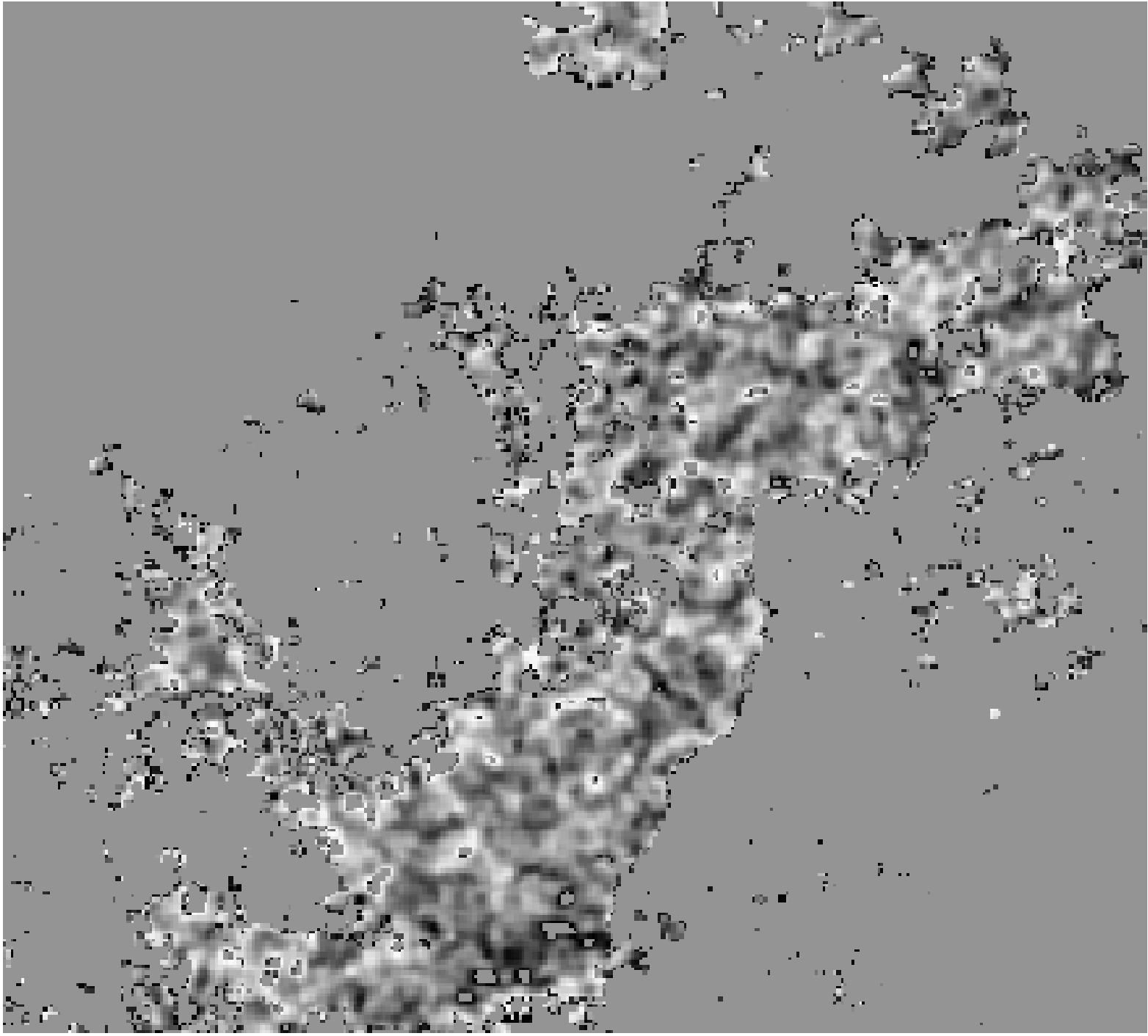}} at 206 -12.5
\axis left label {}
ticks in long unlabeled from -17 to -8 by 2
      short unlabeled from -17 to -8 by 1 /
\axis right label {}
ticks in long numbered from -17 to -8 by 2
      short unlabeled from -17 to -8 by 1 /
\axis bottom label {$l$\,[$^\circ$]}
ticks in long numbered from 202 to 211 by 2
      short unlabeled from 201 to 211 by 1 /
\axis top label {}
ticks in long unlabeled from 202 to 211 by 2
      short unlabeled from 201 to 211 by 1 /
\put {\tiny $\circ$} at 201.96 -12.11  
\put {\tiny $\circ$} at 205.18 -14.13  
\put {\tiny $\circ$} at 205.31 -14.30  
\put {\tiny $\circ$} at 205.34 -14.32  
\put {\tiny $\circ$} at 205.87 -12.61  
\put {\tiny $\circ$} at 206.48 -16.33  
\put {\tiny $\circ$} at 206.85 -16.54  
\endpicture 
\caption{\label{map_oria} As Fig.\,\ref{map_aurigae2} but for the region Ori\,B.
Extinction values are scaled linearly from -0.33 to 10\,mag of optical
extinction.}  
\end{figure*}

\begin{figure*}
\beginpicture
\setcoordinatesystem units <-6.3636mm,6.3636mm> point at 0 0
\setplotarea x from 219 to 208 , y from -21 to -16
\put {\includegraphics[width=7cm]{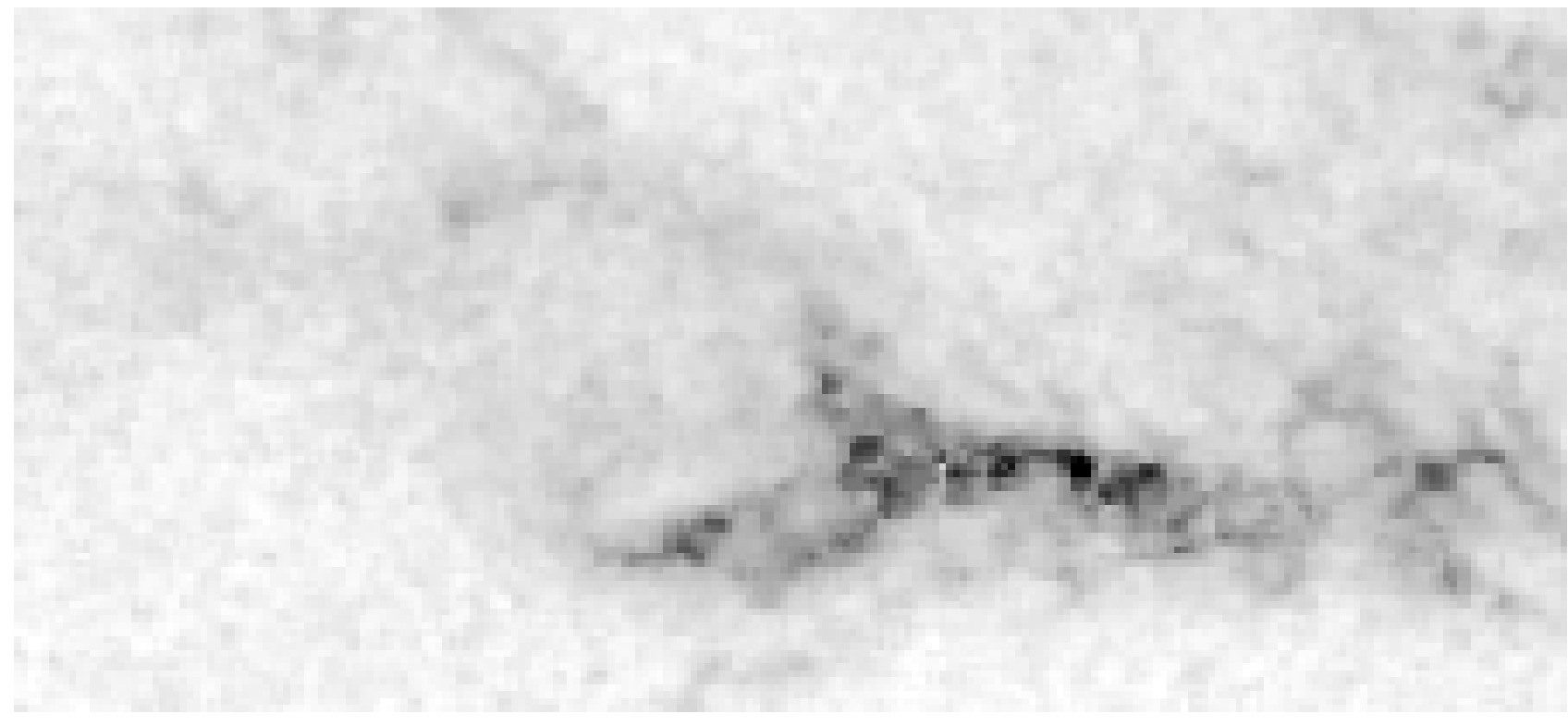}} at 213.5 -18.5
\axis left label {$b$\,[$^\circ$]}
ticks in long numbered from -21 to -16 by 2
      short unlabeled from -21 to -16 by 1 /
\axis right label {}
ticks in long unlabeled from -21 to -16 by 2
      short unlabeled from -21 to -16 by 1 /
\axis bottom label {$l$\,[$^\circ$]}
ticks in long numbered from 208 to 219 by 2
      short unlabeled from 208 to 219 by 1 /
\axis top label {}
ticks in long unlabeled from 208 to 219 by 2
      short unlabeled from 208 to 219 by 1 /
\put {\tiny $+$} at 209.707  -17.853	 
\put {\tiny $+$} at 211.783  -17.368	 
\put {\tiny $+$} at 213.453  -19.856	 
\put {\tiny $+$} at 217.387  -18.717	 
\put {\tiny $\circ$} at 208.07 -18.95  
\put {\tiny $\circ$} at 208.53 -19.16  
\put {\tiny $\circ$} at 208.68 -19.21  
\put {\tiny $\circ$} at 208.73 -19.20  
\put {\tiny $\circ$} at 208.82 -19.25  
\put {\tiny $\circ$} at 209.00 -19.38  
\put {\tiny $\circ$} at 209.52 -19.59  
\put {\tiny $\circ$} at 210.06 -19.59  
\put {\tiny $\circ$} at 210.09 -19.84  
\put {\tiny $\circ$} at 210.12 -19.60  
\put {\tiny $\circ$} at 210.34 -19.14  
\put {\tiny $\circ$} at 210.44 -19.73  
\put {\tiny $\circ$} at 210.80 -19.51  
\put {\tiny $\circ$} at 210.91 -19.34  
\put {\tiny $\circ$} at 210.98 -19.33  
\put {\tiny $\circ$} at 212.23 -19.37  
\put {\tiny $\circ$} at 212.44 -18.99  
\put {\tiny $\circ$} at 212.47 -19.02  
\put {\tiny $\circ$} at 212.98 -19.15  
\put {\tiny $\circ$} at 214.06 -19.62  
\put {\tiny $\circ$} at 214.27 -19.78  
\put {\tiny $\circ$} at 217.39 -18.73  
\setcoordinatesystem units <-6.3636mm,6.3636mm> point at 12.5715 0
\setplotarea x from 219 to 208 , y from -21 to -16
\put {\includegraphics[width=7cm]{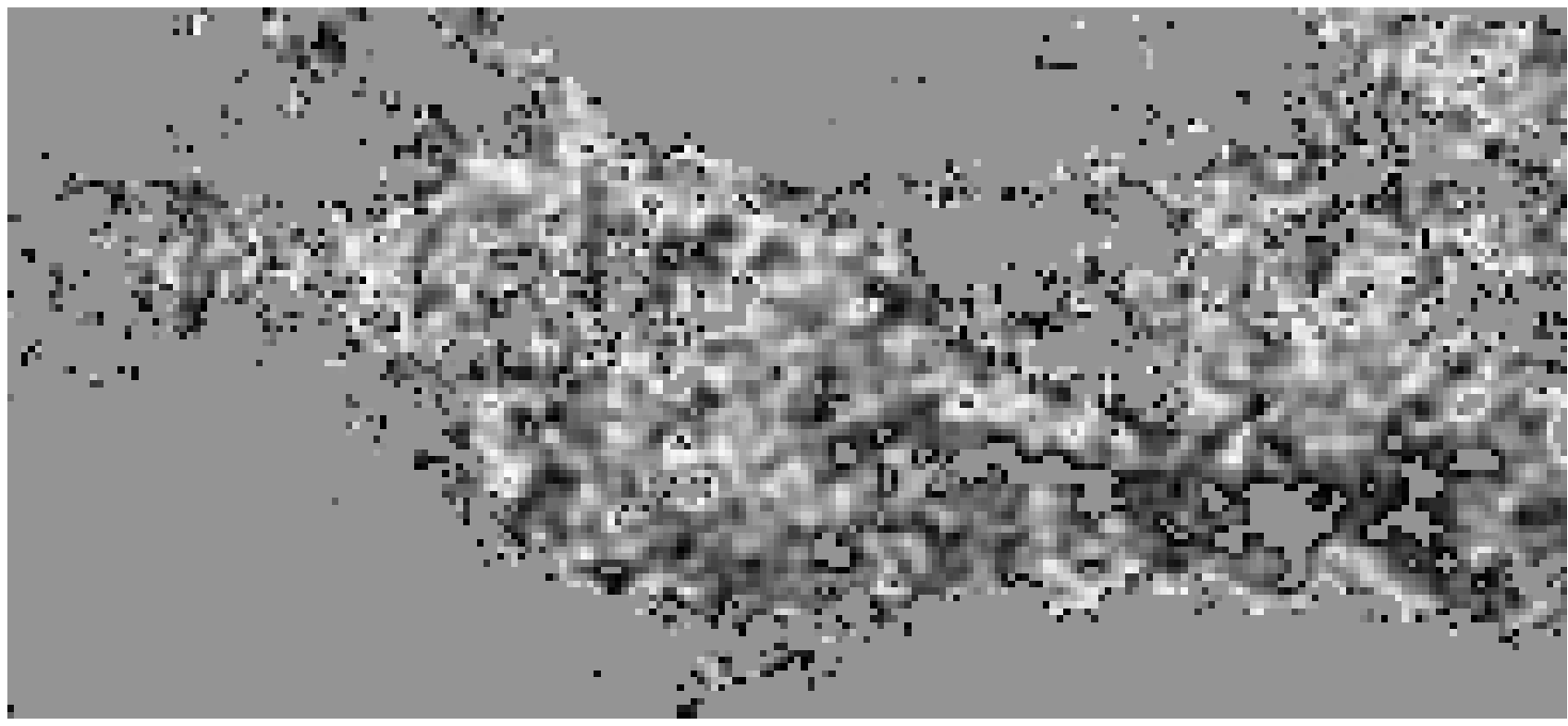}} at 213.5 -18.5
\axis left label {}
ticks in long unlabeled from -21 to -16 by 2
      short unlabeled from -21 to -16 by 1 /
\axis right label {}
ticks in long numbered from -21 to -16 by 2
      short unlabeled from -21 to -16 by 1 /
\axis bottom label {$l$\,[$^\circ$]}
ticks in long numbered from 208 to 219 by 2
      short unlabeled from 208 to 219 by 1 /
\axis top label {}
ticks in long unlabeled from 208 to 219 by 2
      short unlabeled from 208 to 219 by 1 /
\put {\tiny $+$} at 209.707  -17.853	 
\put {\tiny $+$} at 211.783  -17.368	 
\put {\tiny $+$} at 213.453  -19.856	 
\put {\tiny $+$} at 217.387  -18.717	 
\put {\tiny $\circ$} at 208.07 -18.95  
\put {\tiny $\circ$} at 208.53 -19.16  
\put {\tiny $\circ$} at 208.68 -19.21  
\put {\tiny $\circ$} at 208.73 -19.20  
\put {\tiny $\circ$} at 208.82 -19.25  
\put {\tiny $\circ$} at 209.00 -19.38  
\put {\tiny $\circ$} at 209.52 -19.59  
\put {\tiny $\circ$} at 210.06 -19.59  
\put {\tiny $\circ$} at 210.09 -19.84  
\put {\tiny $\circ$} at 210.12 -19.60  
\put {\tiny $\circ$} at 210.34 -19.14  
\put {\tiny $\circ$} at 210.44 -19.73  
\put {\tiny $\circ$} at 210.80 -19.51  
\put {\tiny $\circ$} at 210.91 -19.34  
\put {\tiny $\circ$} at 210.98 -19.33  
\put {\tiny $\circ$} at 212.23 -19.37  
\put {\tiny $\circ$} at 212.44 -18.99  
\put {\tiny $\circ$} at 212.47 -19.02  
\put {\tiny $\circ$} at 212.98 -19.15  
\put {\tiny $\circ$} at 214.06 -19.62  
\put {\tiny $\circ$} at 214.27 -19.78  
\put {\tiny $\circ$} at 217.39 -18.73  
\endpicture 
\caption{\label{map_orib} As Fig.\,\ref{map_aurigae2} but for the region Ori\,A.
Extinction values are scaled linearly from -0.33 to 12\,mag of optical
extinction.}  
\end{figure*}

\begin{figure*}
\beginpicture
\setcoordinatesystem units <-10mm,10mm> point at 0 0
\setplotarea x from 163 to 156 , y from -25 to -15
\put {\includegraphics[width=7cm]{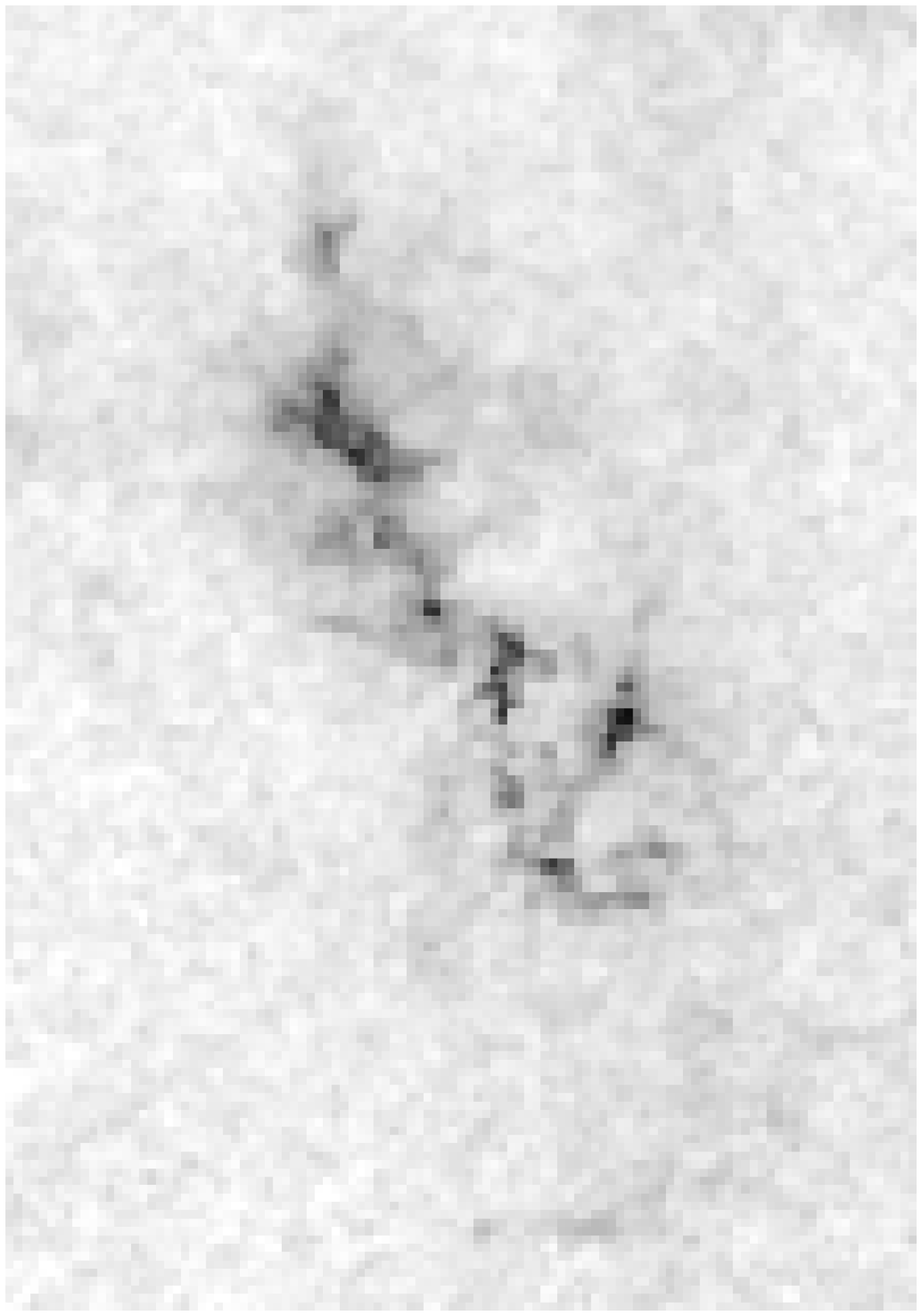}} at 159.5 -20
\axis left label {$b$\,[$^\circ$]}
ticks in long numbered from -24 to -15 by 2
      short unlabeled from -25 to -15 by 1 /
\axis right label {}
ticks in long unlabeled from -24 to -15 by 2
      short unlabeled from -25 to -15 by 1 /
\axis bottom label {$l$\,[$^\circ$]}
ticks in long numbered from 156 to 163 by 2
      short unlabeled from 156 to 163 by 1 /
\axis top label {}
ticks in long unlabeled from 156 to 163 by 2
      short unlabeled from 156 to 163 by 1 /
\put {\tiny $+$} at 158.301  -16.278	 
\put {\tiny $\circ$} at 158.31 -20.47  
\put {\tiny $\circ$} at 158.33 -20.56  
\put {\tiny $\circ$} at 159.24 -17.15  
\put {\tiny $\circ$} at 159.60 -15.92  
\put {\tiny $\circ$} at 160.29 -17.85  
\put {\tiny $\circ$} at 160.33 -17.86  
\put {\tiny $\circ$} at 160.39 -17.87  
\put {\tiny $\circ$} at 160.42 -17.73  
\put {\tiny $\circ$} at 160.44 -17.89  
\put {\tiny $\circ$} at 160.44 -17.99  
\put {\tiny $\circ$} at 160.48 -18.01  
\put {\tiny $\circ$} at 160.49 -17.80  
\put {\tiny $\circ$} at 160.55 -17.92  
\setcoordinatesystem units <-10mm,10mm> point at 8 0
\setplotarea x from 163 to 156 , y from -25 to -15
\put {\includegraphics[width=7cm]{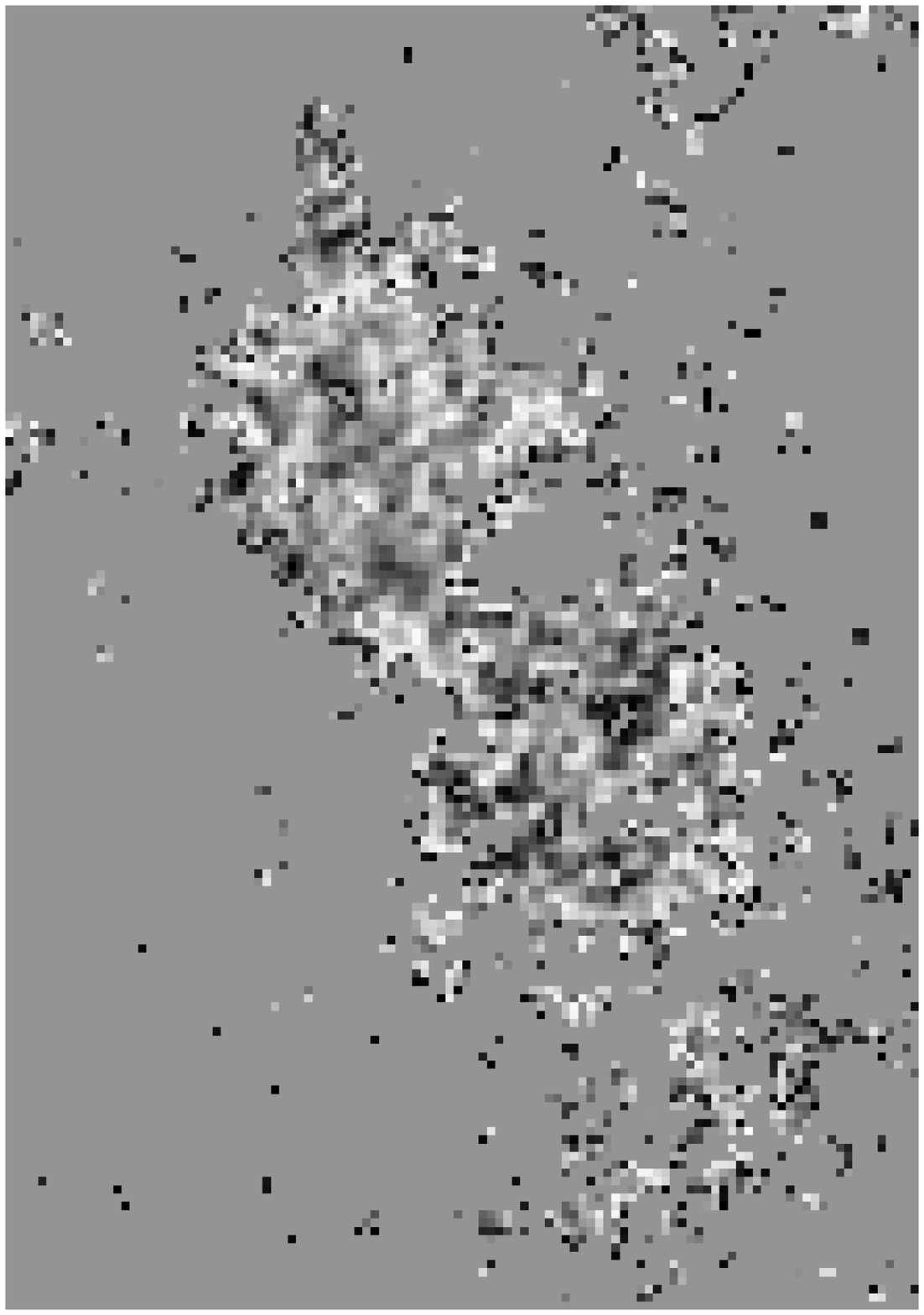}} at 159.5 -20
\axis left label {}
ticks in long unlabeled from -24 to -15 by 2
      short unlabeled from -25 to -15 by 1 /
\axis right label {}
ticks in long numbered from -24 to -15 by 2
      short unlabeled from -25 to -15 by 1 /
\axis bottom label {$l$\,[$^\circ$]}
ticks in long numbered from 156 to 163 by 2
      short unlabeled from 156 to 163 by 1 /
\axis top label {}
ticks in long unlabeled from 156 to 163 by 2
      short unlabeled from 156 to 163 by 1 /
\put {\tiny $+$} at 158.301  -16.278	 
\put {\tiny $\circ$} at 158.31 -20.47  
\put {\tiny $\circ$} at 158.33 -20.56  
\put {\tiny $\circ$} at 159.24 -17.15  
\put {\tiny $\circ$} at 159.60 -15.92  
\put {\tiny $\circ$} at 160.29 -17.85  
\put {\tiny $\circ$} at 160.33 -17.86  
\put {\tiny $\circ$} at 160.39 -17.87  
\put {\tiny $\circ$} at 160.42 -17.73  
\put {\tiny $\circ$} at 160.44 -17.89  
\put {\tiny $\circ$} at 160.44 -17.99  
\put {\tiny $\circ$} at 160.48 -18.01  
\put {\tiny $\circ$} at 160.49 -17.80  
\put {\tiny $\circ$} at 160.55 -17.92  
\endpicture 
\caption{\label{map_perseus} As Fig.\,\ref{map_aurigae2} but for the
region Perseus. Extinction values are scaled linearly from -0.33 to 10\,mag of
optical extinction.} 
\end{figure*}

\begin{figure*}
\beginpicture
\setcoordinatesystem units <-5mm,5mm> point at 0 0
\setplotarea x from 178 to 164 , y from -19 to -10
\put {\includegraphics[width=7cm]{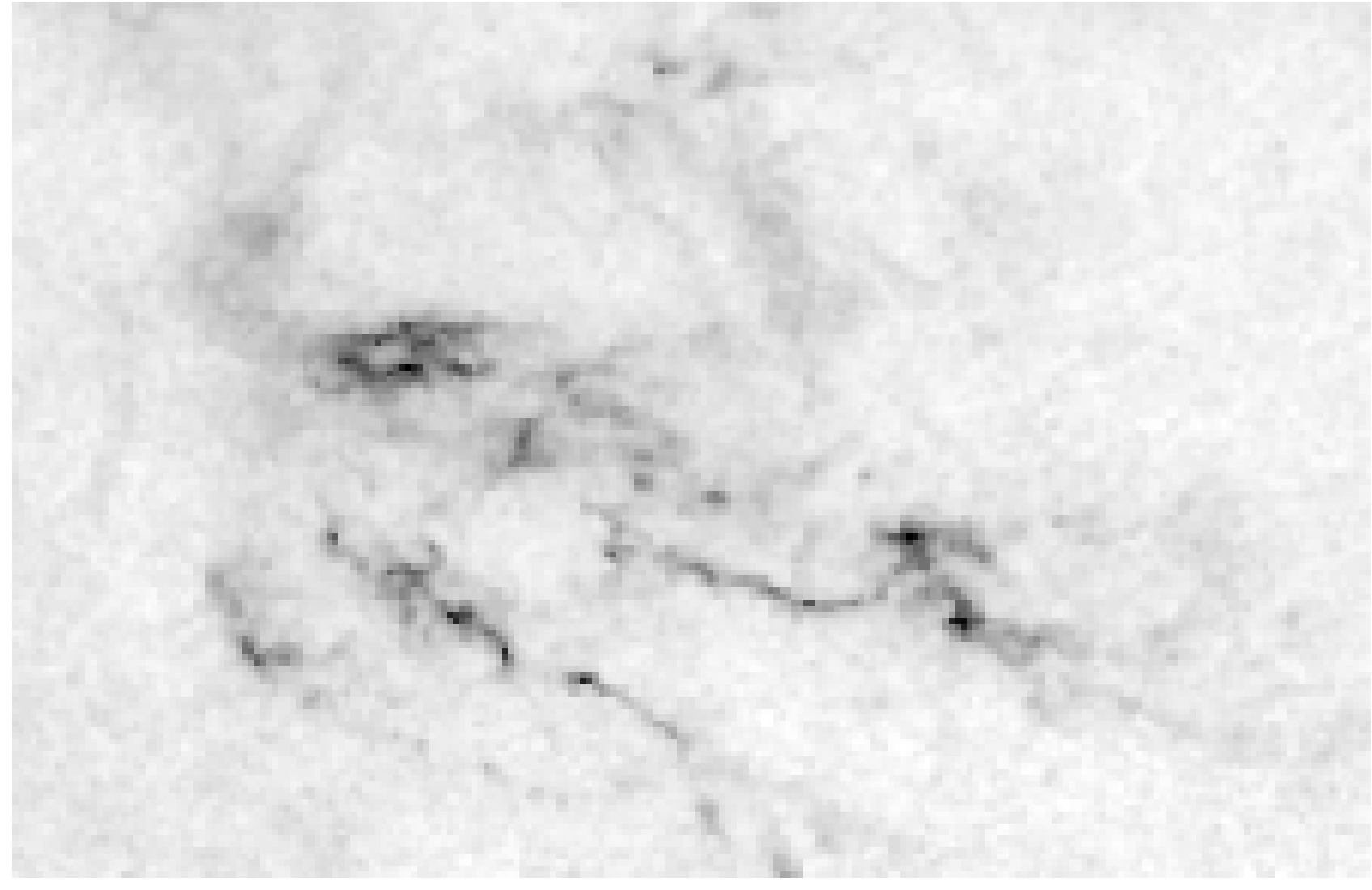}} at 171 -14.5
\axis left label {$b$\,[$^\circ$]}
ticks in long numbered from -18 to -10 by 2
      short unlabeled from -19 to -10 by 1 /
\axis right label {}
ticks in long unlabeled from -18 to -10 by 2
      short unlabeled from -19 to -10 by 1 /
\axis bottom label {$l$\,[$^\circ$]}
ticks in long numbered from 164 to 178 by 4
      short unlabeled from 164 to 178 by 1 /
\axis top label {}
ticks in long unlabeled from 164 to 178 by 4
      short unlabeled from 164 to 178 by 1 /
\put {\tiny $+$} at 165.08   -18.678	
\put {\tiny $+$} at 165.169  -18.602	
\put {\tiny $+$} at 165.668  -11.205	
\put {\tiny $+$} at 168.433  -16.029	
\put {\tiny $+$} at 168.641  -13.696	
\put {\tiny $+$} at 168.737  -18.167	
\put {\tiny $+$} at 170.131  -15.798	
\put {\tiny $+$} at 170.252  -17.155	
\put {\tiny $+$} at 170.303  -17.118	
\put {\tiny $+$} at 170.933  -11.153	 
\put {\tiny $+$} at 172.311  -15.292	 
\put {\tiny $+$} at 175.267  -17.142	 
\put {\tiny $+$} at 175.784  -13.025	 
\put {\tiny $+$} at 176.505  -17.244	 
\put {\tiny $+$} at 176.506  -16.685	 
\put {\tiny $+$} at 176.564  -16.665	 
\put {\tiny $\circ$} at 167.55 -18.08  
\put {\tiny $\circ$} at 168.02 -16.37  
\put {\tiny $\circ$} at 168.15 -16.28  
\put {\tiny $\circ$} at 168.29 -12.30  
\put {\tiny $\circ$} at 168.90 -15.55  
\put {\tiny $\circ$} at 168.95 -16.47  
\put {\tiny $\circ$} at 169.74 -15.60  
\put {\tiny $\circ$} at 169.90 -15.48  
\put {\tiny $\circ$} at 172.02 -15.52  
\put {\tiny $\circ$} at 172.70 -14.77  
\put {\tiny $\circ$} at 173.94 -15.95  
\put {\tiny $\circ$} at 174.13 -14.01  
\put {\tiny $\circ$} at 175.72 -16.25  
\setcoordinatesystem units <-5mm,5mm> point at 16 0
\setplotarea x from 178 to 164 , y from -19 to -10
\put {\includegraphics[width=7cm]{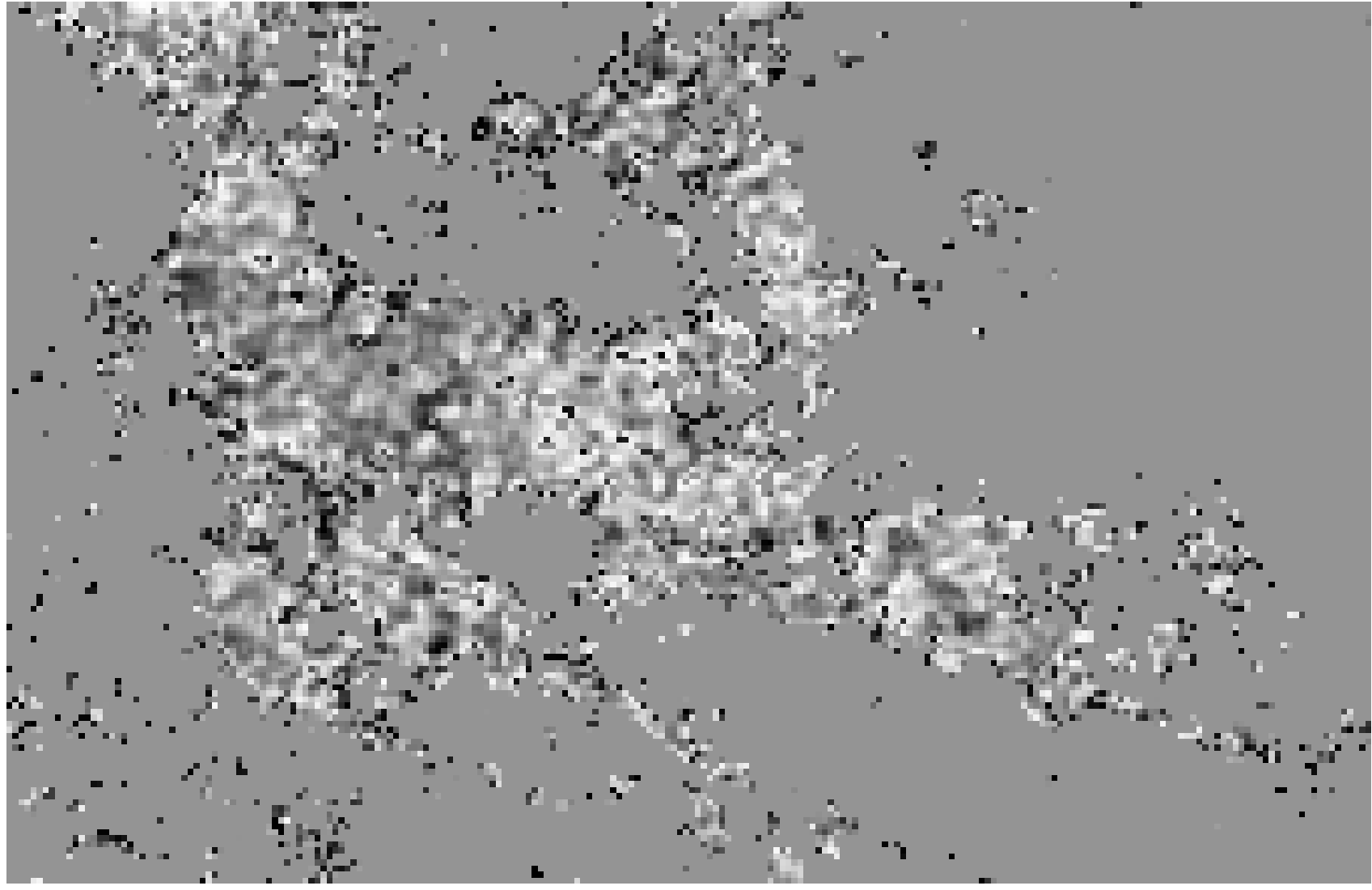}} at 171 -14.5
\axis left label {}
ticks in long unlabeled from -18 to -10 by 2
      short unlabeled from -19 to -10 by 1 /
\axis right label {}
ticks in long numbered from -18 to -10 by 2
      short unlabeled from -19 to -10 by 1 /
\axis bottom label {$l$\,[$^\circ$]}
ticks in long numbered from 164 to 178 by 4
      short unlabeled from 164 to 178 by 1 /
\axis top label {}
ticks in long unlabeled from 164 to 178 by 4
      short unlabeled from 164 to 178 by 1 /
\put {\tiny $+$} at 165.08   -18.678	
\put {\tiny $+$} at 165.169  -18.602	
\put {\tiny $+$} at 165.668  -11.205	
\put {\tiny $+$} at 168.433  -16.029	
\put {\tiny $+$} at 168.641  -13.696	
\put {\tiny $+$} at 168.737  -18.167	
\put {\tiny $+$} at 170.131  -15.798	
\put {\tiny $+$} at 170.252  -17.155	
\put {\tiny $+$} at 170.303  -17.118	
\put {\tiny $+$} at 170.933  -11.153	 
\put {\tiny $+$} at 172.311  -15.292	 
\put {\tiny $+$} at 175.267  -17.142	 
\put {\tiny $+$} at 175.784  -13.025	 
\put {\tiny $+$} at 176.505  -17.244	 
\put {\tiny $+$} at 176.506  -16.685	 
\put {\tiny $+$} at 176.564  -16.665	 
\put {\tiny $\circ$} at 167.55 -18.08  
\put {\tiny $\circ$} at 168.02 -16.37  
\put {\tiny $\circ$} at 168.15 -16.28  
\put {\tiny $\circ$} at 168.29 -12.30  
\put {\tiny $\circ$} at 168.90 -15.55  
\put {\tiny $\circ$} at 168.95 -16.47  
\put {\tiny $\circ$} at 169.74 -15.60  
\put {\tiny $\circ$} at 169.90 -15.48  
\put {\tiny $\circ$} at 172.02 -15.52  
\put {\tiny $\circ$} at 172.70 -14.77  
\put {\tiny $\circ$} at 173.94 -15.95  
\put {\tiny $\circ$} at 174.13 -14.01  
\put {\tiny $\circ$} at 175.72 -16.25  
\endpicture 
\caption{\label{map_taurus} As Fig.\,\ref{map_aurigae2} but for the region
Taurus. Extinction values are scaled linearly from -0.33 to 12\,mag of optical
extinction.}  
\end{figure*}

\begin{figure*}
\beginpicture
\setcoordinatesystem units <-3.8888mm,3.8888mm> point at 0 0
\setplotarea x from 189 to 171 , y from -31 to -17
\put {\includegraphics[width=7cm]{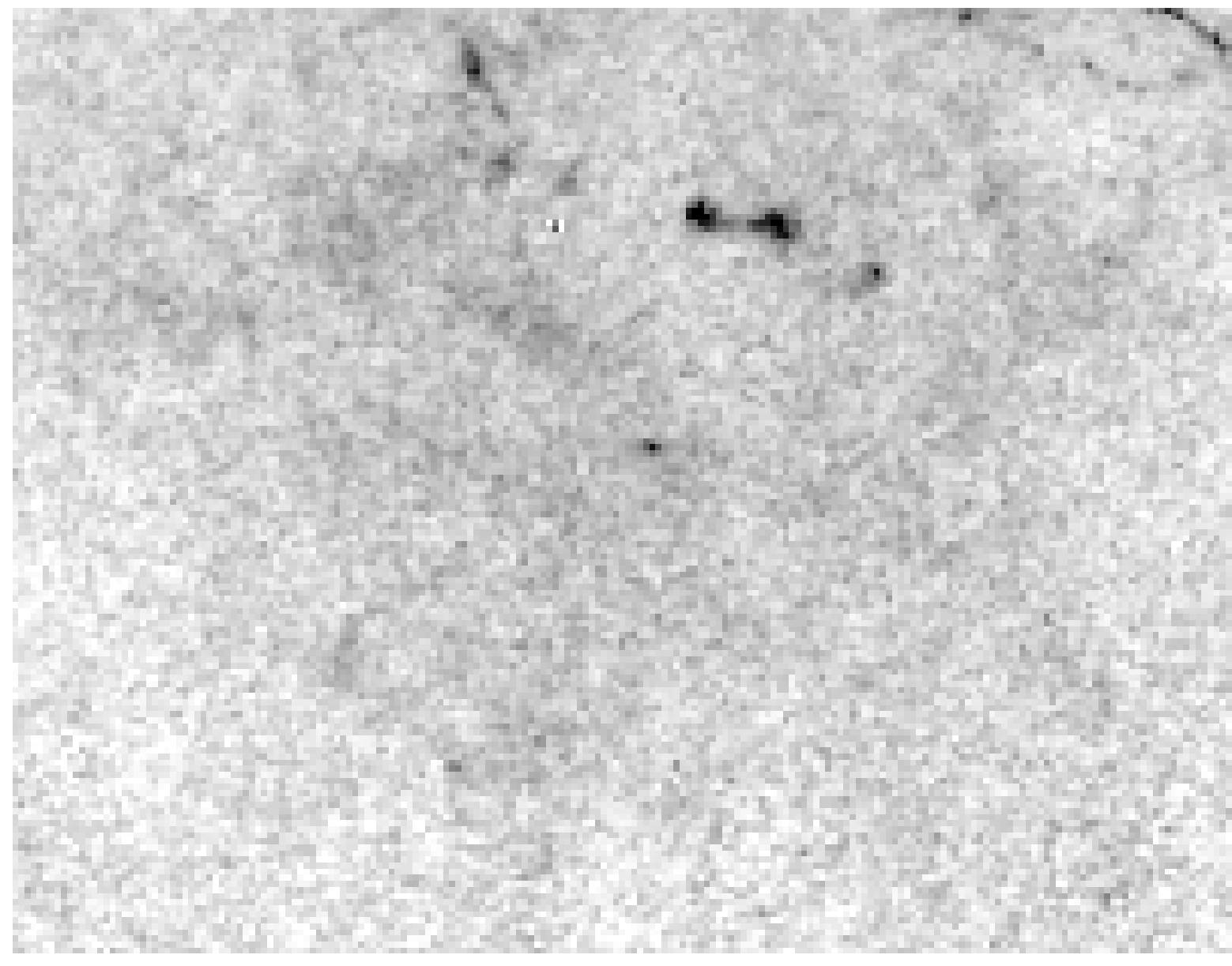}} at 180 -24
\axis left label {$b$\,[$^\circ$]}
ticks in long numbered from -30 to -17 by 4
      short unlabeled from -31 to -17 by 1 /
\axis right label {}
ticks in long unlabeled from -30 to -17 by 4
      short unlabeled from -31 to -17 by 1 /
\axis bottom label {$l$\,[$^\circ$]}
ticks in long numbered from 172 to 189 by 4
      short unlabeled from 171 to 189 by 1 /
\axis top label {}
ticks in long unlabeled from 172 to 189 by 4
      short unlabeled from 171 to 189 by 1 /
\put {\tiny $+$} at 181.238  -19.946	 
\put {\tiny $\circ$} at 178.75 -19.96  
\put {\tiny $\circ$} at 179.38 -22.91  
\put {\tiny $\circ$} at 180.06 -22.34  
\put {\tiny $\circ$} at 182.26 -17.84  
\put {\tiny $\circ$} at 185.92 -19.65  
\put {\tiny $\circ$} at 187.69 -21.13  
\setcoordinatesystem units <-3.8888mm,3.8888mm> point at 20.5719 0
\setplotarea x from 189 to 171 , y from -31 to -17
\put {\includegraphics[width=7cm]{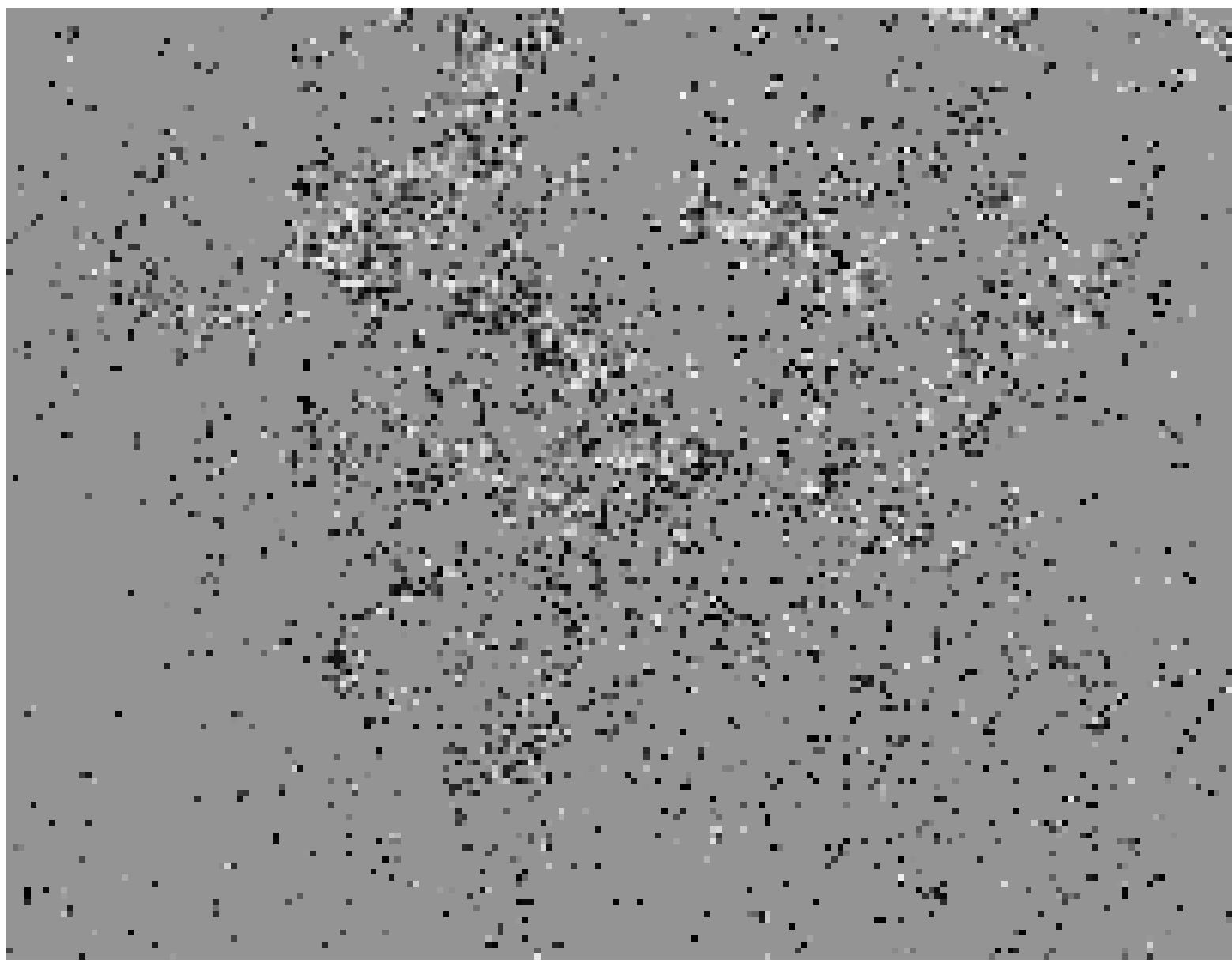}} at 180 -24
\axis left label {}
ticks in long unlabeled from -30 to -17 by 4
      short unlabeled from -31 to -17 by 1 /
\axis right label {}
ticks in long numbered from -30 to -17 by 4
      short unlabeled from -31 to -17 by 1 /
\axis bottom label {$l$\,[$^\circ$]}
ticks in long numbered from 172 to 189 by 4
      short unlabeled from 171 to 189 by 1 /
\axis top label {}
ticks in long unlabeled from 172 to 189 by 4
      short unlabeled from 171 to 189 by 1 /
\put {\tiny $+$} at 181.238  -19.946	 
\put {\tiny $\circ$} at 178.75 -19.96  
\put {\tiny $\circ$} at 179.38 -22.91  
\put {\tiny $\circ$} at 180.06 -22.34  
\put {\tiny $\circ$} at 182.26 -17.84  
\put {\tiny $\circ$} at 185.92 -19.65  
\put {\tiny $\circ$} at 187.69 -21.13  
\endpicture 
\caption{\label{map_taurusextended} As Fig.\,\ref{map_aurigae2} but for the
region Taurusextended. Extinction values are scaled linearly from -0.33 to
5\,mag of optical extinction.} 
\end{figure*}

\clearpage
\newpage

\subsection{$\beta$ and column density distributions}
\label{app_plots}

In the following we show for all individual regions investigated in this paper
the distribution of $\beta$ and column density values.

\begin{figure*}
\includegraphics[height=7.5cm,angle=-90]{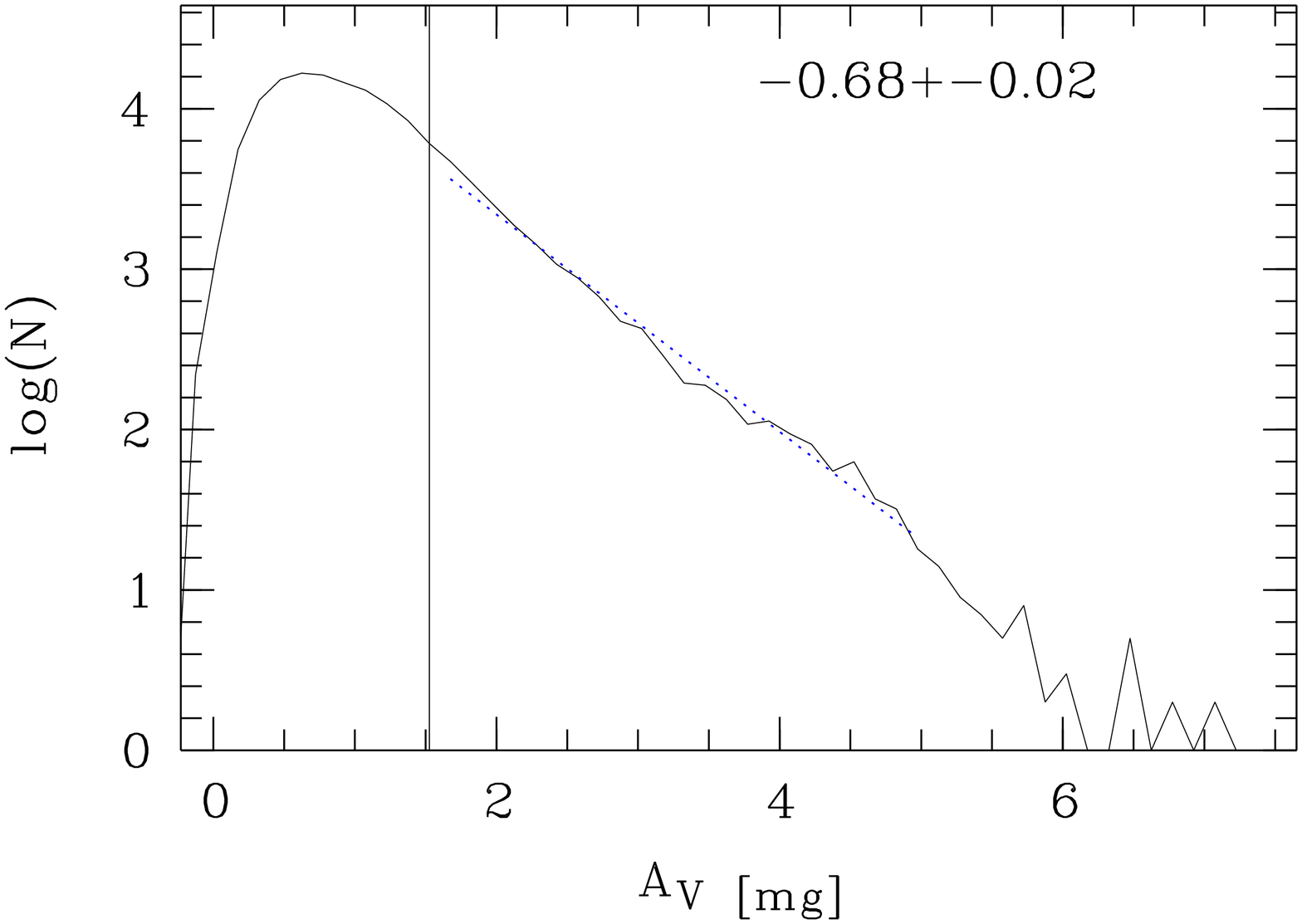}
\hfill
\includegraphics[height=7.5cm,angle=-90]{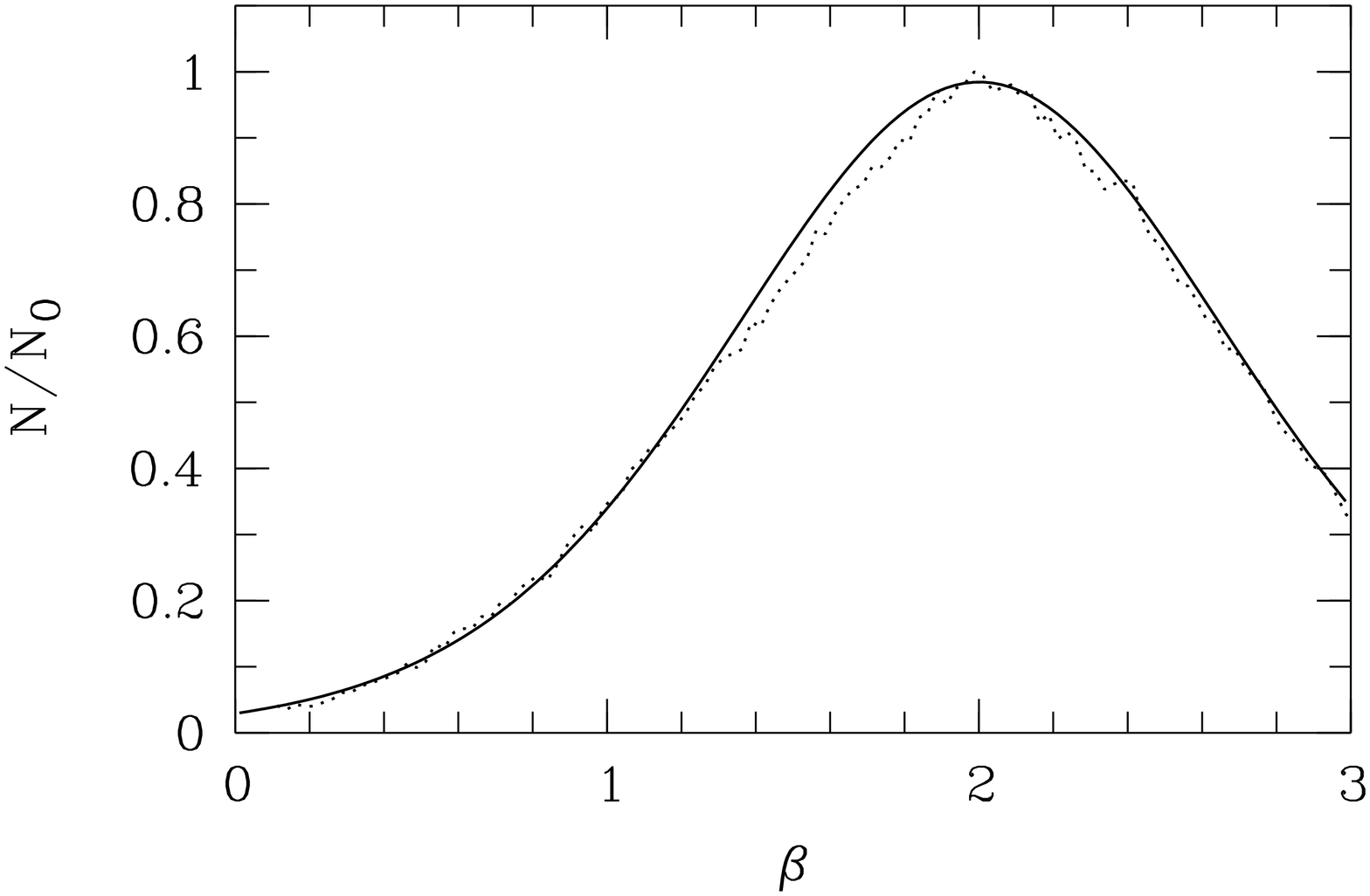}
\caption{\label{plots_aurigae1} {\bf Left:} Measured column density distribution
in the region Auriga\,1 (solid line). As dashed lined we overplot the linear fit
to the area above the 5\,$\sigma$ noise level (vertical line). The slope of this
fit is indicated in the graph. {\bf Right:} Measured distribution of $\beta$
values in the region Auriga\,1 (dotted line) and overplotted model prediction
for the assumption of a single $\beta$ value with internal scatter in the entire
region. The measured distribution includes only pixels where the $<J-H>$ and
$<H-K>$ values are more than 3$\sigma$ above the noise.} 
\end{figure*}

\begin{figure*}
\includegraphics[height=7.5cm,angle=-90]{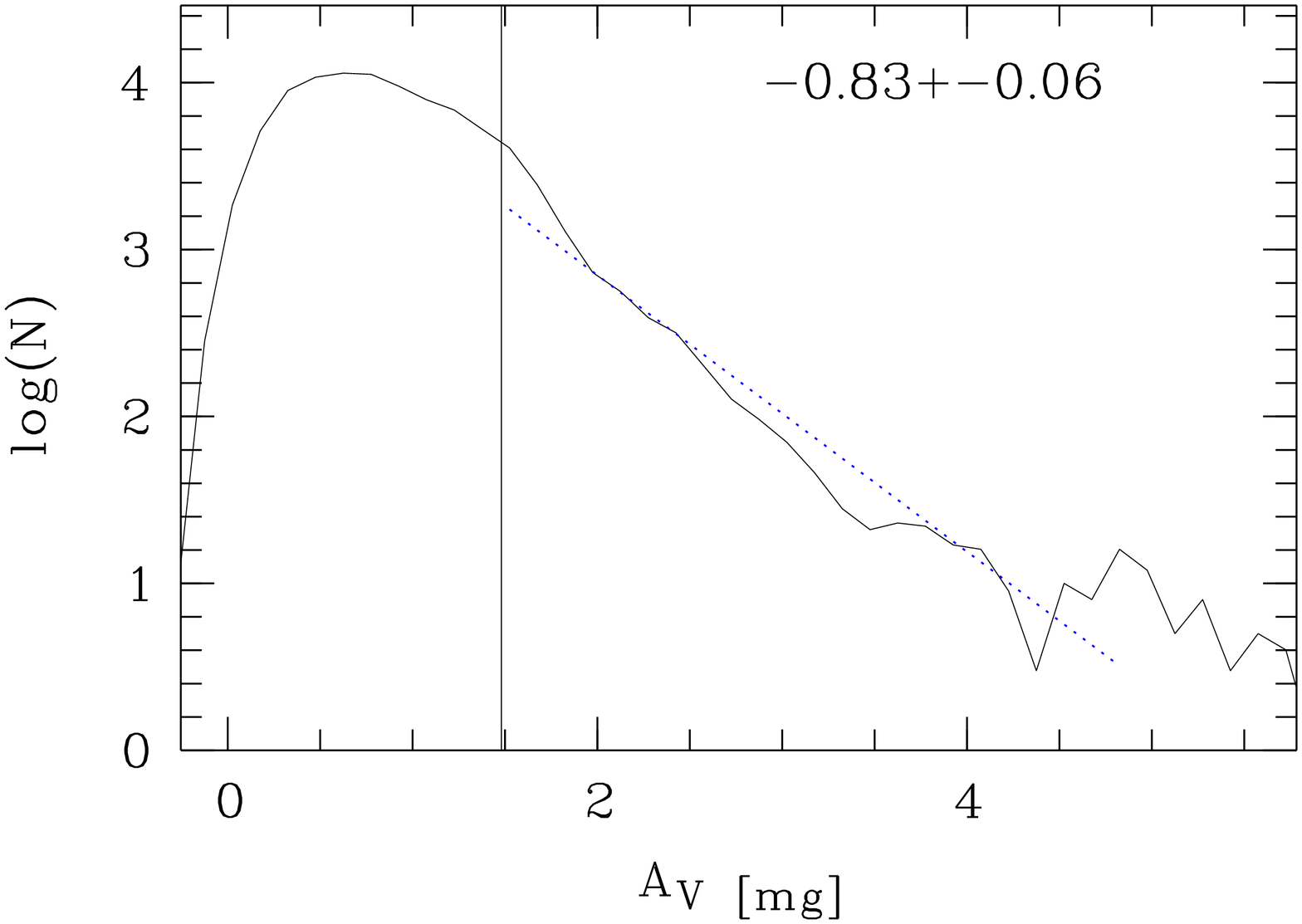}
\hfill
\includegraphics[height=7.5cm,angle=-90]{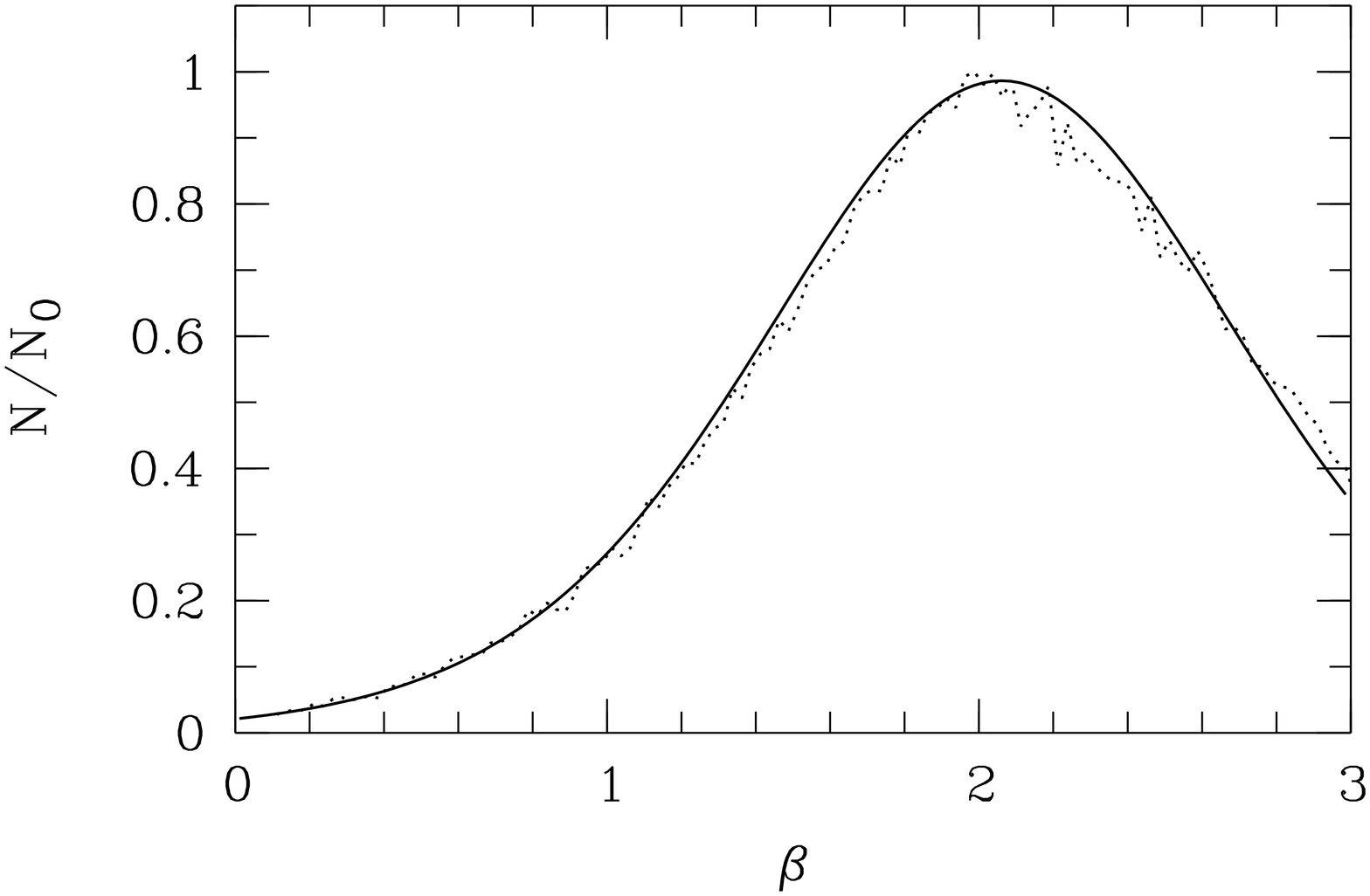}
\caption{\label{plots_aurigae2} As Fig.\,\ref{plots_aurigae1} but for region
Auriga\,2.}  

\end{figure*}

\begin{figure*}
\includegraphics[height=7.5cm,angle=-90]{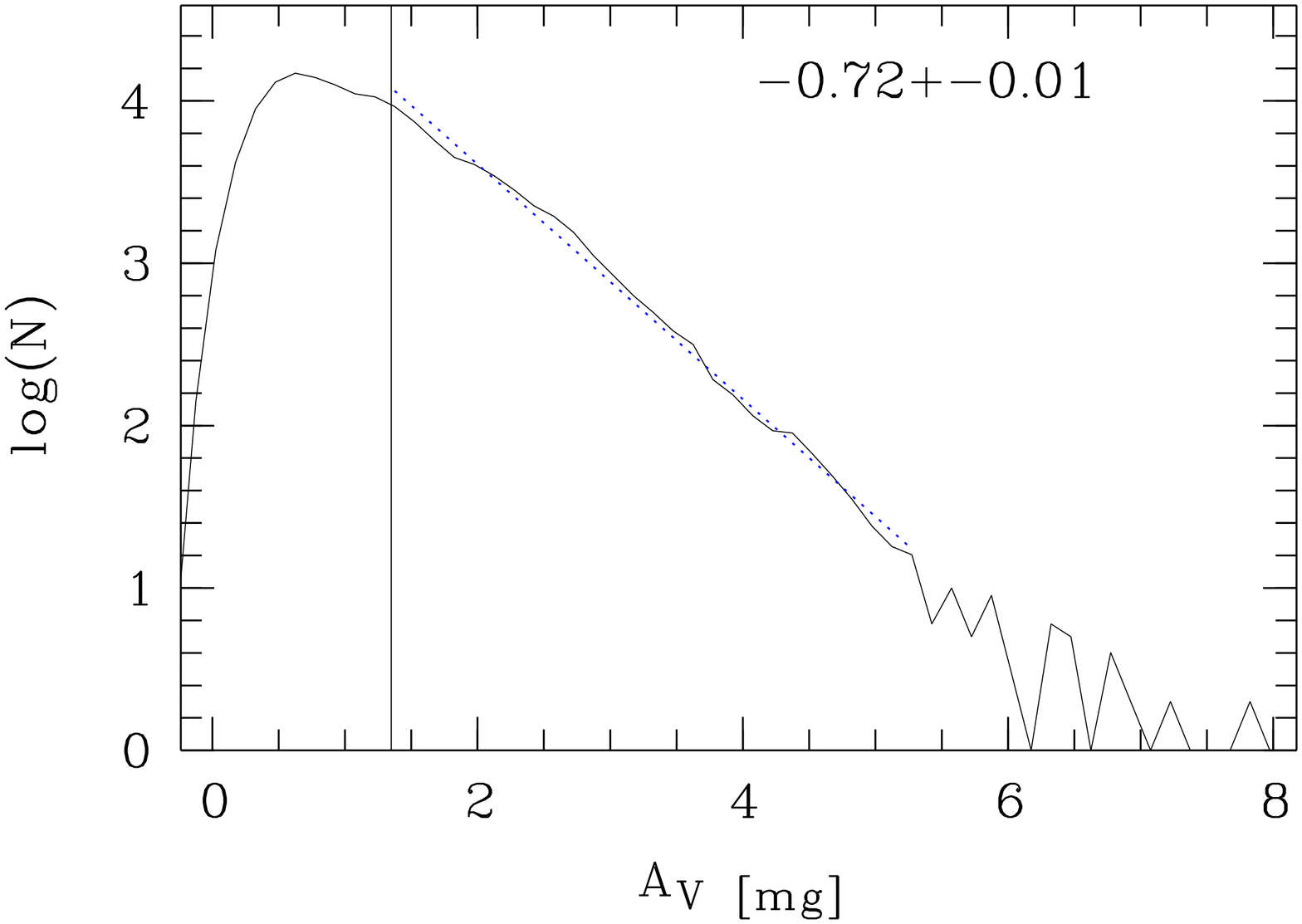}
\hfill
\includegraphics[height=7.5cm,angle=-90]{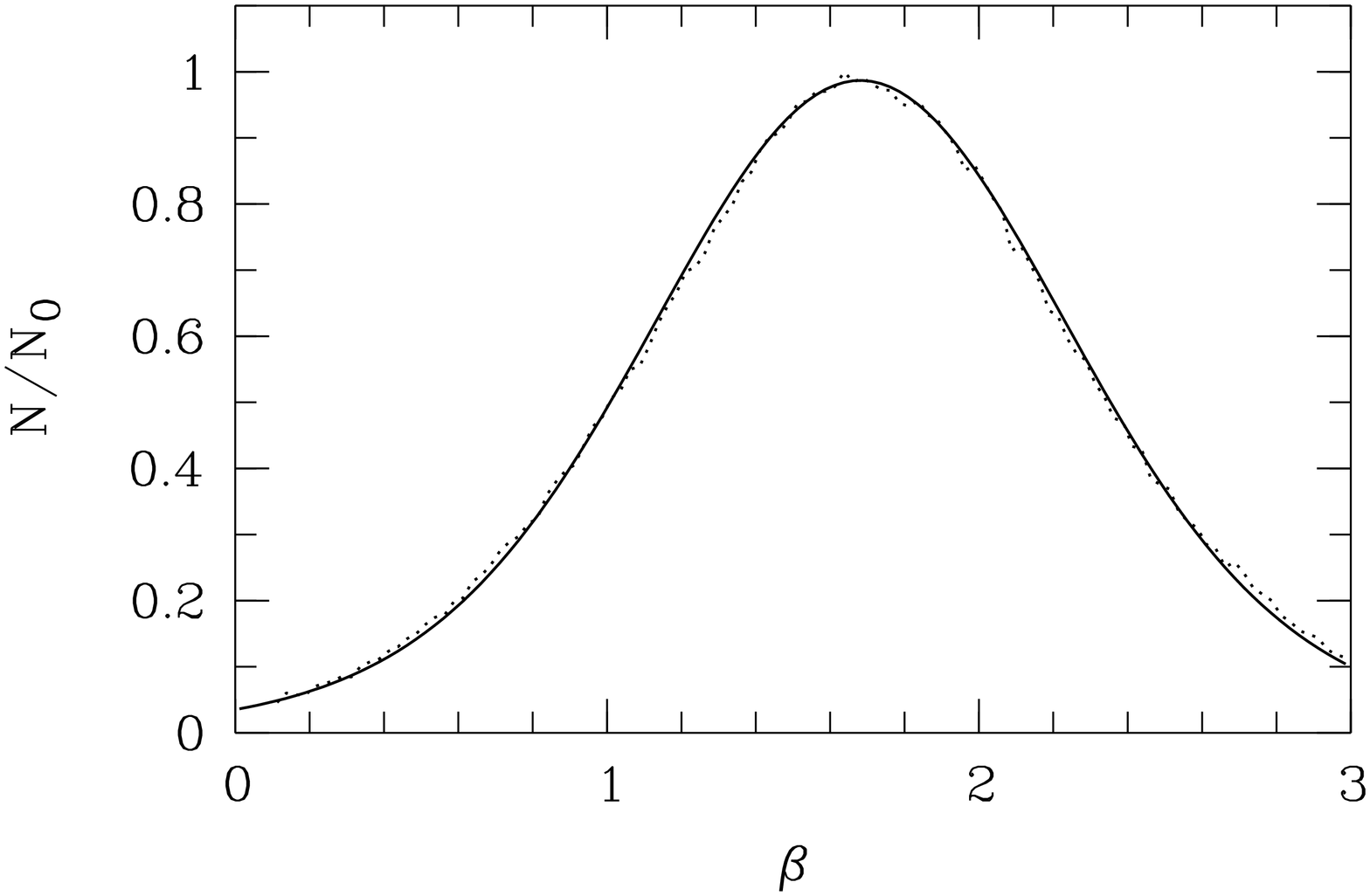}
\caption{\label{plots_cassiopea} As Fig.\,\ref{plots_aurigae1} but for region
Cassiopeia.}  
\end{figure*}

\begin{figure*}
\includegraphics[height=7.5cm,angle=-90]{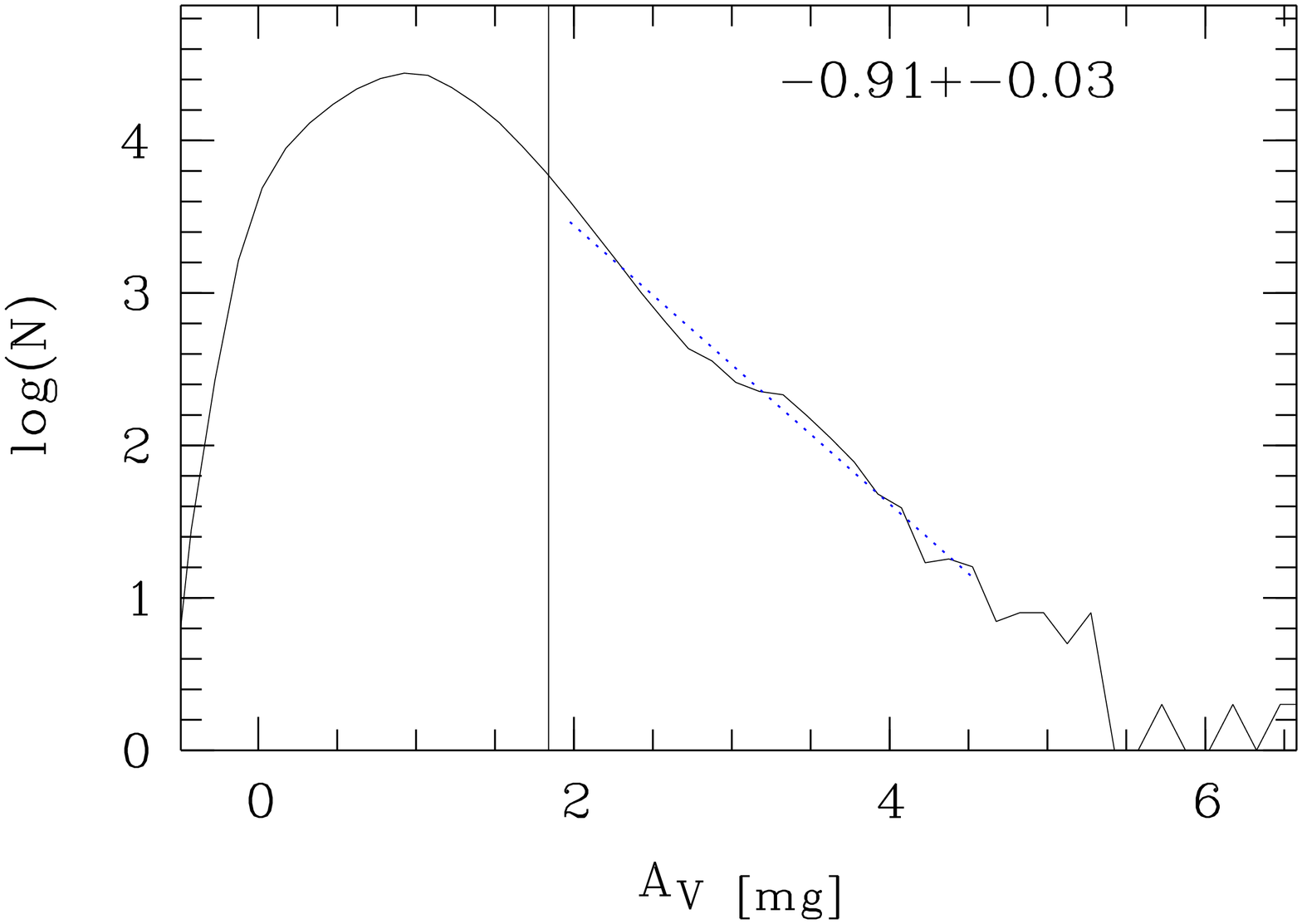}
\hfill
\includegraphics[height=7.5cm,angle=-90]{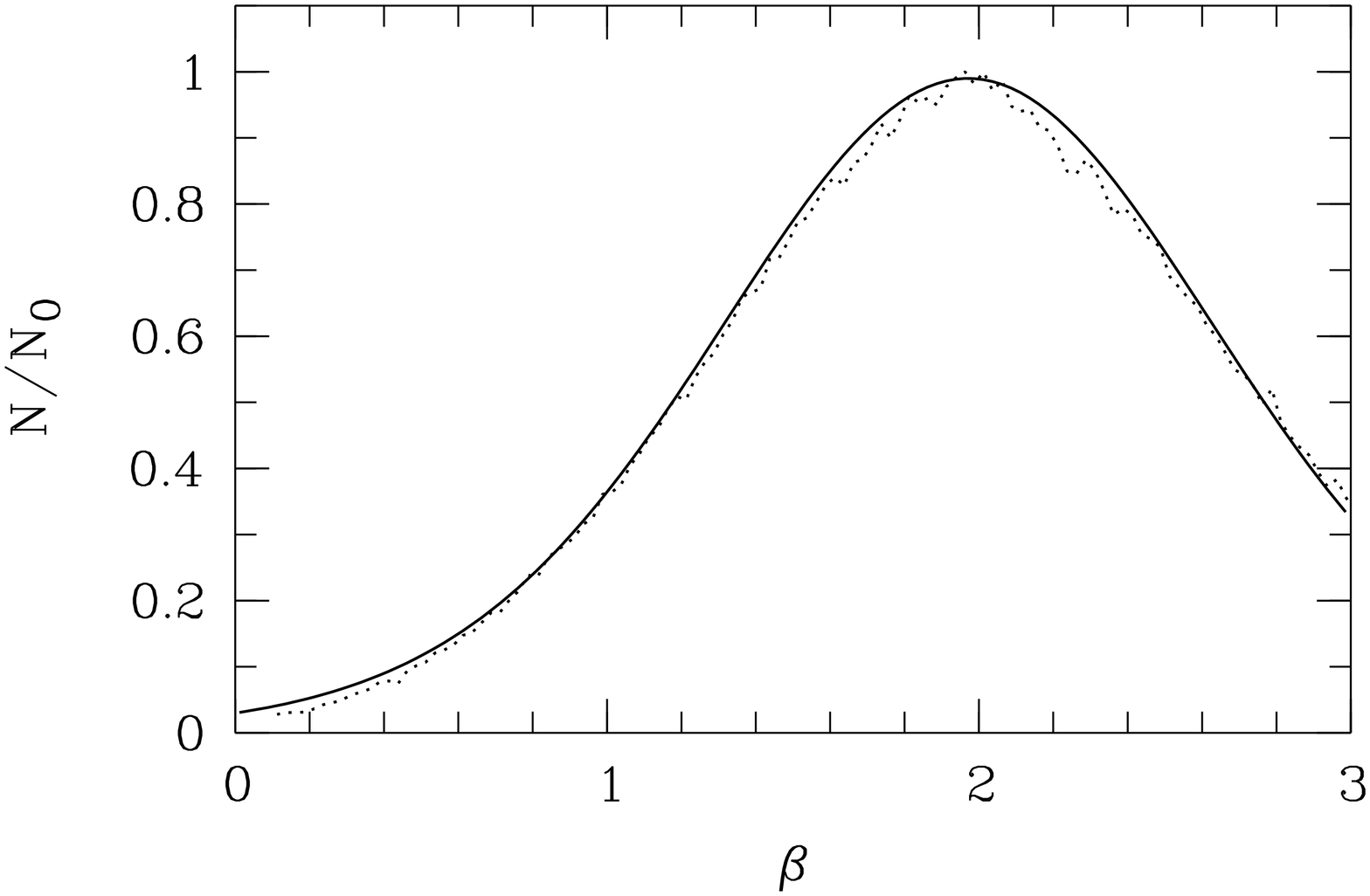}
\caption{\label{plots_chameleopardalis1} As Fig.\,\ref{plots_aurigae1} but for
region Camelopardalis\,1.}   
\end{figure*}

\begin{figure*}
\includegraphics[height=7.5cm,angle=-90]{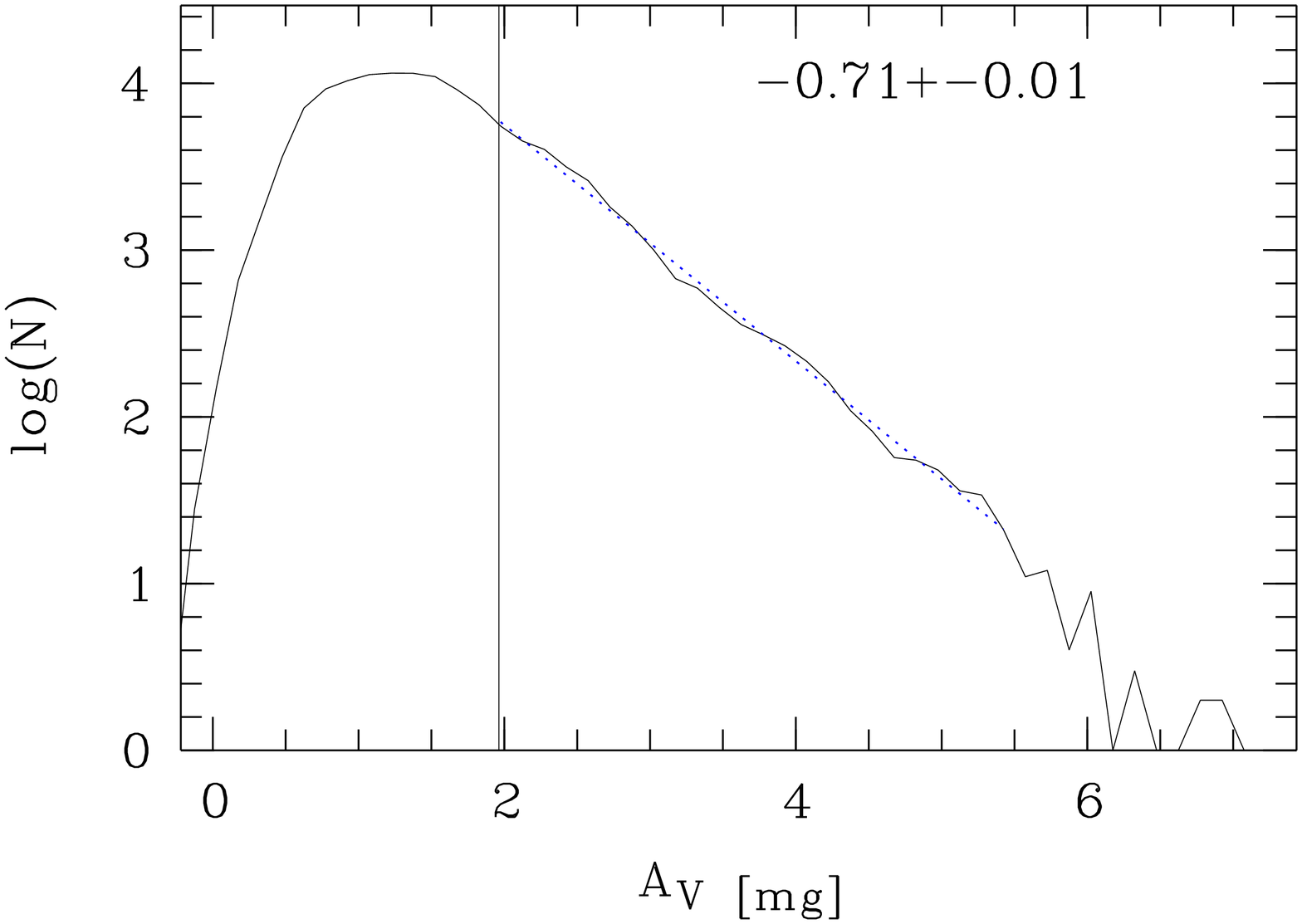}
\hfill
\includegraphics[height=7.5cm,angle=-90]{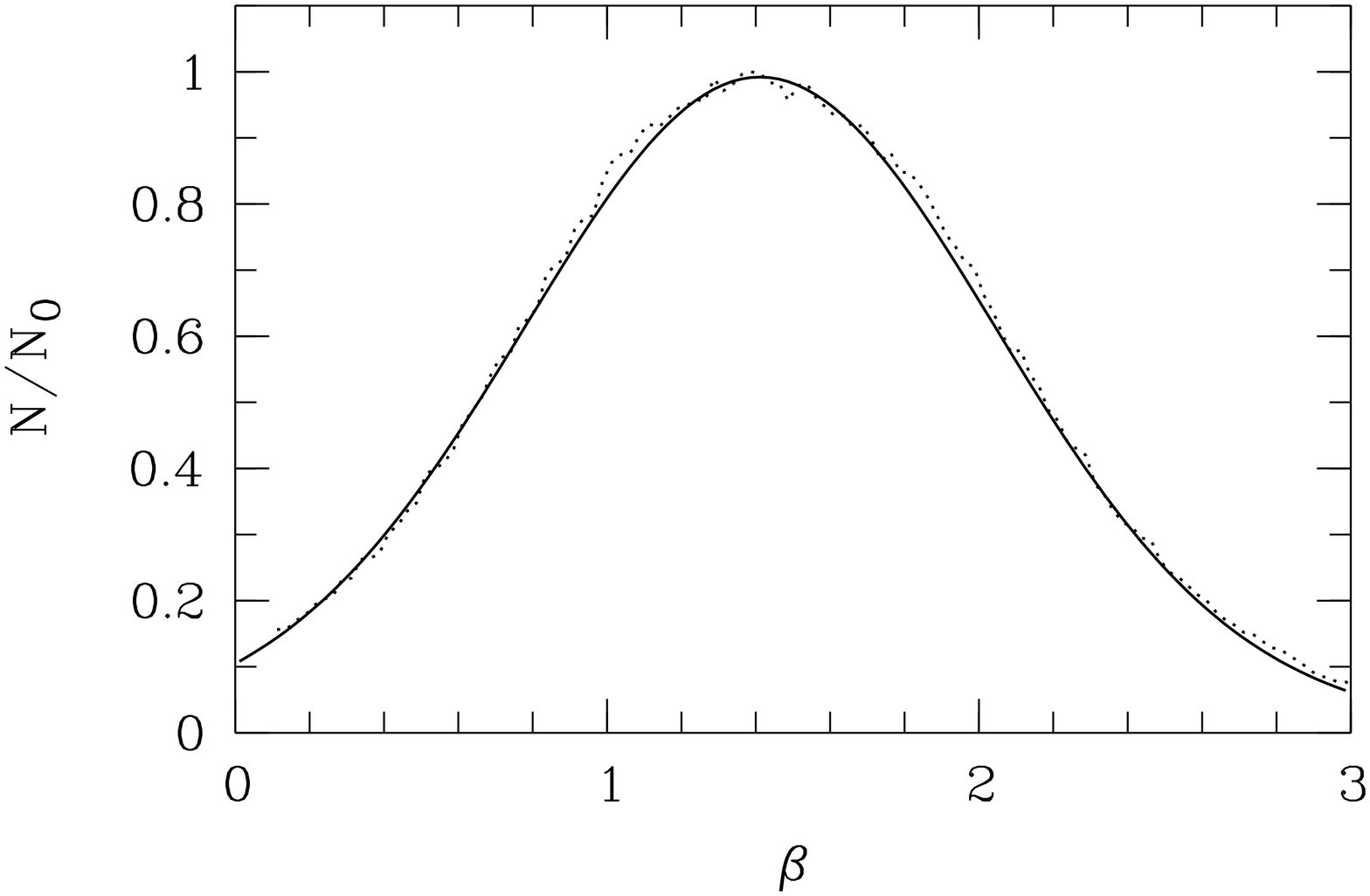}
\caption{\label{plots_chameleopardalis2} As Fig.\,\ref{plots_aurigae1} but for
region Camelopardalis\,2.}   
\end{figure*}

\begin{figure*}
\includegraphics[height=7.5cm,angle=-90]{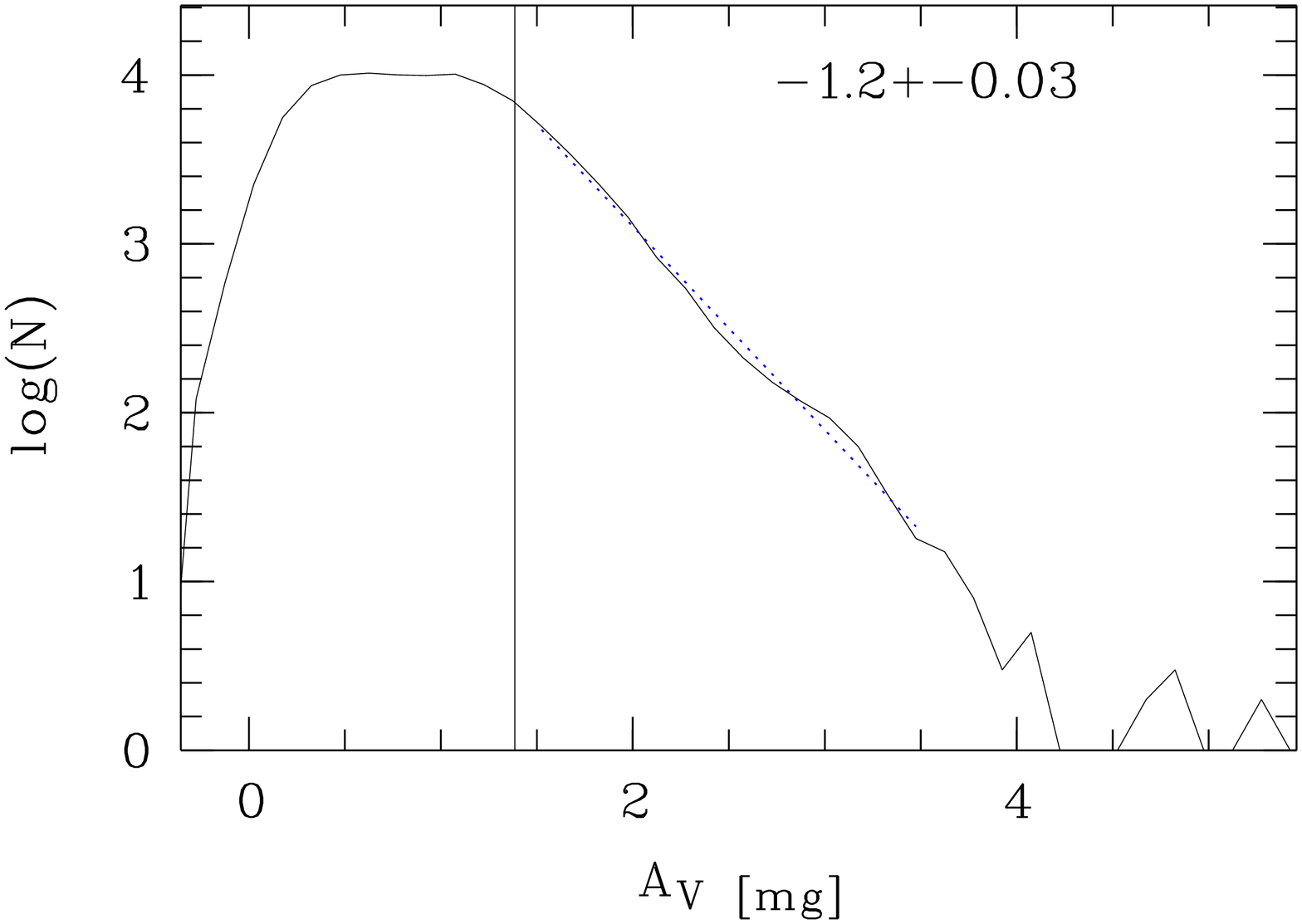}
\hfill
\includegraphics[height=7.5cm,angle=-90]{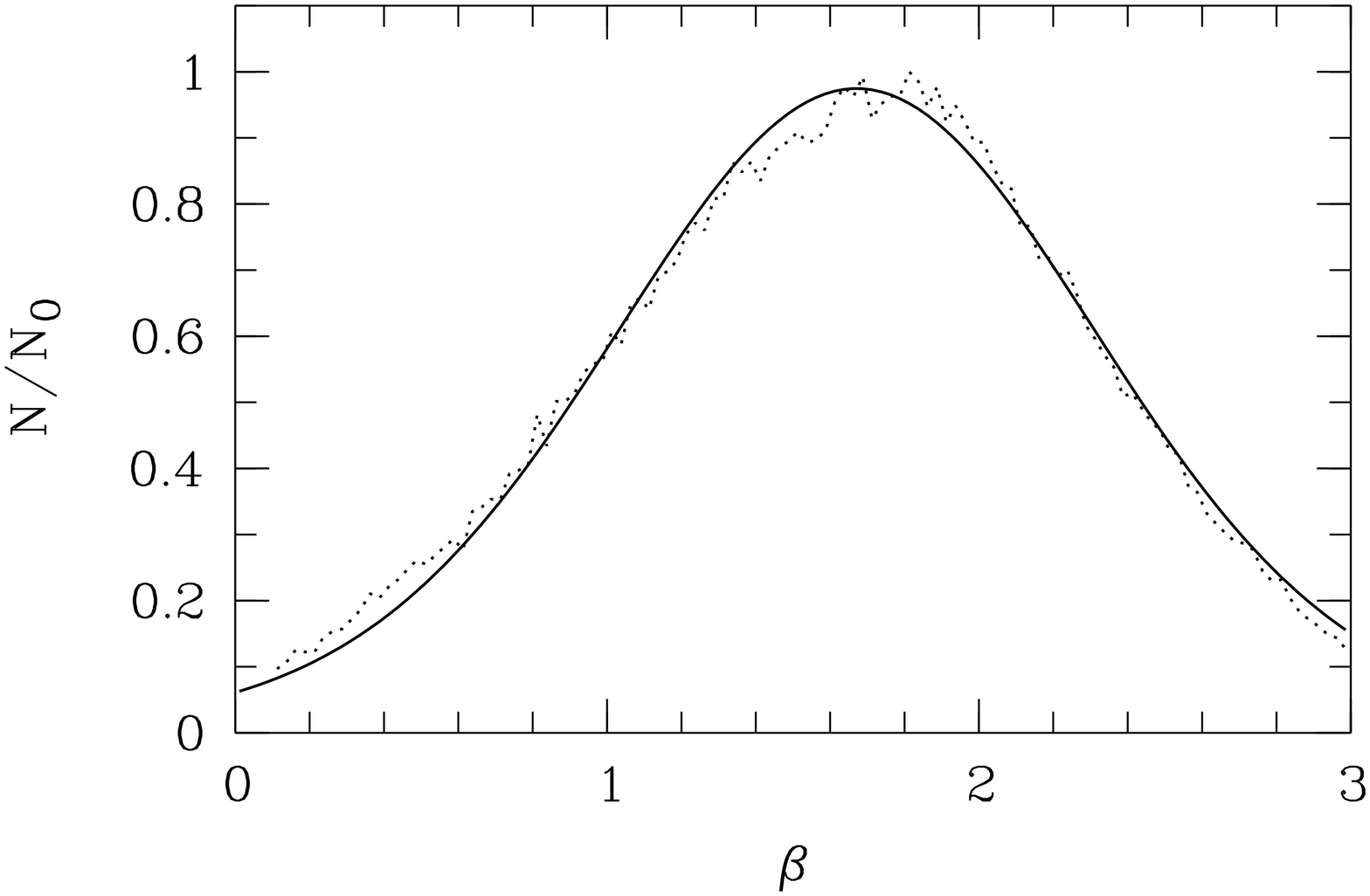}
\caption{\label{plots_chameleopardalis3} As Fig.\,\ref{plots_aurigae1} but for
region Camelopardalis\,3.}   
\end{figure*}

\begin{figure*}
\includegraphics[height=7.5cm,angle=-90]{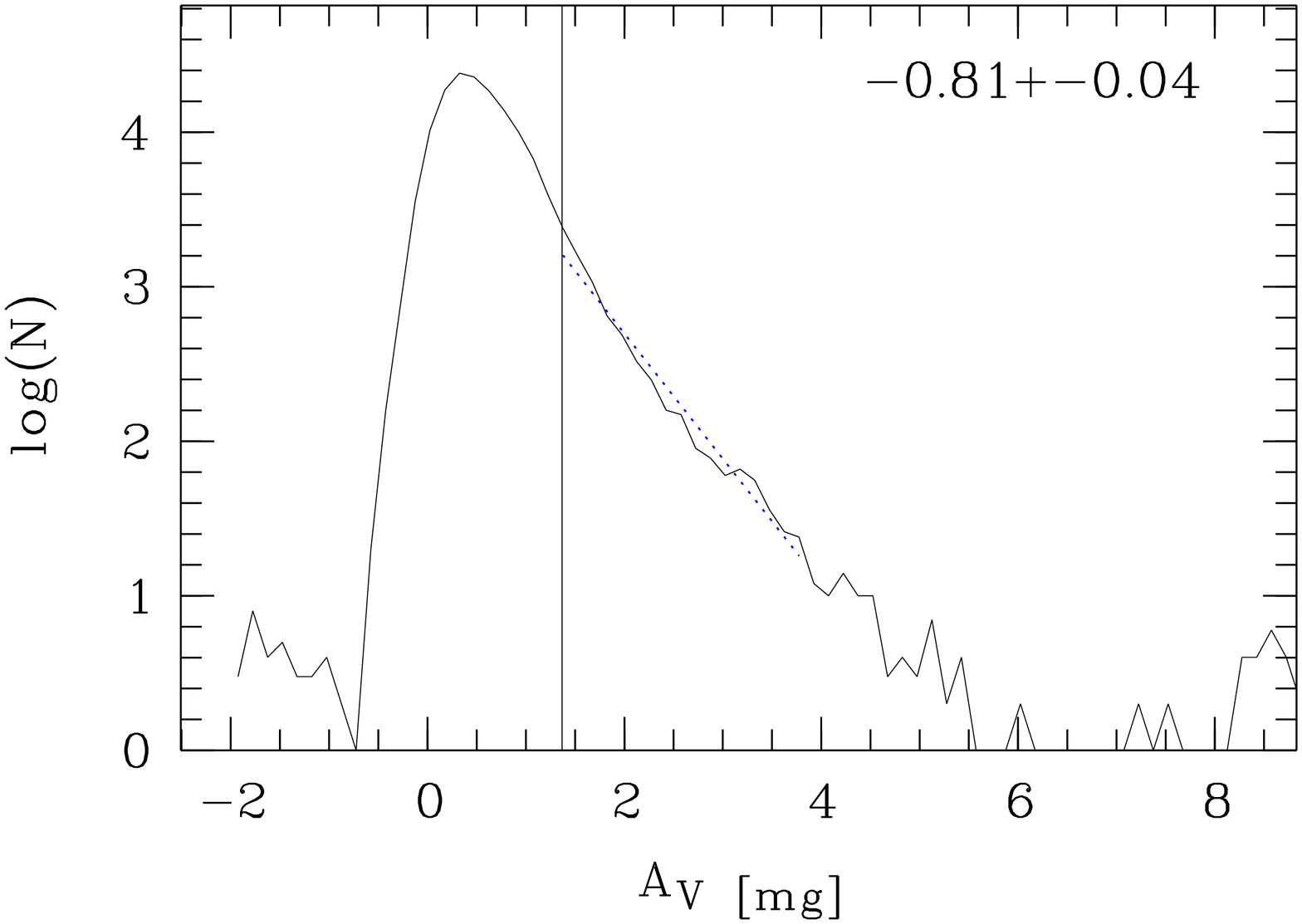}
\hfill
\includegraphics[height=7.5cm,angle=-90]{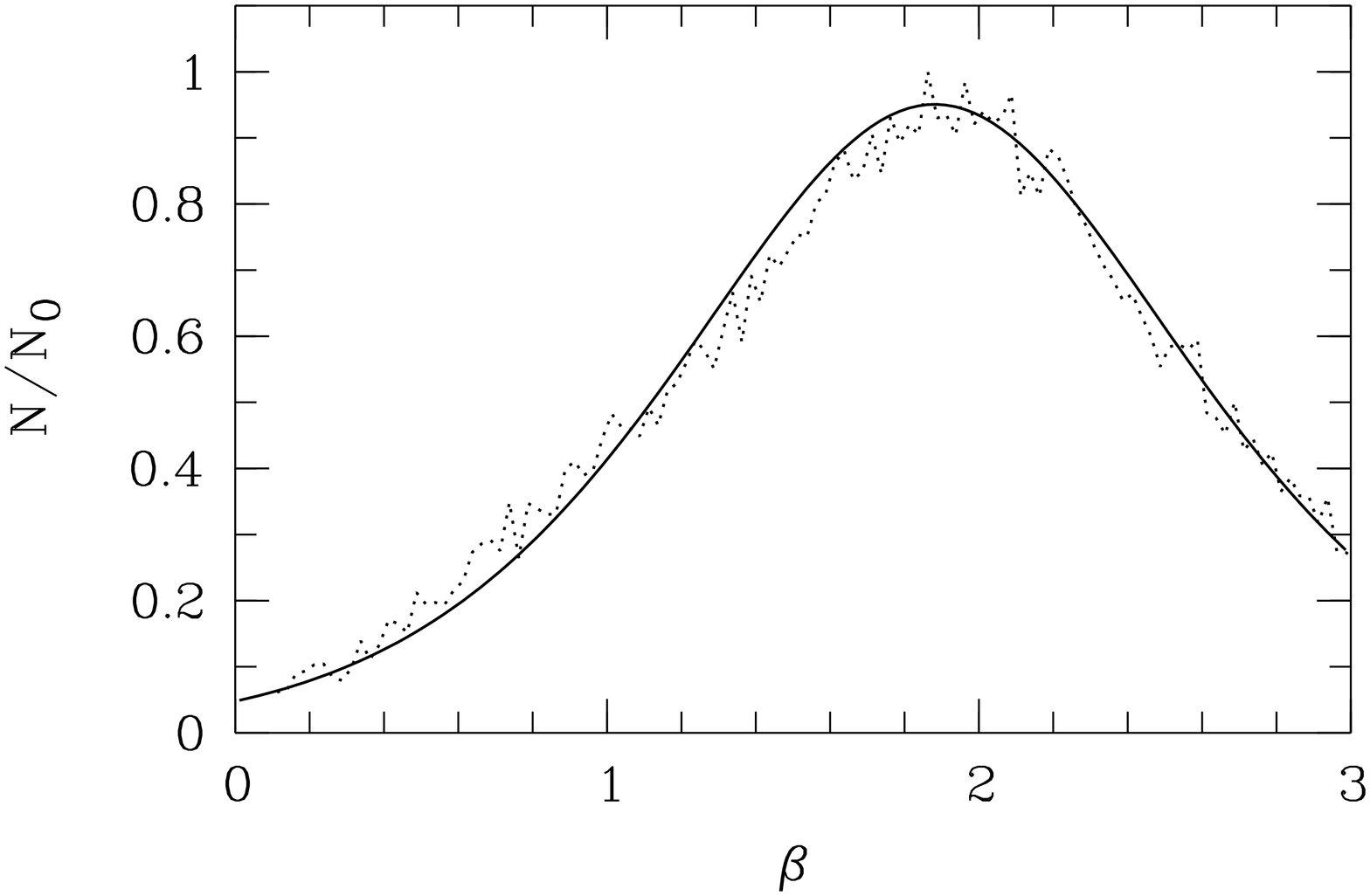}
\caption{\label{plots_lori} As Fig.\,\ref{plots_aurigae1} but for region
$\lambda$-Ori.} 
 
\end{figure*}

\begin{figure*}
\includegraphics[height=7.5cm,angle=-90]{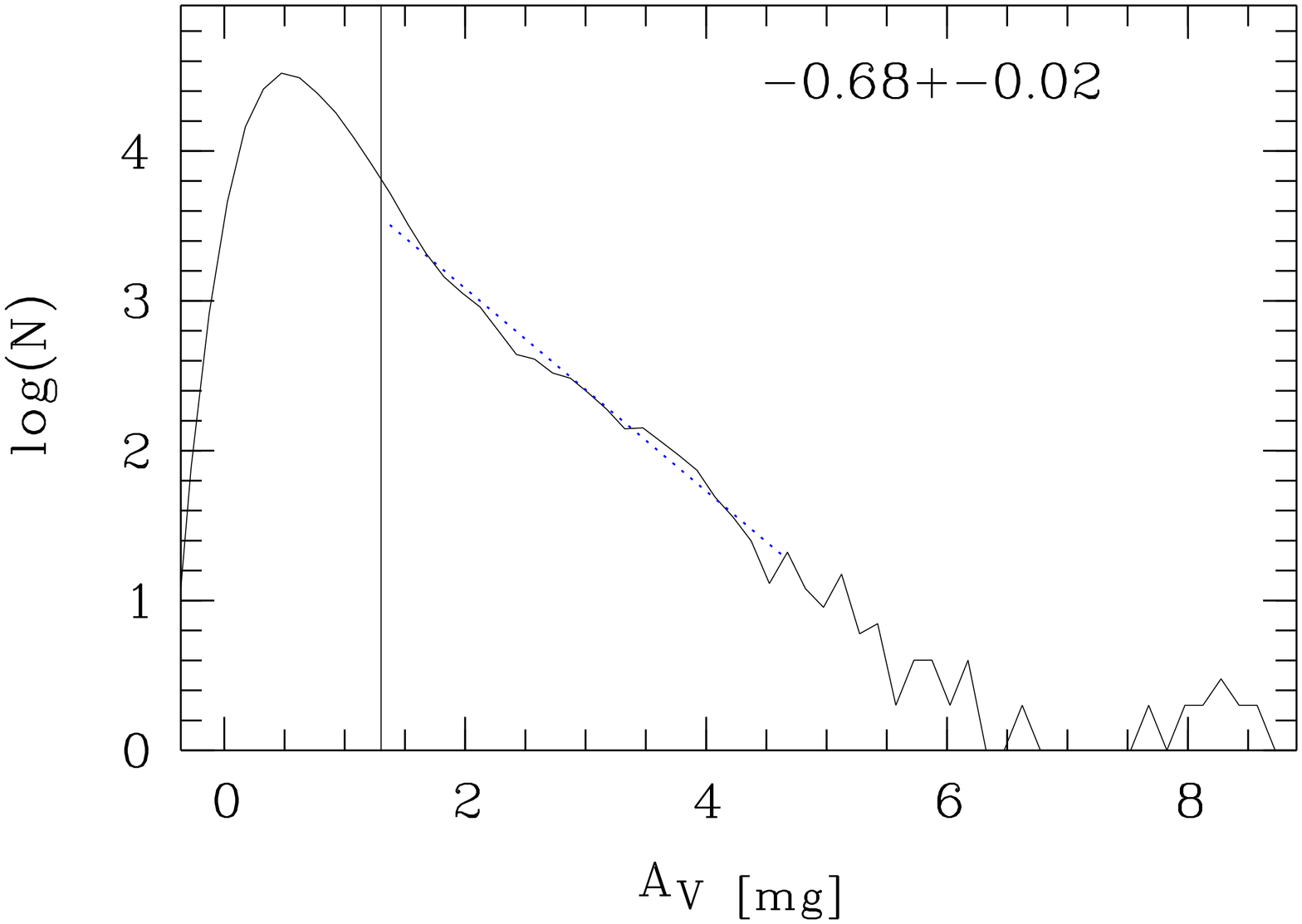}
\hfill
\includegraphics[height=7.5cm,angle=-90]{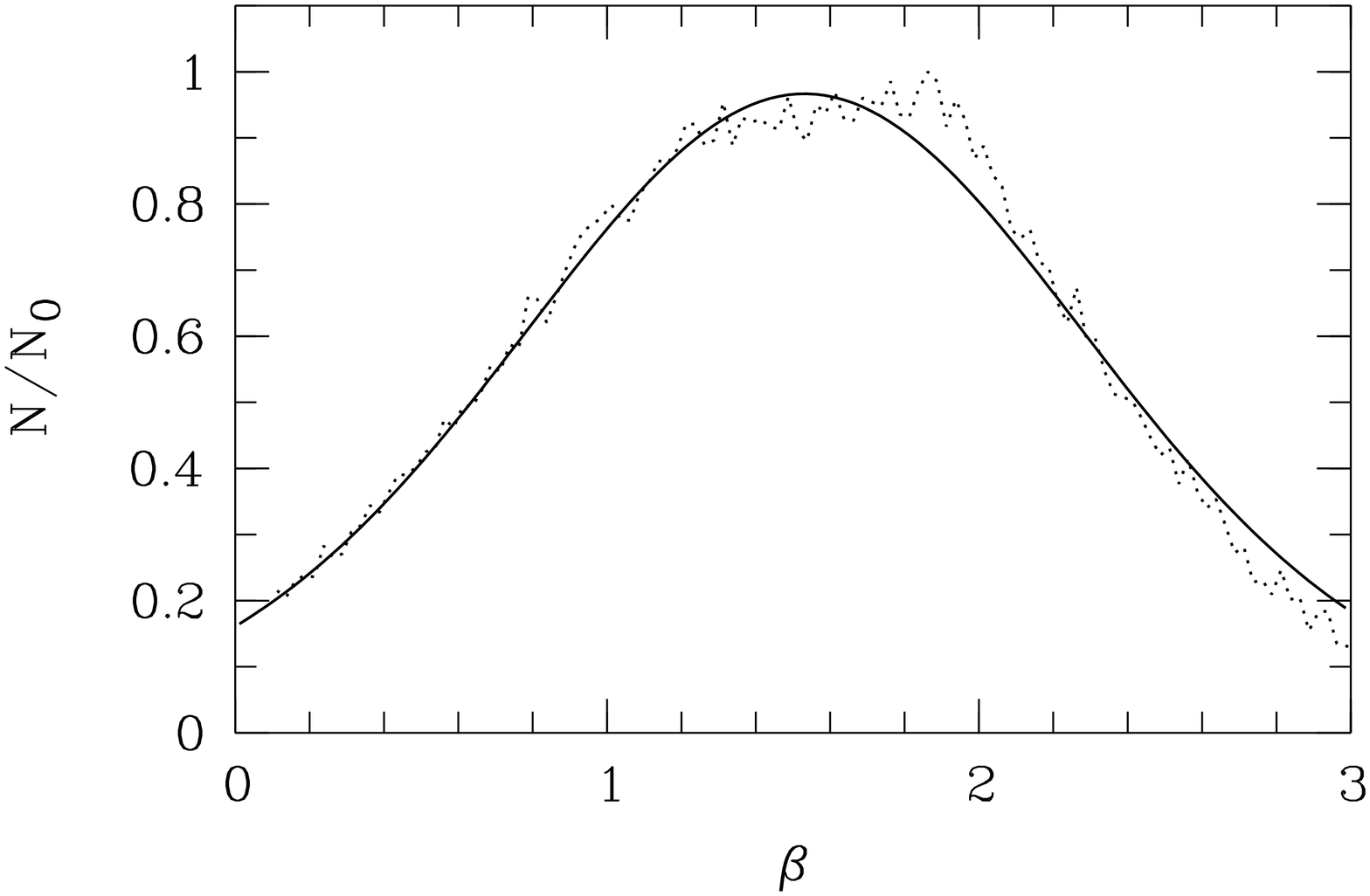}
\caption{\label{plots_monocerus} As Fig.\,\ref{plots_aurigae1} but for region
Monoceros.}  
\end{figure*}

\begin{figure*}
\includegraphics[height=7.5cm,angle=-90]{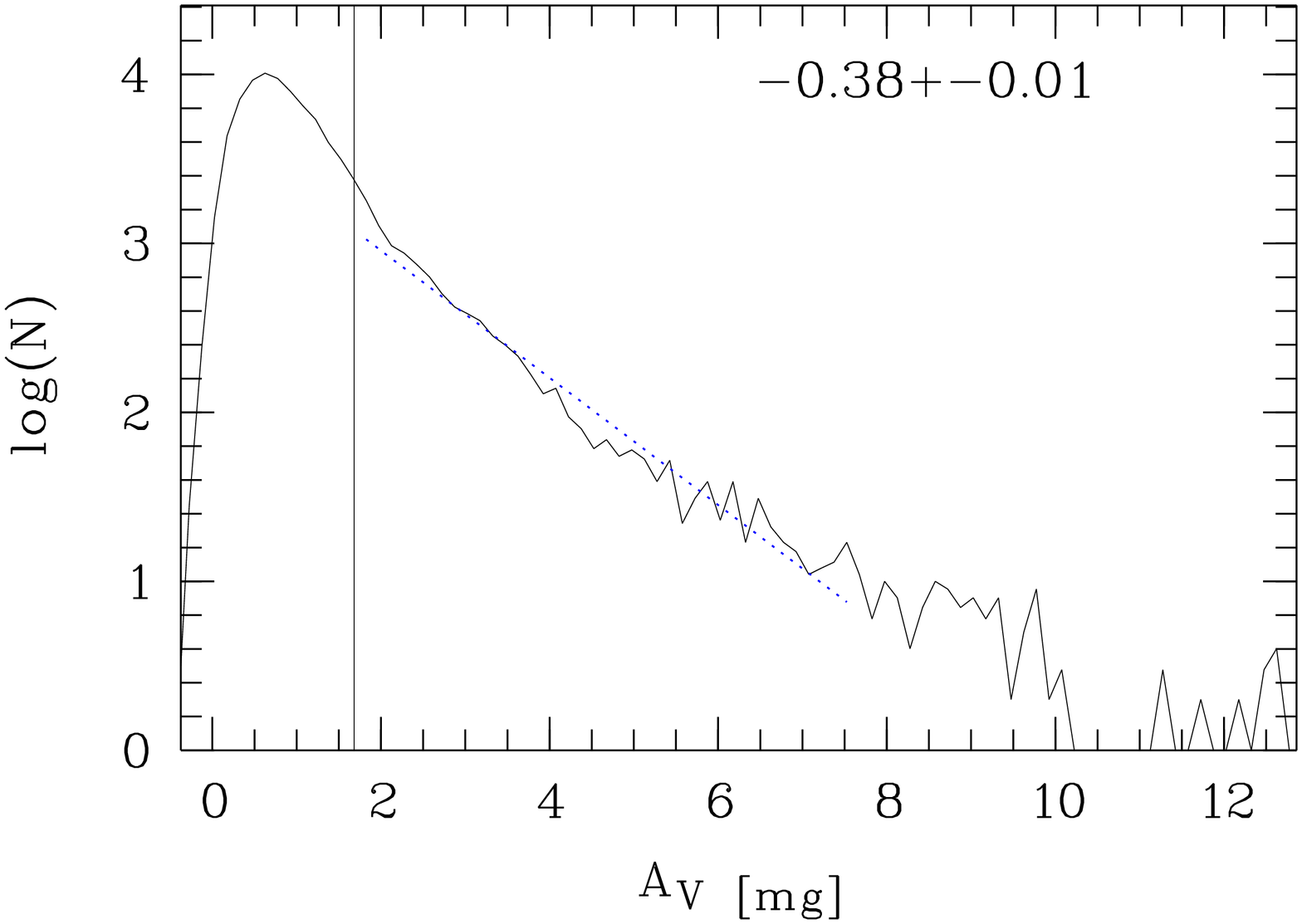}
\hfill
\includegraphics[height=7.5cm,angle=-90]{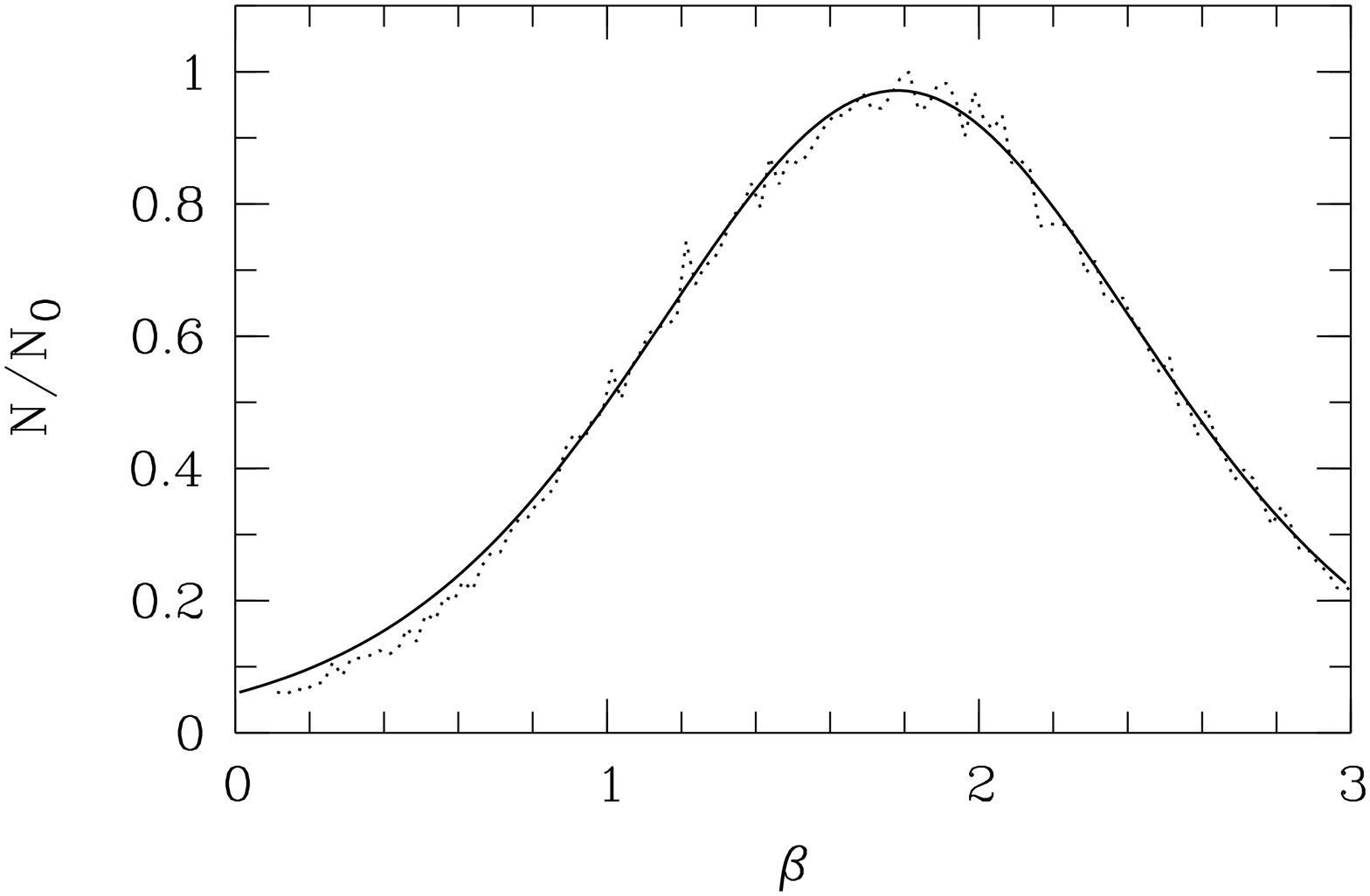}
\caption{\label{plots_oria} As Fig.\,\ref{plots_aurigae1} but for region
Ori\,B.}  

\end{figure*}

\begin{figure*}
\includegraphics[height=7.5cm,angle=-90]{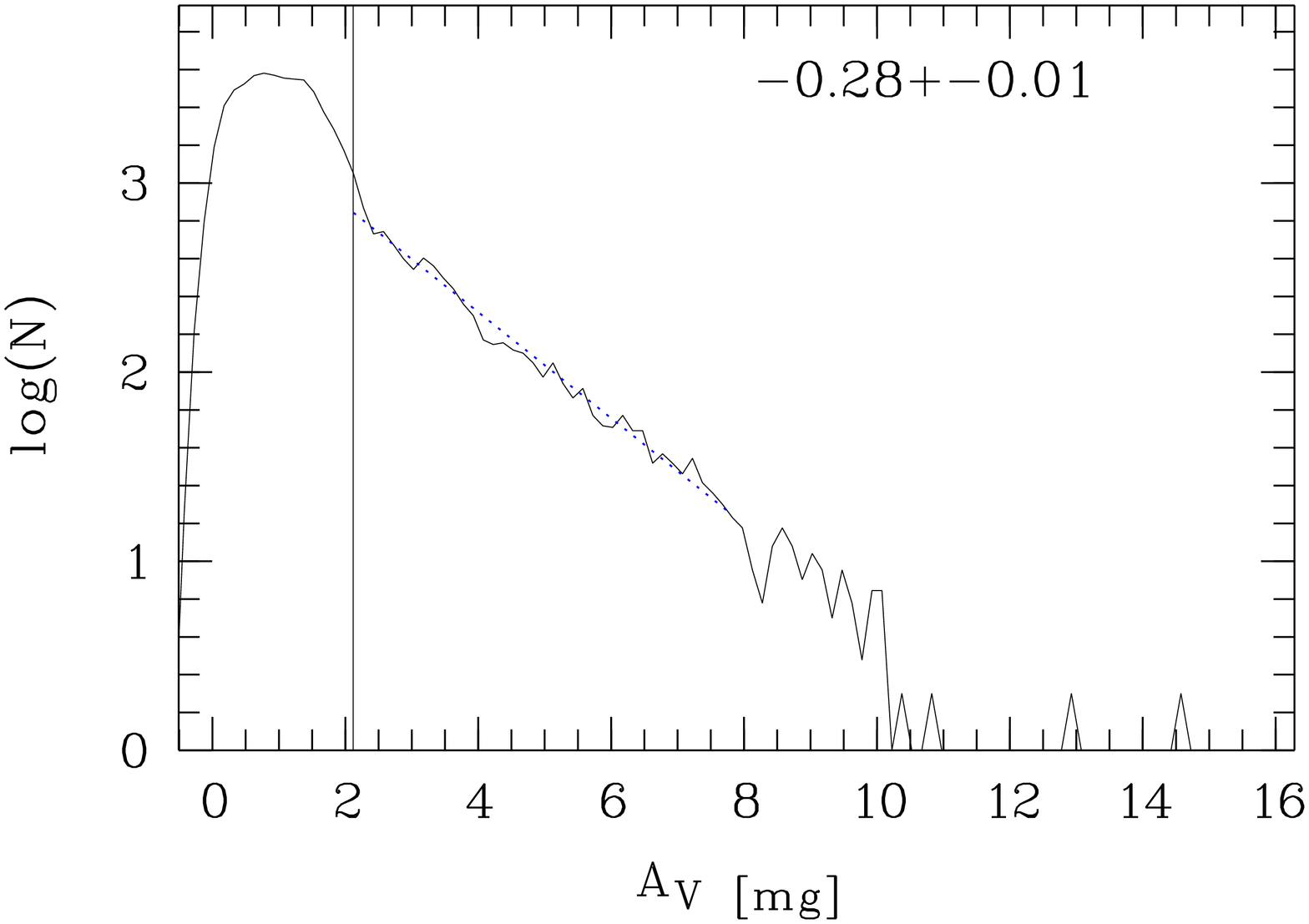}
\hfill
\includegraphics[height=7.5cm,angle=-90]{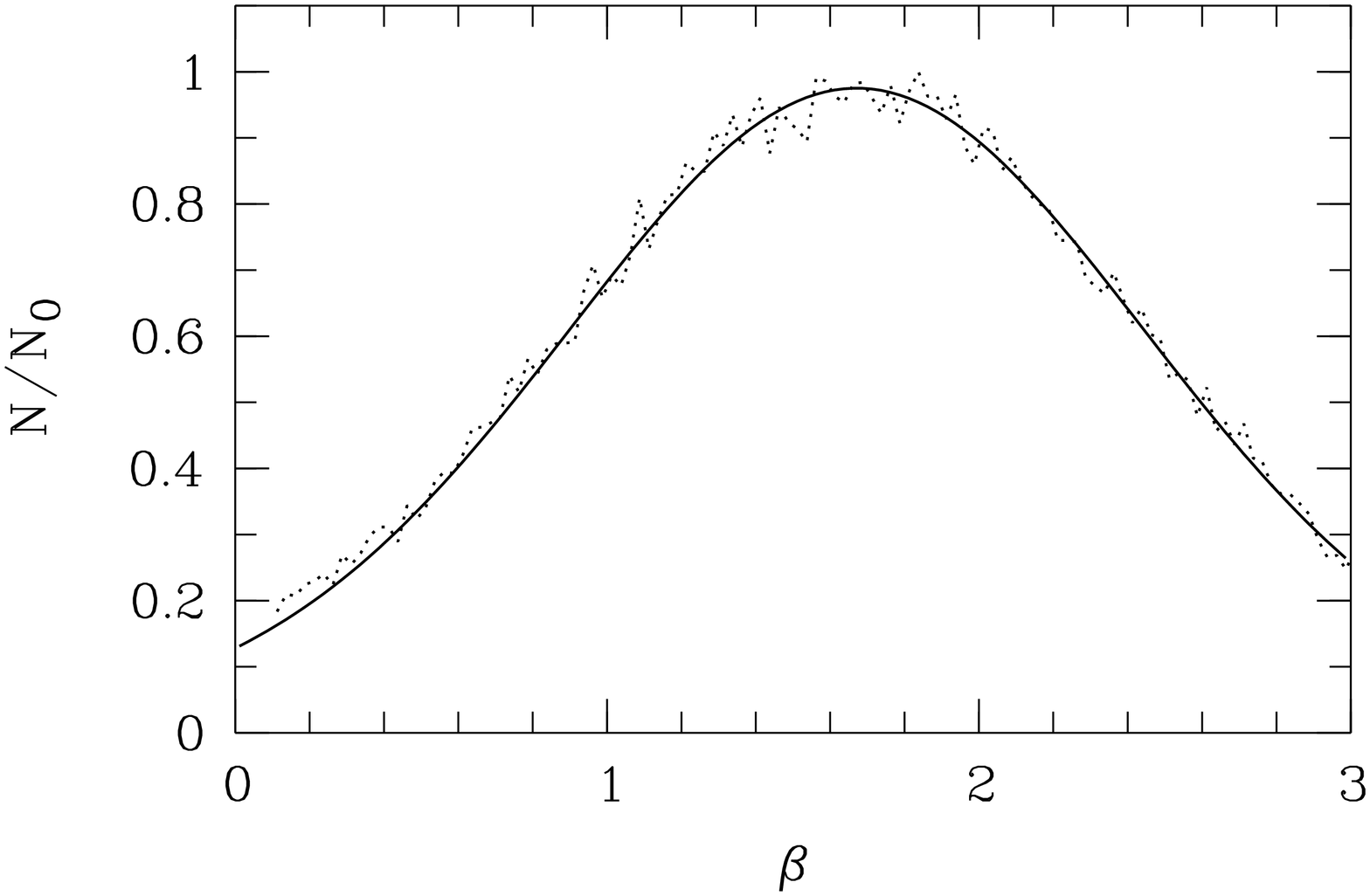}
\caption{\label{plots_orib} As Fig.\,\ref{plots_aurigae1} but for region
Ori\,A.} 
\end{figure*}

\begin{figure*}
\includegraphics[height=7.5cm,angle=-90]{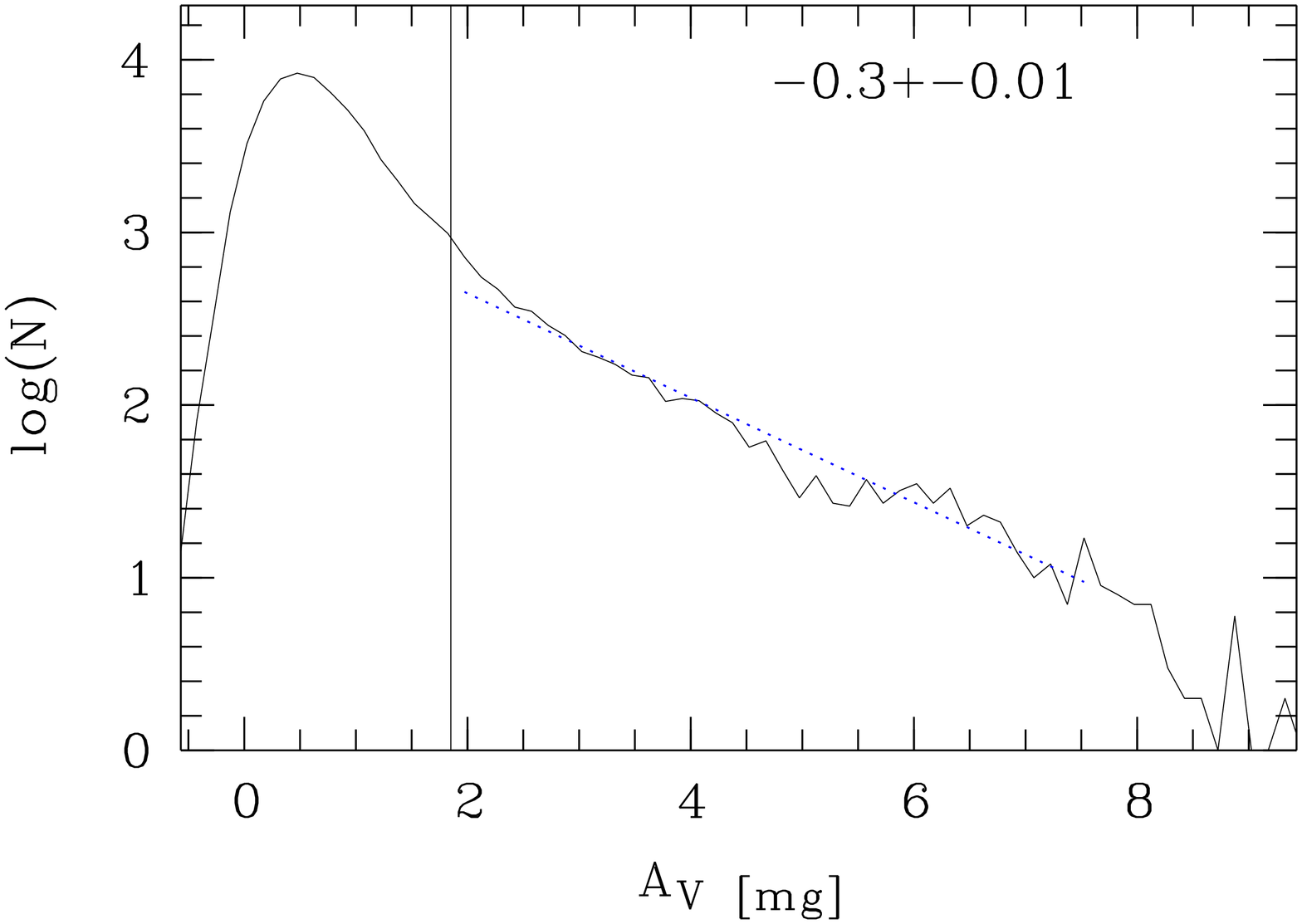}
\hfill
\includegraphics[height=7.5cm,angle=-90]{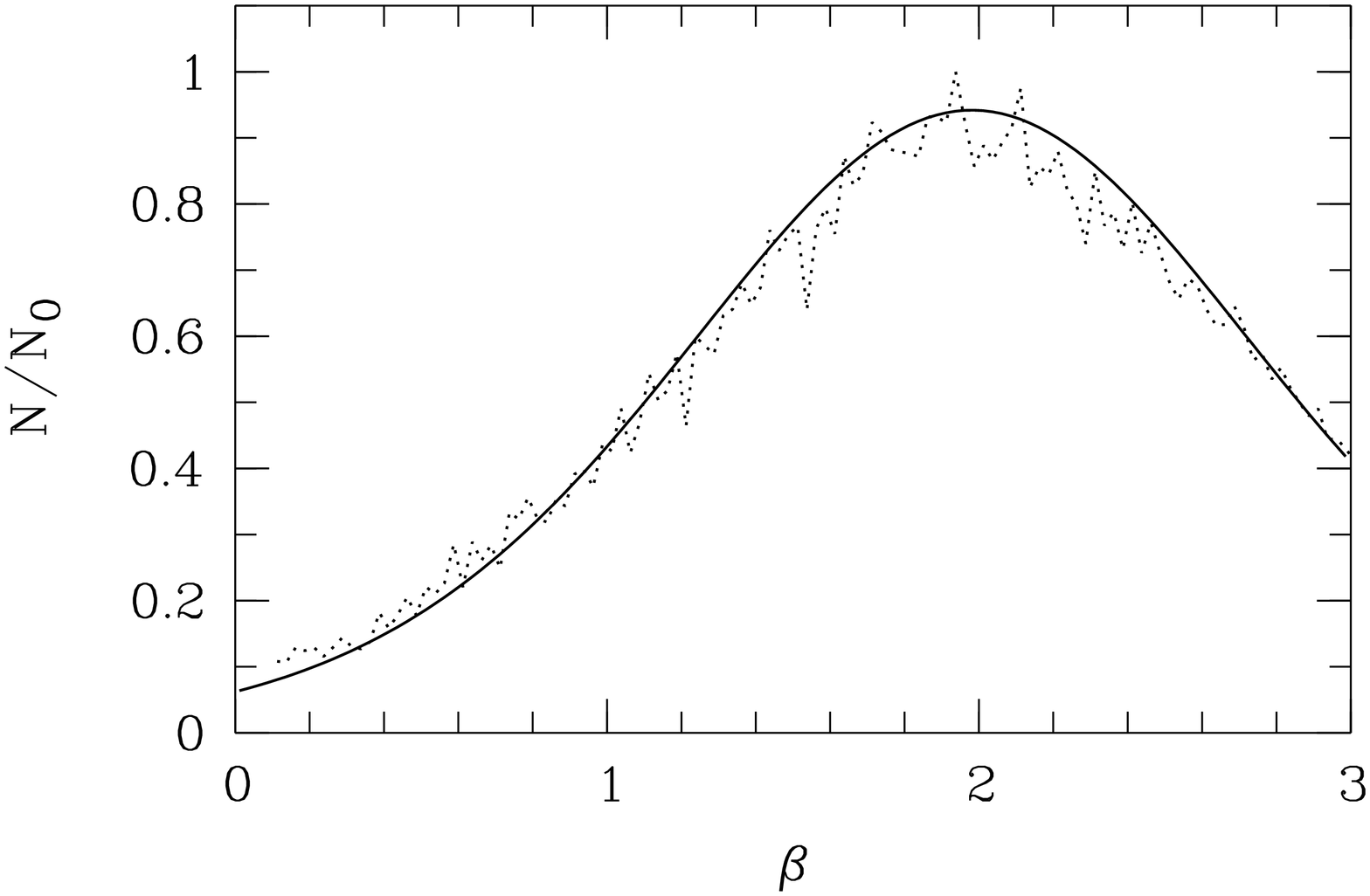}
\caption{\label{plots_perseus} As Fig.\,\ref{plots_aurigae1} but for region
Perseus.}  

\end{figure*}

\begin{figure*}
\includegraphics[height=7.5cm,angle=-90]{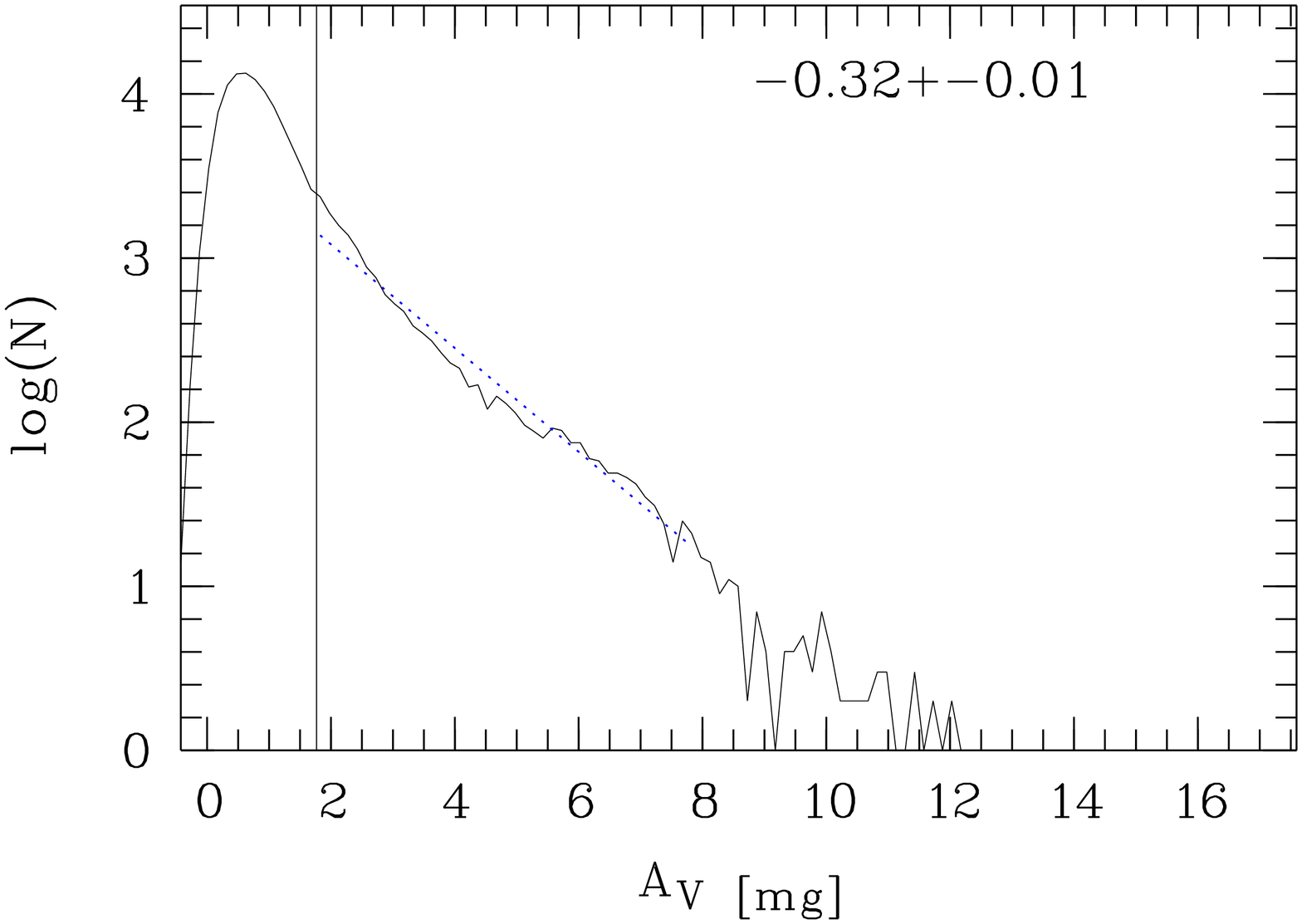}
\hfill
\includegraphics[height=7.5cm,angle=-90]{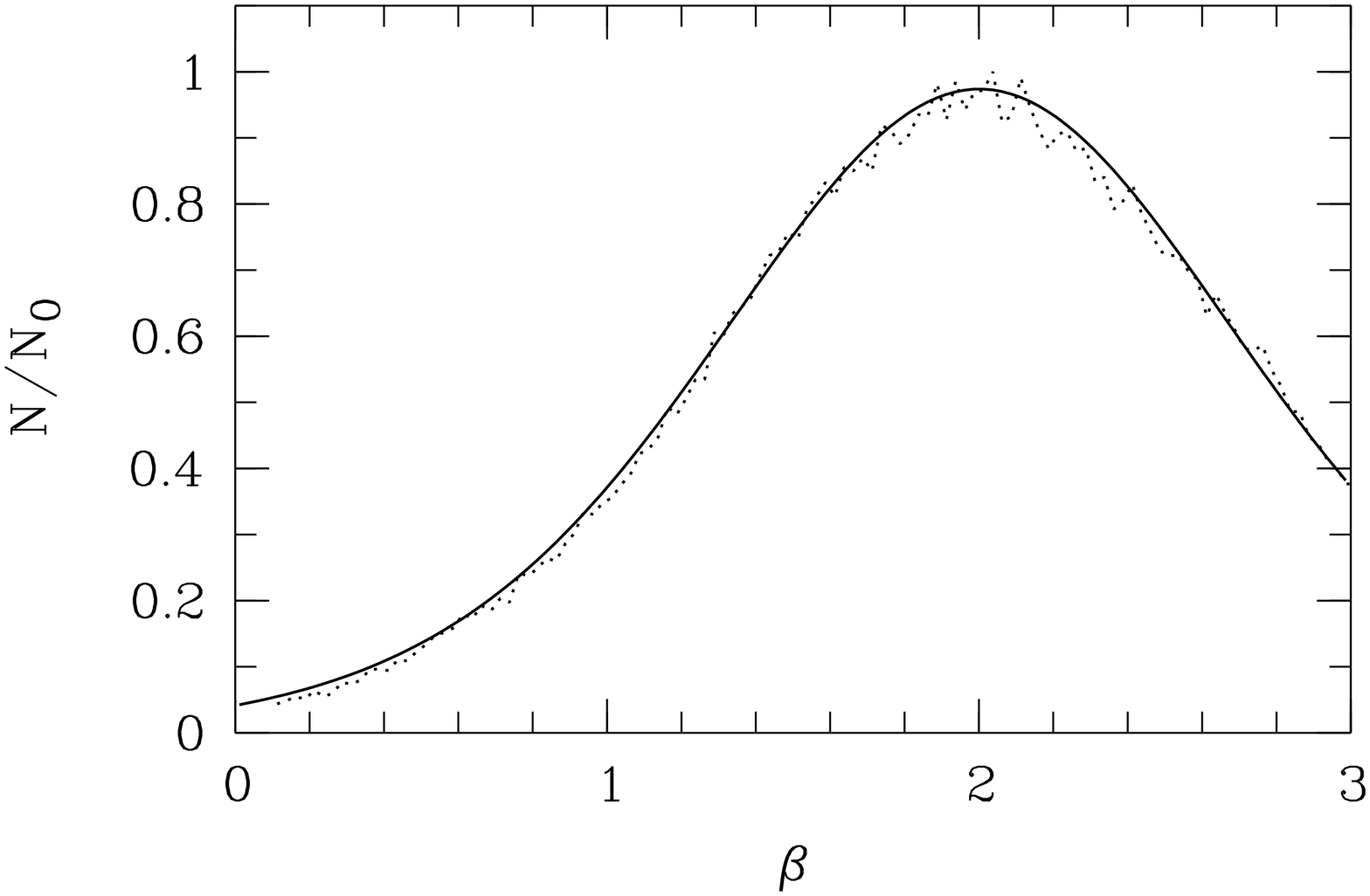}
\caption{\label{plots_taurus} As Fig.\,\ref{plots_aurigae1} but for region
Taurus.} 
\end{figure*}

\begin{figure*}
\includegraphics[height=7.5cm,angle=-90]{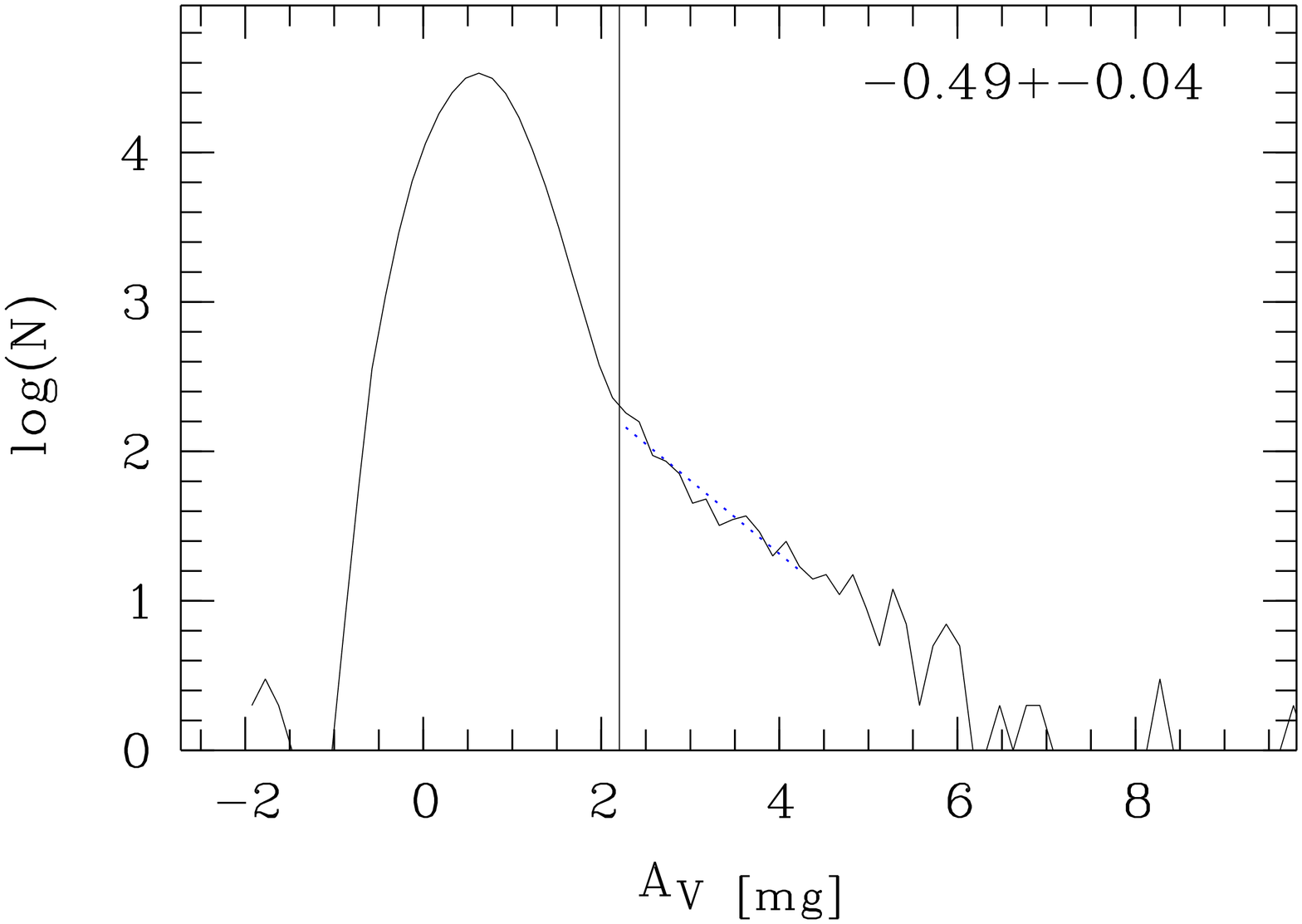}
\hfill
\includegraphics[height=7.5cm,angle=-90]{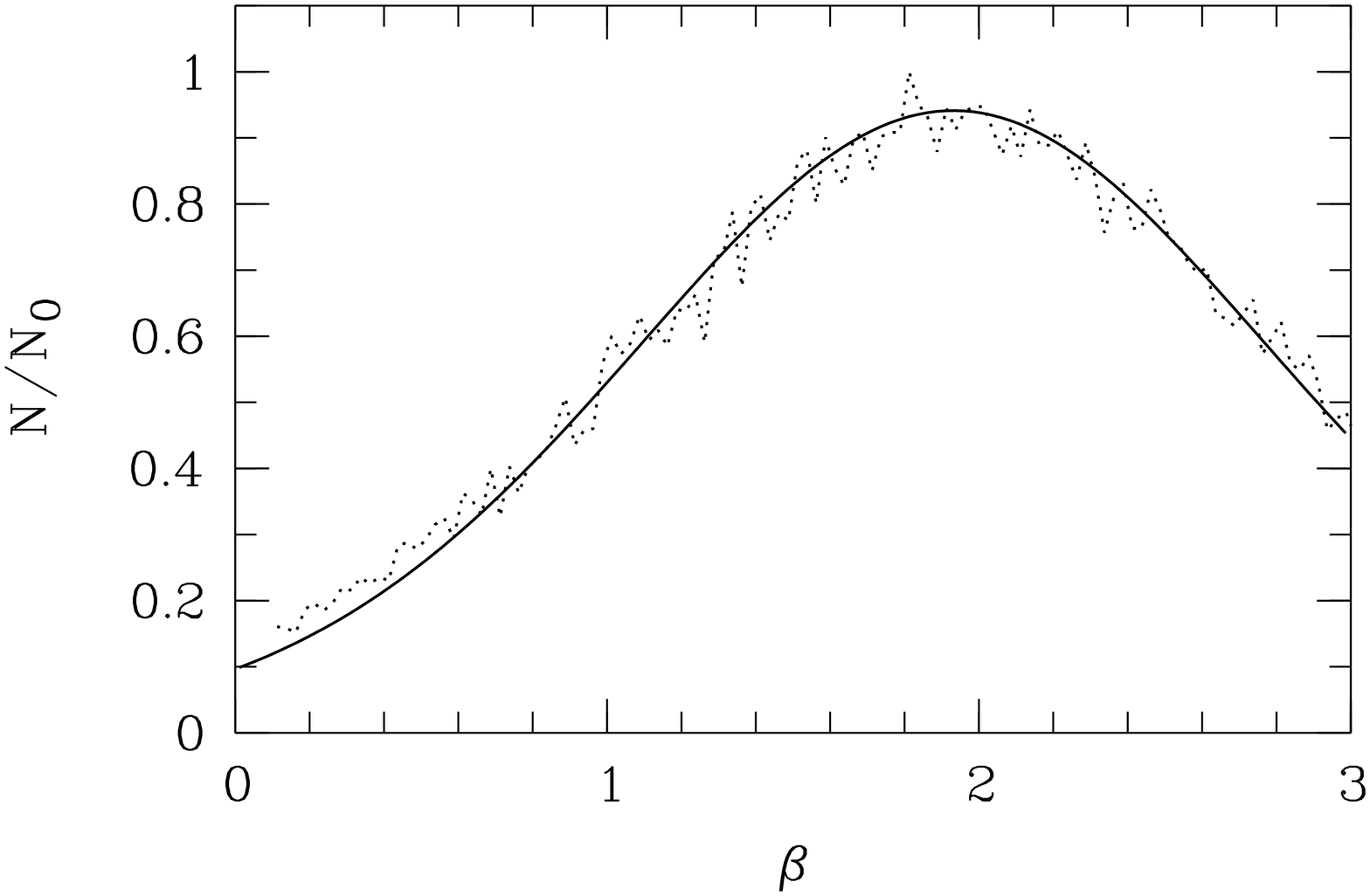}
\caption{\label{plots_taurusextended} As Fig.\,\ref{plots_aurigae1} but for
region Taurusextended.}  
\end{figure*}

\end{appendix}


\begin{thebibliography}{}

\bibitem[1956]{1956AJ.....61..309B} 
Bok, B.J. 1956, AJ, 61, 309 

\bibitem[2005]{Coghlan2.05.EGC}
Coghlan, B.A., Walsh, J. \& D. O'Callaghan 2005, in 'Advances in Grid Computing - EGC 2005', Peter M.A. Sloot, Alfons G. Hoekstra, Thierry Priol, Alexander Reinefeld, and Marian Bubak, editors, LNCS3470, Amsterdam, The Netherlands, February 2005. Springer.

\bibitem[2003]{2003ARA&A..41..241D}
Draine, B.T. 2003, ARA\&A, 41, 241

\bibitem[2007]{2006MNRAS.subm.F}
Froebrich, D., Scholz, A. \& Raftery, C.L. 2007, MNRAS, 374, 399

\bibitem[2006]{2006MNRAS.369.1901F}
Froebrich, D. \& del Burgo, C. 2006, MNRAS, 369, 1901

\bibitem[2005]{2005A&A...432L..67F}
Froebrich, D., Ray, T.P., Murphy, G.C. \& Scholz, A. 2005, A\&A, 432, 67

\bibitem[1998]{1998A&A...338L..33H}
Haas, M., Lemke, D., Stickel, M., Hippelein, H., Kunkel, M., Herbstmeier, U. \& Mattila, K. 1998, A\&A, 338, 33

\bibitem[2001]{2001ApJ...562..852H}
Hartmann, L., Ballesteros-Paredes, J. \& Bergin, E.A. 2001, ApJ, 562, 852

 %
 %

 %
 %

\bibitem[1994]{1994ApJ...429..694L}
Lada, C.J., Lada, E.A., Clemens, D.P. \& Bally, J. 1994, ApJ, 429, 694

\bibitem[2005]{2005A&A...438..169L}
Lombardi, M. 2005, A\&A, 438, 169

\bibitem[2001]{2001A&A...377.1023L}
Lombardi, M. \& Alves, J. 2001, A\&A, 377, 1023

\bibitem[1990]{1990ARA&A..28...37M}
Mathis, J.S. 1990, ARA\&A, 28, 37

\bibitem[1997]{1997MNRAS.288..145P} 
Padoan, P., Nordlund, A. \& Jones, B.J.T. 1997, MNRAS, 288, 145 

 %
 %

 %
 %

\bibitem[2006]{2006AJ....131.1163S}
Skrutskie, M.F., Cutri, R.M., Stiening, R., et al. 2006, AJ, 131, 1163

\bibitem[1994]{1994ApJ...428..693W}
Williams, J.P., de Geus, E,J. \& Blitz, L. 1994, ApJ, 428, 693

\bibitem[1923]{1923AN....219..109W} 
Wolf, M. 1923, AN, 219, 109 

\end{thebibliography}
\end{document}